\documentclass[sigconf]{acmart}

\setcopyright{none} 
\acmYear{2020}

\newcommand{\method}{\textit{VisualPhishNet}}
\newcommand{\dataset}{\textit{VisualPhish}}
\newcommand{\norm}[1]{\lVert\mathbf#1\rVert}

\usepackage{tikz}

\newcommand{\mycircle}[2][black,fill=black]{\tikz[baseline=-0.5ex]\draw[#1,radius=#2] (0,0) circle ;}%

\usepackage{multirow}
\usepackage{multicol}
\usepackage{longtable}
\usepackage{multicol}
\usepackage{caption}
\usepackage{subcaption,graphicx}
\usepackage{dblfloatfix} 
\usepackage{tabularx}
\usepackage{tabulary}
\usepackage{threeparttable}
\usepackage{booktabs}
\usepackage{pifont}
\newcommand*\rot{\rotatebox{90}}
\usepackage{amsmath}
\usepackage{color}
\usepackage[linewidth=0.5pt]{mdframed}
\newcommand{\rpm}{\sbox0{$1$}\sbox2{$\scriptstyle\pm$}\raise\dimexpr(\ht0-\ht2)/2\relax\box2 }

\newcommand{\new}[1]{\textcolor{black}{#1}}

\usepackage{xcolor}

\begin{document}
\title{VisualPhishNet: Zero-Day Phishing Website Detection by \\ Visual Similarity} 
\author{Sahar Abdelnabi}
\affiliation{CISPA Helmholtz Center for Information Security}

\author{Katharina Krombholz}
\affiliation{CISPA Helmholtz Center for Information Security}

\author{Mario Fritz}
\affiliation{CISPA Helmholtz Center for Information Security}

\begin{abstract}
Phishing websites are still a major threat in today's Internet ecosystem. 
Despite numerous previous efforts,  
similarity-based detection methods do not offer sufficient protection for the trusted websites -- in particular against unseen phishing pages.
This paper contributes \method{}, a new similarity-based phishing detection framework, based on a triplet Convolutional Neural Network (CNN). 
\method{} learns profiles for websites in order to detect phishing websites by a \new{similarity metric} that can generalize to pages with new visual appearances.
We furthermore present \dataset{}, the largest dataset to date that facilitates visual phishing detection in an ecologically valid manner. 
We show that our method outperforms previous visual similarity phishing detection approaches by a large margin while being robust against a range of evasion attacks.
\end{abstract}



\maketitle

\section{Introduction} \label{intro}
Phishing pages impersonate legitimate websites without permission~\cite{whittaker2010large} to steal sensitive data from users causing major financial losses and privacy violations~\cite{corona2017deltaphish,jain2017phishing,thomas2017data,khonji2013phishing}.
Phishing attacks have increased due to the advances in creating phishing kits that enabled the deployment of phishing pages on larger scales~\cite{corona2017deltaphish,oest2018inside}. According to the Anti-Phishing Working Group (APWG)~\cite{APWG}, an international association aiming at fighting phishing attacks, 266,387 attempts have been reported in the third quarter of 2019, which is a high level that has not been witnessed since 2016~\cite{APWG}.

There have been numerous attempts to combat the threats imposed by phishing attacks by automatically detecting phishing pages. Modern browsers mostly rely on blocklisting~\cite{sheng2009empirical} as a fundamentally reactive mechanism. However, in a recent empirical study~\cite{oest2019phishfarm}, the new phishing pages that used cloaking techniques were found to be both harder and slower to get detected by blocklists which motivates the development of proactive solutions. An example of the latter is using heuristics that are based on monitored phishing pages~\cite{khonji2013phishing}. These heuristics can be extracted from URL strings~\cite{blum2010lexical,nguyen2014novel,zouina2017novel} or HTML~\cite{chou2004client,li2019stacking} to detect anomalies between the claimed identity of a webpage and its features~\cite{pan2006anomaly}. However, since phishing attacks are continuously evolving, these heuristics are subject to continuous change and might not be effective in detecting future attacks~\cite{zhang2007cantina,jain2017phishing} (e.g. more than two-thirds of phishing sites in 3Q 2019 used SSL protection~\cite{APWG}, its absence formerly was used as a feature to detect phishing pages~\cite{pan2006anomaly}). 
\begin{figure}[!t]
\begin{center}
\includegraphics [width=\linewidth,height=3.5cm,keepaspectratio]{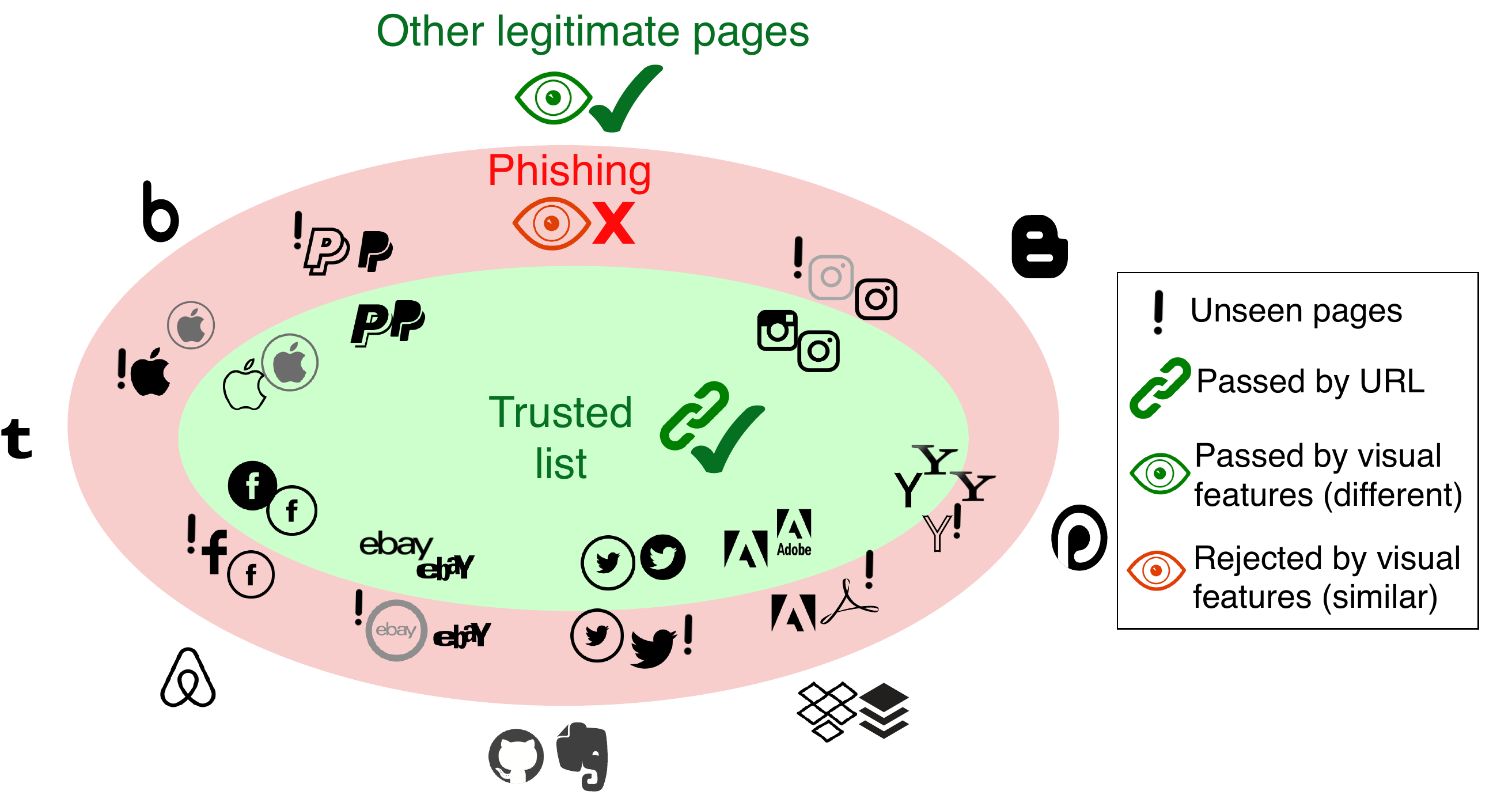} \end{center} \caption{
   Trusted pages are granted based on their URLs. The remaining pages are compared to the trusted pages by a learnt visual similarity metric. Pages that that are too similar are rejected, which even allows detecting phishing pages with new visual appearances.}
\label{fig:whitelist_teaser}
\end{figure}

Since the key factor in deceiving users is the high visual similarity between phishing pages and their corresponding legitimate ones, detecting such similarity was used in many previous detection studies~\cite{jain2017phishing}. In these methods, a list of commonly attacked pages is maintained, (domain names and screenshots), to protect users from the potential impersonation of such pages; whenever a user visits a page that is not in the trusted-list, its content is compared against the trusted ones. If a high visual similarity is detected, it is then classified as a phishing page as it impersonates one of the trusted pages. Similarity-based methods have the advantage of not relying on heuristics that are likely to evolve and instead they rely on the strong incentive of the adversary to design pages that are similar to trustworthy websites. This makes them less prone to an arms race between defenders and attackers. 

These efforts still have limitations. First, their trusted-list are too small in both the number of websites and pages per website (e.g. 4-14 websites in~\cite{mao2013baitalarm,mao2017phishing,dalgic2018phish,dunlop2010goldphish,chen2009fighting}, less than 10 pages in~\cite{fu2006detecting,wenyin2005detection,zhang2011textual}, 41 pages in~\cite{medvet2008visual}) which makes them able to detect attacks against these few pages only. Second, existing approaches fall short in detecting phishing pages that target the same trusted websites but with new unseen visual appearances, as they perform a page-to-page image matching between a previously found phishing page and its legitimate counterpart~\cite{jain2017phishing,fu2006detecting,lam2009counteracting,rao2015computer,bozkir2016use}. Consequently, attackers can bypass detection by using other pages from the targeted websites or by crafting partially similar phishing pages with different background pictures, advertisements, or layout~\cite{chen2010detecting,rao2015computer,fu2006detecting}.

\paragraph{\textbf{Contribution}} Our work targets the above limitations and focuses on improving and generalizing the image-based visual similarity detection.
First, we present \dataset{}\footnote{https://s-abdelnabi.github.io/VisualPhishNet/}, the \textbf{largest dataset to date} (155 trusted websites with 9363 pages), that we constructed \textbf{to mitigate the limitations} of previously published datasets, facilitate \textbf{visual phishing detection} and improve the ecological validity when evaluating phishing detection frameworks. 

Second, we propose \method{}, a similarity-based detection model that is the first to utilize a deep learning approach (in particular, triplet convolutional neural networks) to learn a \textbf{robust visual similarity} metric between any two same-website webpages' screenshots, instead of relying on one-to-one matching, which \textbf{outperforms prior work by a large margin}. 
A conceptual overview of our method is depicted in~\autoref{fig:whitelist_teaser}; we show a trusted-list of websites in a learnt feature space in which same-website pages have higher proximity. Additionally, phishing webpages have high visual similarity and closer embeddings to the trusted-list, thus, they would be classified as phishing. Contrarily, websites that are outside the list have genuine identities and relatively different features. 

\section{Preliminaries} \label{pre}
In this section, we briefly summarize the related similarity-based phishing detection approaches, then we introduce our threat model.
\subsection{Related Work} 
\subsubsection{Page-based similarity approaches}
The similarity between phishing and trusted pages can be inferred by comparing HTML features; Huang et al.~\cite{huang2010mitigate} extracted features that represent the text content and style (e.g. most frequent words, font name and color, etc.), which they used to compare pages against trusted identities. Similarly, Zhang et al.~\cite{zhang2007cantina} used TF-IDF to find lexical signatures which they used to find the legitimate website domain by a search engine. Besides, Liu et al.~\cite{liu2006antiphishing} segmented a webpage to blocks based on HTML visual cues and compared the layout of two pages by matching blocks. Also, Rosiello et al.~\cite{rosiello2007layout} used Document Object Model (DOM) comparison, and Mao et al.~\cite{,mao2017phishing} used Cascading Style Sheet (CSS) comparison. However, these methods fail if attackers used images or embedded objects instead of HTML text~\cite{fu2006detecting}. They are also vulnerable to code obfuscation techniques where a different code produces similar rendered images~\cite{fu2006detecting,lam2009counteracting}.

\subsubsection{Image-based similarity approaches}
Consequently, another line of work (which we adopt) infers similarity directly from rendered screenshots. As examples, Fu et al.~\cite{fu2006detecting} used Earth Mover’s Distance (EMD) to compute the similarity between low-resolution screenshots, which Zhang et al.~\cite{zhang2011textual} also used along with textual features. However, this required the images to have the same aspect ratio~\cite{lam2009counteracting}, which is a constraint we do not impose. Also, Lam et al.~\cite{lam2009counteracting} used layout similarity by matching the screenshots' segmentation blocks. However, the proposed segmentation approach is limited when segmenting pages with complex backgrounds~\cite{bozkir2016use}. Our approach does not suffer from these limitations since we use an end-to-end framework to represent images rather than a heuristic-based one. In addition, Chen et al.~\cite{chen2010detecting} approximated human perception with Gestalt theory to determine the visual similarity of two webpages' layouts with slight differences (e.g. an addition or removal of a block). They evaluated their approach on only 12-16 legitimate pages and their corresponding spoofed ones. In contrast to these approaches, we generalize the similarity detection and show that our method is not limited to phishing pages with a similar layout to the corresponding trusted ones.

Discriminative keypoint features were often used in phishing detection. As examples, Afroz et al.~\cite{afroz2011phishzoo} used Scale-Invariant Feature Transform (SIFT) to match logos, while Rao et al.~\cite{rao2015computer} used Speeded-Up Robust Features (SURF) to match screenshots. Similarly, Bozkir et al.~\cite{bozkir2016use} used Histogram of Oriented Gradients (HOG), Chen et al.~\cite{chen2009fighting} used Contrast Context Histogram (CCH), and Malisa et al.~\cite{malisa2017detecting} used Oriented FAST and rotated BRIEF (ORB) to detect mobile applications spoofing. Besides, Medvet et al.~\cite{medvet2008visual} used color histograms and 2D Haar wavelet transform of screenshots. However, in recent years, CNNs were shown to significantly outperform local and hand-crafted features in computer vision tasks~\cite{krizhevsky2012imagenet,sharif2014cnn}. Thus, our work is the first to use deep learning in pixel-based visual similarity phishing detection and to study the adversarial perturbations against such models.

Chang et al.~\cite{chang2013phishing} and Dunlop et al.~\cite{dunlop2010goldphish} used logo extraction to determine a website's identity and then used the Google search engine to find corresponding domains. These approaches assumed a fixed location for the website logo which could be bypassed. Contrary to these approaches, we use a learning-based identification of the discriminating visual cues and study the performance against shifts in location.

Woodbridge et al.~\cite{woodbridge2018detecting} used Siamese CNNs to detect visually similar URLs by training on URLs rendered as images. In contrast, we propose a visual similarity metric based on screenshots instead of URL pairs, with further optimizations adapting to the harder problem, which goes beyond homoglyph attacks.

Additionally, despite previous efforts, our work explores new territory in similarity detection research with more generalization and fewer constraints; previous methods aim to form a match between a found phishing attempt and its correspondent real page assuming a highly similar layout and content. Therefore, a phishing page targeting the same website but is different from the trusted pages could go undetected. In addition, same-website pages show a lot of variations in background pictures and colors which attackers might exploit to continuously create new pages. Thus, our model and dataset collection do not rely on page-to-page matching, but on learning a similarity metric between any two same-website pages, even with different contents, to proactively generalize to partially similar, obfuscated, and unseen pages.

\subsection{Threat Model}
We consider phishing pages targeting the collected large list of trusted websites. We assume that the attacker would be motivated to target websites that are widely known and trusted, therefore, high coverage of phishing pages could be achieved by the collected trusted-list. We assume that the attacker could craft the phishing page to be fully or partially similar to any page from the targeted websites (not only to pages in the trusted-list), therefore, we relax the page-to-page matching and test on phishing pages that were not seen in the trusted websites' training pages.
We study other evasion techniques (hand-crafted and white-box adversarial perturbations) that introduce small imperceptible noise to the phishing page to reduce the similarity to the targeted page that might be contained in the trusted-list. 
For all these attempts, we assume that the adversary has an incentive to create seemingly trusted pages by not introducing very perceptible noise on the page that might affect the perceived design quality or the website's identity (e.g. large changes to logos and color themes).

\section{Analyses and Limitations of Published Datasets}
\label{prevdatasets}
In this section, we discuss public datasets and their limitations along with the contributions of the \dataset{}{} dataset.

Unfortunately, only a small number of datasets for the phishing detection task using screenshots are publicly available. One of these is \textit{DeltaPhish}~\cite{corona2017deltaphish} for detecting phishing pages hosted within compromised legitimate websites. The dataset consists of groups having the same domain, where each group contains one phishing page and a few other benign pages from the compromised hosting website. Thus, the legitimate examples only cover the hosting websites, not the websites spoofed by the phishing pages. Consequently, this dataset is not suitable for similarity-based detection. 
Moreover, we observed that a large percentage of phishing pages' screenshots in this dataset are duplicates since PhishTank\footnote{https://www.phishtank.com/} reports do not necessarily contain unique screenshots. We also found that the legitimate and phishing examples had different designs as phishing examples generally consisted of login forms with few page elements, while legitimate examples contained more details.
This could cause the trained model to be biased to these design changes and, thus, could fail when tested with legitimate pages with login forms.

The Phish-IRIS dataset~\cite{dalgic2018phish} for similarity-based detection consists of phishing pages collected from PhishTank targeting 14 websites and an ``other'' class collected from the Alexa top 300 websites\footnote{https://www.alexa.com} representing legitimate examples outside the \new{trusted-list}. However, this dataset has a limited number of \new{trusted} websites, and the screenshots of the \new{trusted-list} were taken only from phishing reports which skews the dataset towards poorly designed phishing pages.

\paragraph{\dataset{} contributions} Based on the previously mentioned limitations, we collected the \dataset{} dataset that facilitates similarity-based detection approaches and closes the following gaps: 1) we increased the size of the \new{trusted-list} to detect more phishing attacks. 2) we collected a phishing webpage corpus with removing duplicity in screenshots. 3) instead of only training on phishing pages, we also collected legitimate pages of the targeted websites with different page designs and views (i.e. training \new{trusted-list}). 4) the dataset is not built on a page-to-page basis but on a per-website basis; the \new{trusted-list} contains screenshots from the whole website, phishing pages that target the \new{trusted} website are considered even if their counterparts are not found in the \new{trusted-list}. 5) we collected a legitimate test set of websites (i.e. \new{different from trusted domains}) that limits bias as far as possible (e.g. login forms should also be well represented in this test set). 

\begin{figure}[!b]
\begin{center}
\includegraphics [width=\linewidth,height=3.5cm,keepaspectratio]{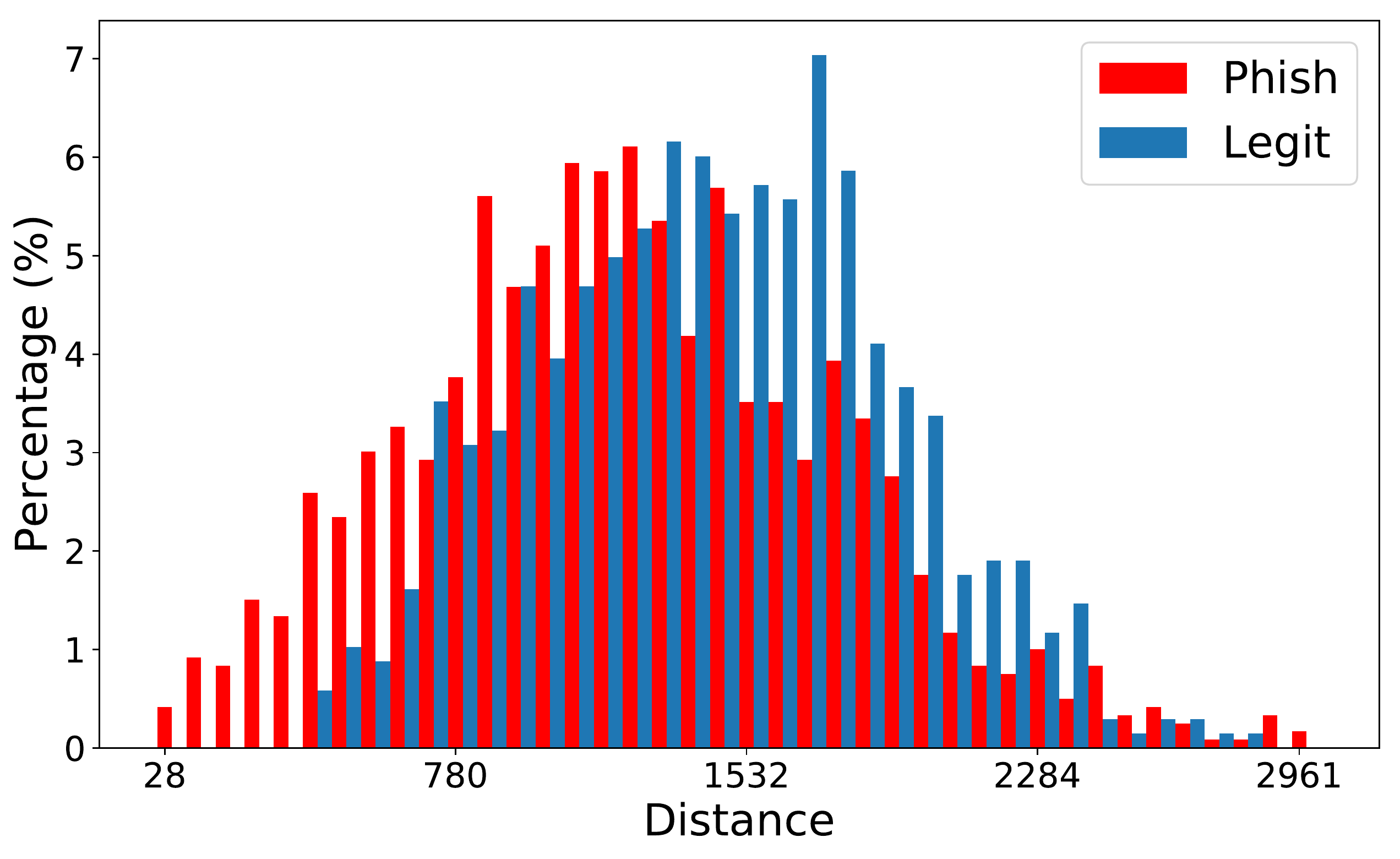}
\end{center}
   \caption{The distances histogram between the pre-trained VGG16 features of the phishing test set and the targeted website in the training \new{trusted-list} (red), in comparison with the ones between the benign test set and the \new{trusted-list} (blue).} \label{fig:vgg_features}
\end{figure} 

Unlike previous work, we extend the visual similarity to phishing pages that target the \new{trusted} websites but were not seen in the training \new{trusted-list}. Thus, we checked that the collected phishing pages are different in terms of simple pixel-wise similarity from the targeted \new{trusted} websites' pages. To denote pixel-wise similarity, we used the distances between the pre-trained VGG16 visual representation instead of naive pixel comparison. We computed the minimum distances between the phishing pages and the corresponding targeted website. As a reference, we compared them to the distances between the legitimate test set (other websites) and the \new{trusted-list}. If the phishing pages had similar counterparts in the \new{trusted-list}, they would have considerably smaller distances compared to other benign pages. However, as can be seen from the two histograms in~\autoref{fig:vgg_features}, the distance ranges in both sets are comparable with high overlap. Hence, the phishing pages are different from the training \new{trusted} websites' ones and can be used to evaluate the performance on future \new{unseen} phishing pages.

\section{The \dataset{} Dataset} \label{dataset} 
In this section, we show how we constructed \dataset{}.
\paragraph{Phishing pages} To collect the phishing examples, we crawled and saved the screenshots of the active verified phishing pages from PhishTank which yielded 10250 pages. We observed that the same phishing screenshot design could be found with multiple URLs, so we manually inspected the saved screenshots to remove duplicates in addition to removing not found and broken URLs. Having non-duplicated screenshots (i.e. unique visual appearance) is important to have an accurate error estimate and to have a disjoint and non-overlapping training and test splits. After filtering, the phishing set contained 1195 phishing pages targeting 155 websites. We observed that phishing pages targeting one website have differences in elements' locations, colors, scales, text languages and designs (including previous websites' versions), therefore, the phishing set can be used to test the model's robustness to these variations. We also found that some phishing pages are poorly designed with little similarity to the overall design of the targeted website, in addition to having templates that cannot be found in the website but in other applications (e.g. Microsoft Word or Excel files). Such dissimilar examples were excluded from previous work (such as~\cite{mao2017phishing}), however, we included all found pages for completeness and to provide a rich dataset for future research. Examples of these variations are in~\autoref{appendix_dataset}. Additionally, the majority of the crawled pages targeted a small subset \new{of the trusted} websites (a histogram is in~\autoref{appendix_dataset}), therefore, even though \new{similarity} methods cannot detect attacks against \new{non-listed} websites, high coverage of phishing pages could be achieved by including a few websites in the \new{trusted-list}.

\paragraph{Targeted legitimate websites' pages} Besides collecting phishing webpages, we collected legitimate pages from those 155 targeted websites to work as a visual \new{trusted-list}. Instead of only gathering the legitimate counterparts of the found phishing pages as typically done in previous work, we crawled all internal links that were parsed from the HTML file of the homepage. As a result, not all phishing pages have corresponding similar legitimate pages in this \new{trusted-list}. We saved all webpages from the website to get different page designs, possible login forms, and different languages to make the similarity model trained with this dataset robust against these differences. For these 155 websites, we collected 9363 screenshots, where the number of collected screenshots for each website depends on the number of hyperlinks found in the homepage.

\paragraph{Top-ranked legitimate websites' pages}
Furthermore, we queried the top 500 ranked websites from Alexa, the top 100 websites from SimilarWeb\footnote{https://www.similarweb.com/}, in addition to the top 100 websites in categories most prone to phishing such as banking, finance, and governmental services. In total, we collected a list of 400 websites from SimilarWeb. From these lists, we excluded the 155 websites we collected from the phishing pages' targets, and then we downloaded the screenshots of the top $\approx$60 websites (non-overlapping) from each list.

\paragraph{Training and test pages split}
We have three data components: a training \new{trusted-list} of legitimate pages, phishing pages targeting the websites in the \new{trusted-list}, and legitimate/benign test examples of websites outside the \new{trusted-list} (i.e. different domains). Our objective is to differentiate the phishing pages from other benign examples based on their similarity to the \new{trusted-list}. 

To train the model, we used the first legitimate set that we built from the phishing pages' targets (155 websites) as a \new{trusted-list} that is used only in training. We used a subset of the phishing examples in training as a form of augmentation in order to learn to associate the dissimilar examples to their targets. We do not train on any other legitimate websites (i.e. domains) outside the \new{trusted-list}.

To test the model, we used the rest of the phishing set. In addition, we constructed a legitimate test set of 683 benign examples from the top-ranked websites' pages that we crawled (with domains different from the \new{trusted-list}); we selected 3-7 screenshots from each website. In order not to have a biased dataset with inherent differences between the legitimate and phishing test sets that might give optimistic or spurious results, we rigorously constructed the legitimate test set such that it contains an adequate number of forms and categories that are used in phishing attacks (e.g. banks, Software as a Service (SaaS), and payment~\cite{APWG}). With a well-balanced test set, we can accurately evaluate the similarity model performance and whether it can find the website identity instead of relying on other unrelated features such as the page layout (e.g. having forms). Additionally, we included other categories (histogram in~\autoref{appendix_dataset}) to have high coverage of websites users might face.

\begin{figure}[!b]
\begin{center}
\includegraphics [width=\linewidth,height=3.5cm,keepaspectratio]{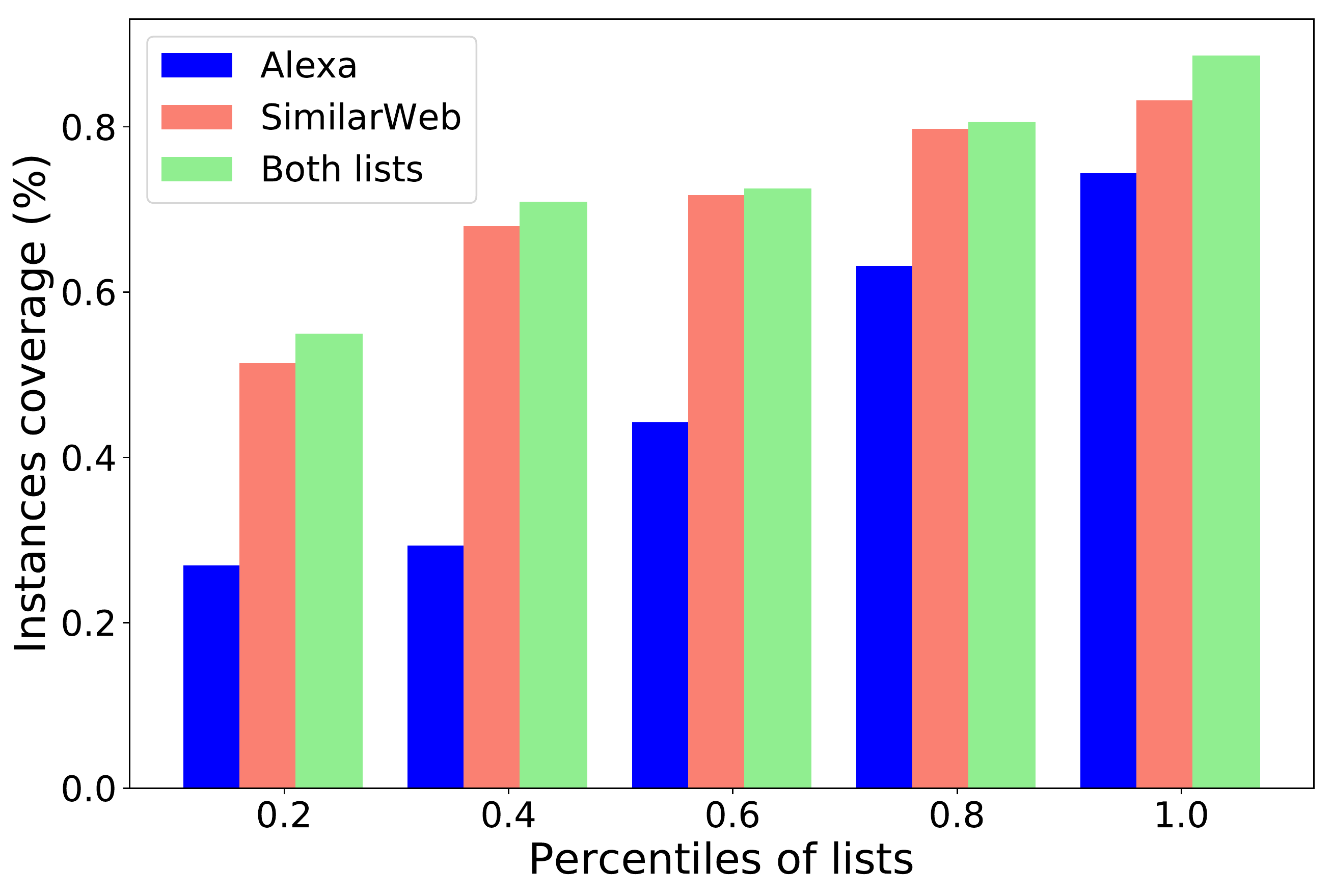}
\end{center}
   \caption{Percentage of phishing instances whose targets are covered by ascending percentiles of other lists.}
\label{fig:alexa_sw_hist}
\end{figure}

\paragraph{\new{Trusted-list} analysis}
In addition to the \new{trusted-list} we built from PhishTank, we also examined other sources for building \new{trusted-lists} without needing to crawl phishing data. This could help in taking proactive steps to protect websites that might be attacked in the future if the adversary decided to avoid detection by targeting other websites than the ones which have been already known to be vulnerable. In order for the attacks to succeed, attackers have an incentive to target websites that are trusted and known for a large percentage of users, therefore, we built our analysis on the top 500 websites from Alexa, and the top 400 websites from SimilarWeb in categories most prone to phishing. To evaluate whether or not these lists can represent the targets that might be susceptible to attacks, we computed the intersection between them and the PhishTank \new{trusted-list}. \autoref{fig:alexa_sw_hist} shows cumulative percentages of phishing instances whose targets are included in ascending percentiles of the Alexa, SimilarWeb, and the concatenation of both lists. We found that including both lists covered around 88\% of the phishing instances we collected from PhishTank, which indicates that the top-ranked websites are relevant for constructing \new{trusted-lists}. Additionally, SimilarWeb list covered more instances than Alexa list, we accounted that for the fact that the former was built from categories such as banks, SaaS and payment, in addition to the general top websites. We, therefore, conclude that this categorization approach is more effective in forming potential \new{trusted-lists} since important categories are less likely to change in future attacks.
\begin{figure*}[!t]
\begin{center}
\includegraphics [width = \textwidth, height=3.8cm,keepaspectratio]{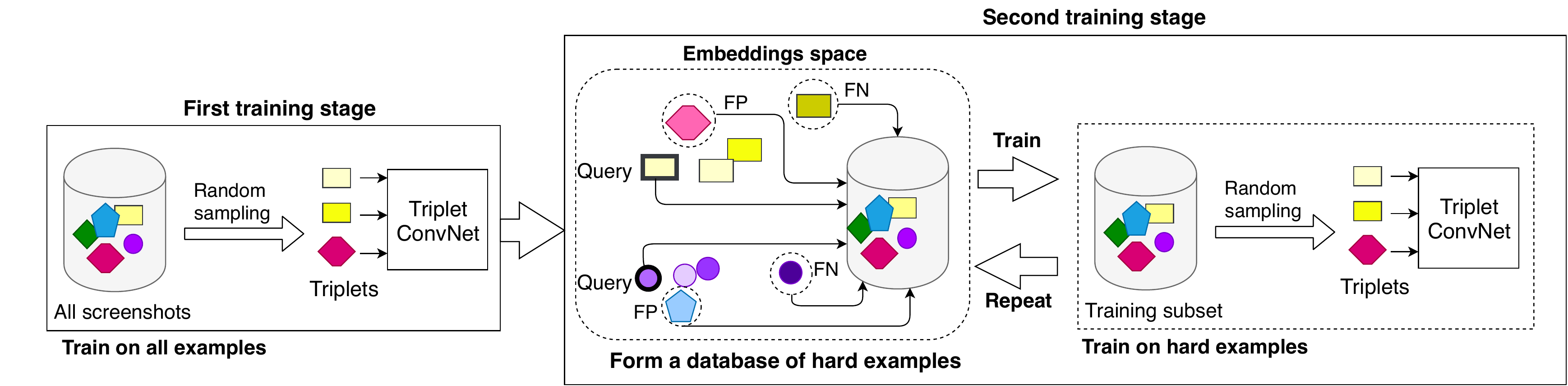}
\end{center}
\caption{An overview of \method{}. We utilize triplet networks with convolutional sub-networks to learn similarity between same-website screenshots (same shaped symbols), and dissimilarity between different-website screenshots. Our network has two training stages; first, training is performed with uniform random sampling from all \new{trusted-list}'s screenshots. Second, training is performed by iteratively finding hard examples according to the model's latest checkpoint. } 
\label{fig:model}
\end{figure*}

\section{VisualPhishNet}
As we presented in~\autoref{fig:whitelist_teaser}, similarity-based phishing detection is based on whether there
is a high visual similarity between a visited webpage to any of the \new{trusted} websites,
while having a different domain. If the visited page was found to be not similar enough
to the \new{trusted-list}, it would be classified as a legitimate page with a genuine identity. Therefore, our objective can be considered as a similarity learning problem rather than a multi-class classification between \new{trusted-list}'s websites and an ``other'' class. Including
a subset of ``other'' websites in training with a multi-class classification method could cause the model to fail at test time when testing with new websites. Additionally, instead of the typically used page-to-page correspondence, we aim to learn the similarity between any two same-website pages despite having different contents. 

Motivated by these reasons, we treated the problem as a similarity learning problem with deep learning using Siamese or triplet networks which have been successfully used in applications such as face verification~\cite{taigman2014deepface}, signature verification~\cite{dey2017signet}, and character recognition~\cite{koch2015siamese}. In each of these applications, the identity of an image is compared against a database and the model verifies if this identity is matched with any of those in the database. They have been also used in the tasks of few-shots learning or one-shot learning~\cite{koch2015siamese} by learning a good representation that encapsulates the identity with few learning examples. These reasons make this deep learning paradigm suitable for similarity-based phishing detection. 

Our network, \method{}, adopts the triplet network paradigm with three shared convolutional networks. We show an overview of the training of \method{} in~\autoref{fig:model} which consists of two stages: in the first stage, training is performed on all screenshots with a random sampling of examples. The second training stage fine-tunes the model weights by iteratively training on hard examples that were wrongly classified by the model's last checkpoint according to the distance between the embeddings. By learning these deep embeddings, we build a profile for each website that encapsulates its identity, which would enable us to \new{generalize to new webpages} that are not contained in the \new{trusted-list} database. The rest of this section illustrates in more detail each aspect of the \method{} model.
\subsection{Triplet Networks}
The Siamese networks are two networks with shared weights trained with the goal of learning a feature representation of the input such that similar images have higher proximity in the new feature space than different images. The sub-networks shares weights and parameters and the weight updates are mirrored for each of them, the sub-networks are then joined with a loss function that minimizes the distance of similar objects' embeddings while maximizing the distance of dissimilar objects' ones~\cite{dey2017signet}. 

The triplet network, which we used in \method{}, extends this approach; it was initially used in the FaceNet system~\cite{schroff2015facenet} to learn an embedding for the face verification task. This type of architectures performs the training on three images, an anchor image, a positive image whose identity is the same as the anchor, and a negative image with a different identity than the anchor. The overall objective of the network is to learn a feature space in which the distance between the positive and anchor images' embeddings is smaller than the distance between the anchor and negative images' ones. This is achieved by minimizing the loss function that is 
\resizebox{0.47\textwidth}{!}{
$\text{Loss}=\sum\limits_{\substack{i}}^N\;\max(\; \norm{f(x_i^a) - f(x_i^p)}_2^2 - \norm{f(x_i^a) - f(x_i^n)}_2^2 + \alpha,0 \;)$ }

where: $f(x)$ represents the embedding space (produced by a shared network), $(x_i^a,x_i^p,x_i^n)$ is a set of possible triplets (anchor, positive, and negative), and $\alpha$ is a margin that is enforced between positive and negative pairs which achieves a relative distance constraint. The loss penalizes the triplet examples in which the distance between the anchor and positive images is not smaller by at least the margin $\alpha$ than the distance between the anchor and negative images. In our problem, the positive image is a screenshot of the same website as the sampled anchor, and similarly, the negative image is a screenshot of a website that is different from the anchor.

For the shared network, we used the VGG16 (as a standard architecture) with ImageNet pre-training initialization~\cite{simonyan2014very}. We used all layers excluding the top fully connected layers, we then added a new convolution layer of size 5x5 with 512 filters, with ReLU activations, and initialized randomly with HE initialization~\cite{he2015delving}. Instead of using a fully connected layer after the convolution layers, we used a Global Max Pooling (GMP) layer that better fits the task of detecting possible local discriminating patterns in patches such as logos.
To match the VGG image size, all screenshots were resized to 224x224 with the RGB channels. 

\subsection{Triplet Sampling}
Since there are a large number of possible combinations of triplets, the training is usually done based on sampling or mining of triplets instead of forming all combinations. However, random sampling could produce a large number of triplets that easily satisfy the condition due to having zero or small loss which would not contribute to training. Therefore, mining of hard examples was previously used in FaceNet to speed-up convergence~\cite{schroff2015facenet}.

Therefore, as we show in~\autoref{fig:model}, our training process has two training stages. In the first stage, we used a uniform random sampling of triplets to cover most combinations. 
After training the network with random sampling, we then fine-tuned the model by iteratively finding the hard examples to form a new training subset. First, we randomly sample a query set representing one screenshot from each website, then with the latest model checkpoint, we compute the L2 distance between the embeddings of the query set and all the rest of training screenshots. In this feature space, the distance between a query image and any screenshot from the same website should ideally be closer than the distance from the same query image to any image from different websites. Based on this, we can find the examples that have the largest error in distance. Hence, we retrieve the one example from the same website that had the largest distance to the query (hard positive example), and the one example from a different website that had the smallest distance to the query (hard negative example). We then form a new training subset by taking the hard examples along with the sampled query set altogether, and we continue the training process with triplet sampling on this new subset. For the same query set, we repeat the process of finding a new subset of hard examples for a defined number of iterations for further fine-tuning. Finally, to avoid overfitting to a query set that might have outliers, we repeat the overall process by sampling a new query set and selecting the training subsets for this new query set accordingly. 

This hard example mining framework can be considered as an approximation to a training scheme where a query image is paired with screenshots from all websites and a Softmin function is applied on top of the pairwise distances with a supervised label, however, this would not scale well with the number of websites in the \new{trusted-list}, and therefore it is not tractable in our case as a single training example would have 155 pairs (\new{trusted} websites).

\subsection{Prediction}
At test time, the closest screenshot in distance to a phishing test page targeting a website should ideally be a screenshot of the same website. Therefore, the decision is not done based on all triplets comparison but it can be done by finding the screenshot with the minimum distance to the query image. To this end, we use the shared network to compute the embeddings then we compute the L2 distance between the embeddings of the test screenshot and all training screenshots. After computing the pairwise distances, the test screenshot is assigned to the website of the screenshot that has the minimum distance. This step could identify the website targeted if the test page is a phishing page. 

As depicted in~\autoref{fig:whitelist_teaser}, if the minimum distance between a page and the \new{trusted-list} is smaller than a defined threshold, the page would be classified as a phishing page that tries to impersonate one of the \new{trusted} websites by having a high visual similarity. On the other hand, if the distance is not small enough, the page would be classified as a legitimate page with a genuine identity. Therefore, we apply a threshold on the minimum distance for classification. 

\section{Evaluation}
In this section, we first show the implementation details of \method{} and its performance, then we present further experiments to evaluate the robustness of \method{}.

\subsection{\method{}: Final Model} \label{perf}
\paragraph{Evaluation metrics} Since our method is based on the visual similarity of a phishing page to websites in the \new{trusted-list}, we computed the percentage of correct matches between a phishing page and its targeted website.
We also calculated the overall accuracy of the binary classification between legitimate test pages and phishing pages at different distance thresholds to calculate the Receiver Operating Characteristic (ROC) curve area. 

\paragraph{Implementation details} To train the network, we used Adam optimizer~\cite{kingma2014adam} with momentum values of $\beta_1=0.9$, $\beta_2=0.999$ and a learning rate of 0.00002 with a decay of 1\% every 300 mini-batches where we used a batch size of 32 triplets. We set the margin ($\alpha$) in the triplet loss to 2.2. The first stage of triplet sampling had 21,000 mini-batches, followed by hard examples fine-tuning, which had 18,000 mini-batches divided as follows: we sampled 75 random query sets, for each, we find a training subset which will be used for 30 minibatches, then we repeat this step 8 times. We used 40\% of the phishing examples in training (added to the targeted website pages and used normally in triplet sampling) and used the other 60\% for the test set. We used the same training/test split in the two phases of training. We tested the model with the legitimate test set consisting of 683 screenshots; these domains were only used in testing since we train the model on \new{trusted} domains only (and partially their spoofed pages). 

\paragraph{Performance} Using \method{}, 81\% of the phishing test pages were matched to their correct website using the top-1 closest screenshot, while the top-5 match is 88.6\%. After computing the correct matches, we computed the false positive and true positive rates at different thresholds (where the positive class is phishing) which yielded a ROC curve area of 0.9879 (at a cut-off of 1\% false positives, the partial ROC area is 0.0087) outperforming the examined models and re-implemented visual similarity approaches which we show in the following sections.

\subsection{Ablation Study} \label{ablation}
Given the results of \method{}, this sub-section investigates the effects of different parameters in the model, we summarize our experiments in~\autoref{tab:summary} which shows the top-1 match and the ROC area for each model in comparison with the final one (see~\autoref{appendix_results} for the ROC curves). We first evaluated the triplet network by experimenting with Siamese network as an alternative. We used a similar architecture to the one used in~\cite{koch2015siamese} with two convolutional networks and a supervised label of 1 if the two sampled screenshots are from the same website, and 0 otherwise. The network was then trained with binary cross-entropy loss. We also examined both L1 and L2 as the distance function used in the triplet loss. 
Besides, we inspected different architecture's parameters regarding the shared sub-network including the added convolution layer, and the final layer that is used as the embedding vector where we experimented with Global Average Pooling (GAP)~\cite{lin2013network}, fully connected layer, and taking all spatial locations by flattening the final feature map. In addition to VGG16, we evaluated ResNet50 as well~\cite{he2016deep}. We also studied the effect of the second training phase of hard examples training by comparing it with a model that was only trained by random sampling. 
\begin{table} [!b]
\centering
\resizebox{0.93\linewidth}{!}{%
\begin{tabular}{lllllll | ll}
\toprule
\rot{Sub-network} & \rot{Added Layer} & \rot{Last Layer} & \rot{Network type} & \rot{Distance} & \rot{Sampling} & \rot{\%Phishing} & \rot{Top-1 Match} & \rot{ROC Area} \\  \midrule
VGG16 & Conv 5x5(512) & GMP & Triplet & L2 & 2 stages & 40\% & \textbf{81.03\%} & \textbf{0.9879}\\ \hline 
\mycircle{1.5pt}& \mycircle{1.5pt} & \mycircle{1.5pt} & Siamese & \mycircle{1.5pt} & \mycircle{1.5pt} & \mycircle{1.5pt} & 75.31\% & 0.8871\\
\mycircle{1.5pt} & \mycircle{1.5pt} & FC (1024) & Siamese & L1 & \mycircle{1.5pt} & \mycircle{1.5pt} & 64.8\% & 0.655\\
\mycircle{1.5pt} & \mycircle{1.5pt} & \mycircle{1.5pt} & \mycircle{1.5pt} & L1 & \mycircle{1.5pt} & \mycircle{1.5pt} & 73.91\% & 0.9739\\
\mycircle{1.5pt} & \mycircle{1.5pt} & GAP & \mycircle{1.5pt} & \mycircle{1.5pt} & \mycircle{1.5pt} & \mycircle{1.5pt} & 68.61\% & 0.6449\\
\mycircle{1.5pt} & \mycircle{1.5pt} & FC (1024) & \mycircle{1.5pt} & \mycircle{1.5pt} & \mycircle{1.5pt} & \mycircle{1.5pt} & 78.94\% & 0.8517\\
\mycircle{1.5pt} & \mycircle{1.5pt} & Flattening & \mycircle{1.5pt} & \mycircle{1.5pt} & \mycircle{1.5pt} & \mycircle{1.5pt} & 80.05\% & 0.8721\\
\mycircle{1.5pt} & Conv 3x3(512) & \mycircle{1.5pt} & \mycircle{1.5pt} & \mycircle{1.5pt} & \mycircle{1.5pt} & \mycircle{1.5pt} & 80.19\% & 0.9174\\
\mycircle{1.5pt} & No new layer  & \mycircle{1.5pt} & \mycircle{1.5pt} & \mycircle{1.5pt} & \mycircle{1.5pt} & \mycircle{1.5pt} & 79.91\% & 0.8703\\
ResNet50 & No new layer  & \mycircle{1.5pt} & \mycircle{1.5pt} & \mycircle{1.5pt} & \mycircle{1.5pt} & \mycircle{1.5pt} & 78.52\% & 0.8526\\
\mycircle{1.5pt} & \mycircle{1.5pt}  & \mycircle{1.5pt} & \mycircle{1.5pt} & \mycircle{1.5pt} & Random & \mycircle{1.5pt} & 75.3\% & 0.9477\\ \hline
\mycircle{1.5pt} & \mycircle{1.5pt}  & \mycircle{1.5pt} & \mycircle{1.5pt} & \mycircle{1.5pt} & \mycircle{1.5pt} & 20\% & 74.37\% &0.9899\\
\bottomrule
\end{tabular}}
\caption{A summary of the ablation study. Row 1 is the finally used model, cells indicated by "\mycircle{1.5pt}" denotes the same cell value of row 1 (\method{}).} \label{tab:summary}
\end{table}
As can be seen from~\autoref{tab:summary}, the triplet network outperformed the Siamese network. Also, the second training phase of hard examples improved the performance, which indicates the importance of this step to reach convergence as previously reported in~\cite{schroff2015facenet}. We also show that the used parameters in \method{} outperform the other studied parameters.
Motivated by the observation that some phishing pages had poor quality designs and were different from their targeted websites (see~\autoref{appendix_dataset} for examples), we studied the robustness of \method{} to the ratio of phishing examples seen in training. We, thus, reduced the training phishing set to only 20\% and tested with the other 80\%, which slightly decreased the top-1 match (mostly on these different examples).
\subsection{\new{Trusted-list} Expansion}
\begin{figure}[!b]
    \centering
    \includegraphics [width =\linewidth, height=3.5cm,keepaspectratio]{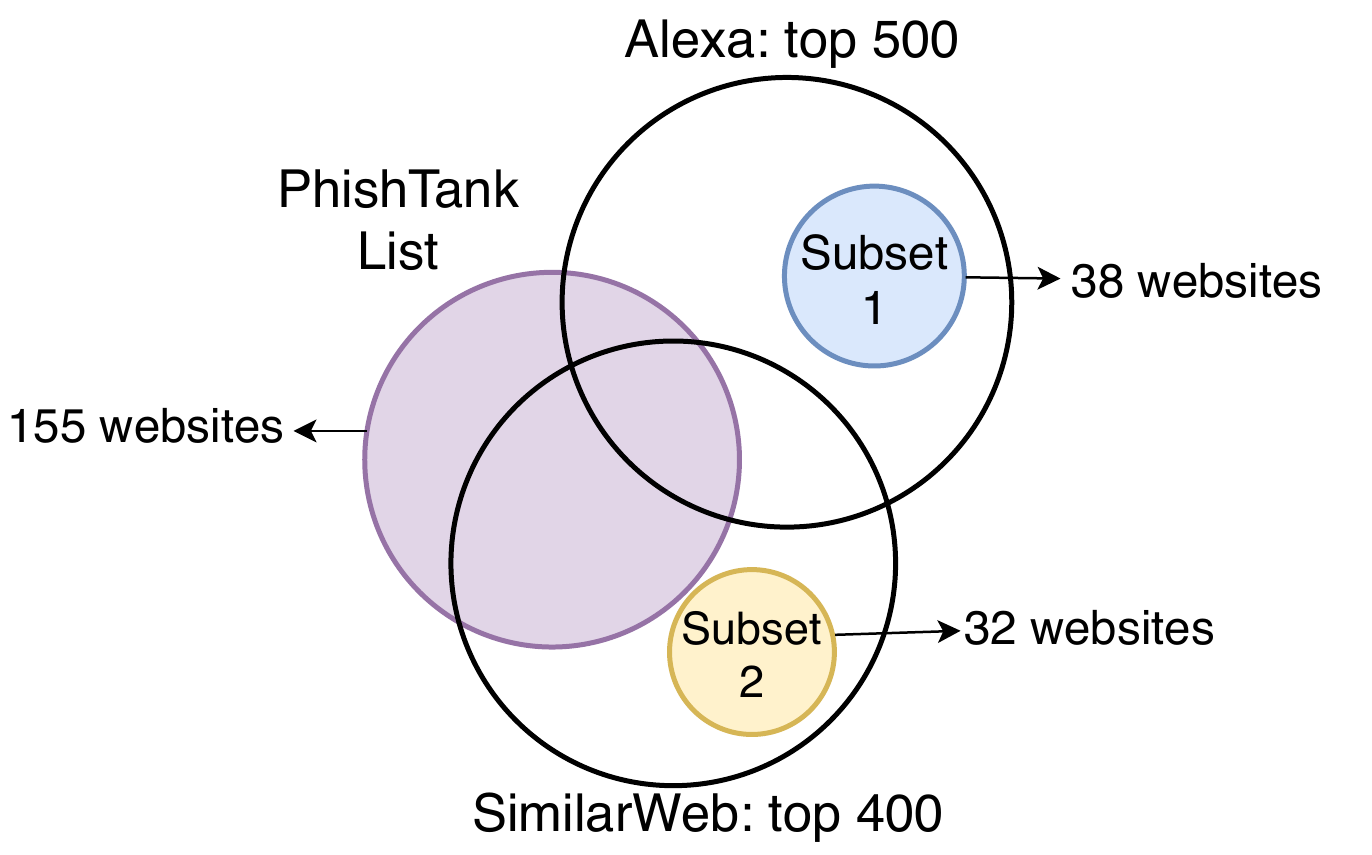}
    \caption{The three main lists used in training, the \new{list} collected from PhishTank, a subset of Alexa list, and a subset of SimilarWeb list.} \label{fig:lists}
\end{figure}

In addition to the PhishTank \new{list} gathered from phishing reports, we studied other sources of \new{trusted-lists} as per the analysis presented earlier in our dataset collection procedure. We then studied the robustness of \method{}'s performance when adding new websites to the training \new{trusted-list}. To that end, we categorized the training websites to three lists (as shown in~\autoref{fig:lists}), the PhishTank \new{list}, a subset containing 32 websites from SimilarWeb top 400 list (418 screenshots), a subset containing 38 websites (576 screenshots) from Alexa top 500 list. Since we have phishing pages for the websites in the PhishTank \new{list} only, the other two lists can be used in training as distractors to the performance on the phishing examples. When training on one of these additional lists, we remove its websites from the legitimate test set yielding test sets of 562 and 573 screenshots in the case of adding SimilarWeb and Alexa lists respectively.

\begin{table} [!b]
\begin{center}
\resizebox{0.85\linewidth}{!}{%
\begin{tabular}{l|ll}
\toprule
Experiment & Top-1 Match & ROC Area \\  \midrule 
PhishTank list (155 websites) & 81.03\% & 0.9879 \\ 
Add SimilarWeb list (32+155 websites) & 78.3\% & 0.9764 \\ 
Add Alexa list (38+155 websites) & 78.1\% & 0.9681\\ \bottomrule
\end{tabular}}
\caption{A summary of our experiments when adding more websites from Alexa and SimilarWeb lists to training.} \label{tab:distractors}
\end{center}
\end{table}
As shown in~\autoref{tab:distractors}, when adding new websites to the training \new{trusted-list}, the performance of the classification (indicated by the ROC area and the top-1 match) decreased. However, this decrease in performance was relatively slight, which indicates the robustness of \method{} to adding a few more websites to training.
\begin{table} [!b]
\begin{center}
\resizebox{0.55\linewidth}{!}{%
\begin{tabular}{l|ll}
\toprule
Method & Top-1 Match  & ROC Area \\  \midrule 
\textbf{\method{}} & \textbf{81.03\%} & \textbf{0.9879} \\ \hline
VGG16 & 51.32\% & 0.8134 \\
ResNet50 & 32.21\% & 0.7008 \\ \hline
ORB &  24.9\% & 0.6922 \\
HOG &  27.61\% & 0.58 \\ 
SURF & 6.55\% & 0.488 \\ \bottomrule
\end{tabular}}
\caption{Our experiments to compare \method{}'s performance against prior methods and alternative baselines.} \label{tab:baselines}
\end{center}
\end{table}
\begin{figure*}[!ht]
  \centering
  \begin{minipage}{0.45\textwidth}
  \centering
    \includegraphics[width=\textwidth,height=3.2cm,keepaspectratio]{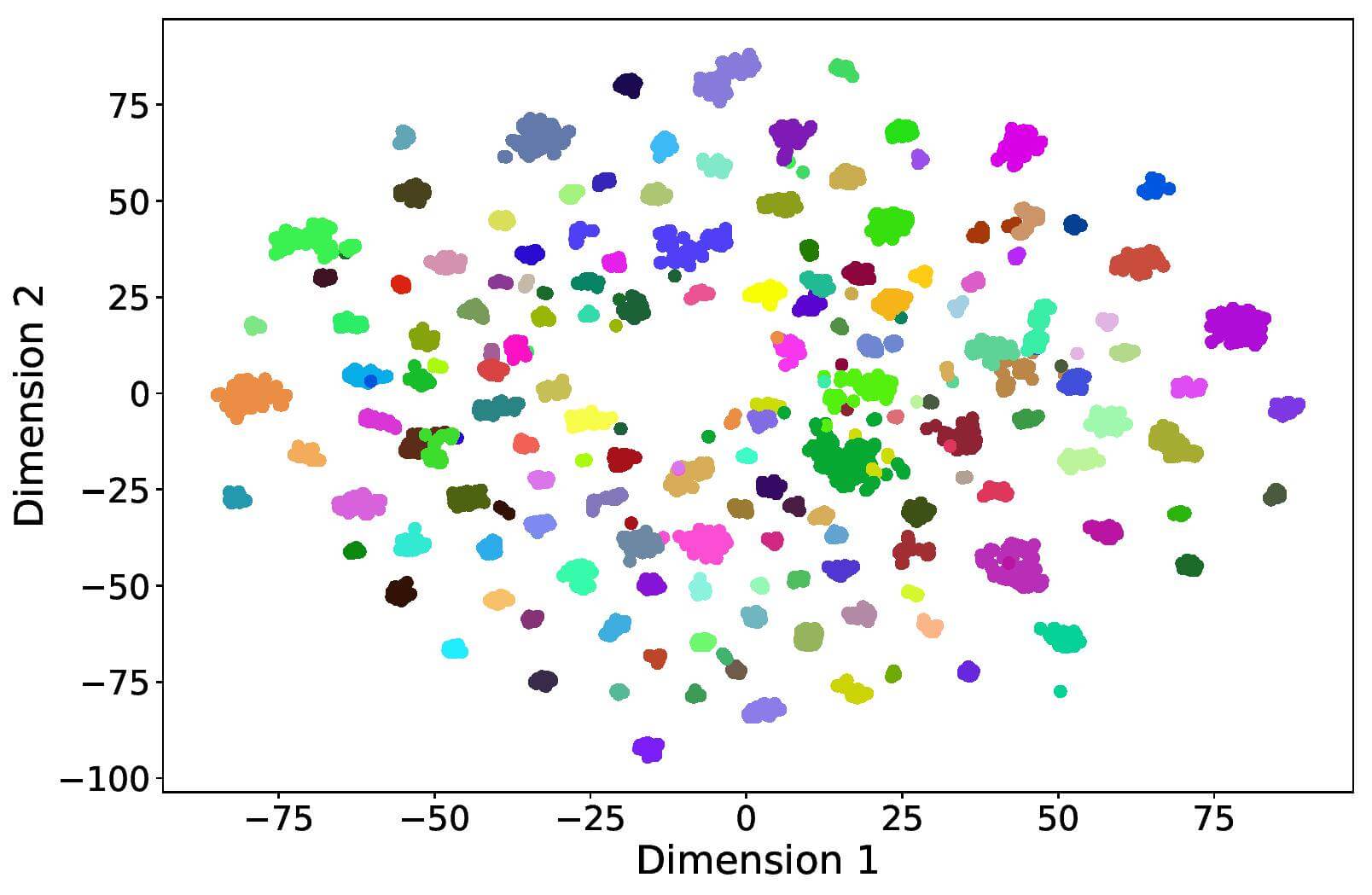}
    \subcaption{\method{}} \label{fig:white_websitess}
  \end{minipage}
  \hfill
  \begin{minipage}{0.51\textwidth}
  \centering
    \includegraphics[width=\textwidth,height=3.2cm,keepaspectratio]{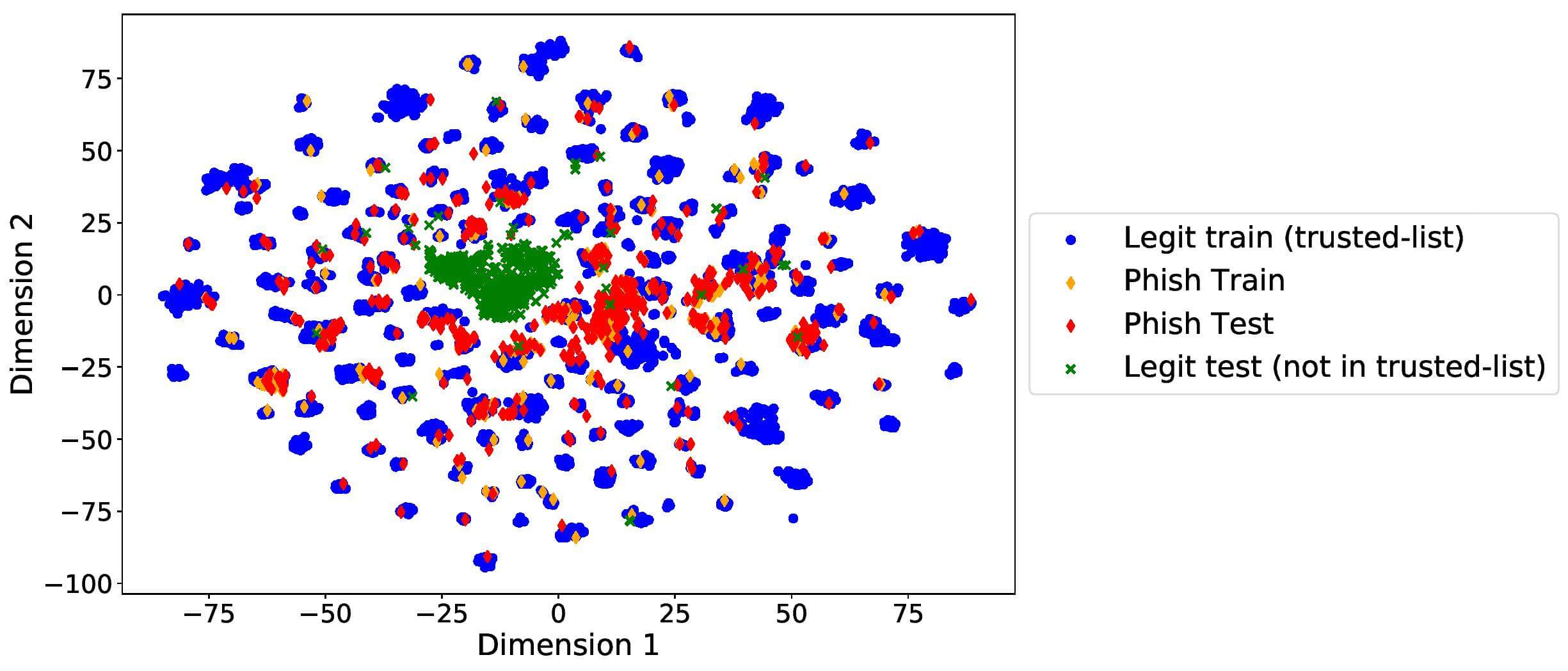}
    \subcaption{\method{}} \label{fig:white_sets}
  \end{minipage}
  \allowbreak
  \begin{minipage}{0.45\textwidth}
  \centering
    \includegraphics[width=\textwidth,height=3.2cm,keepaspectratio]{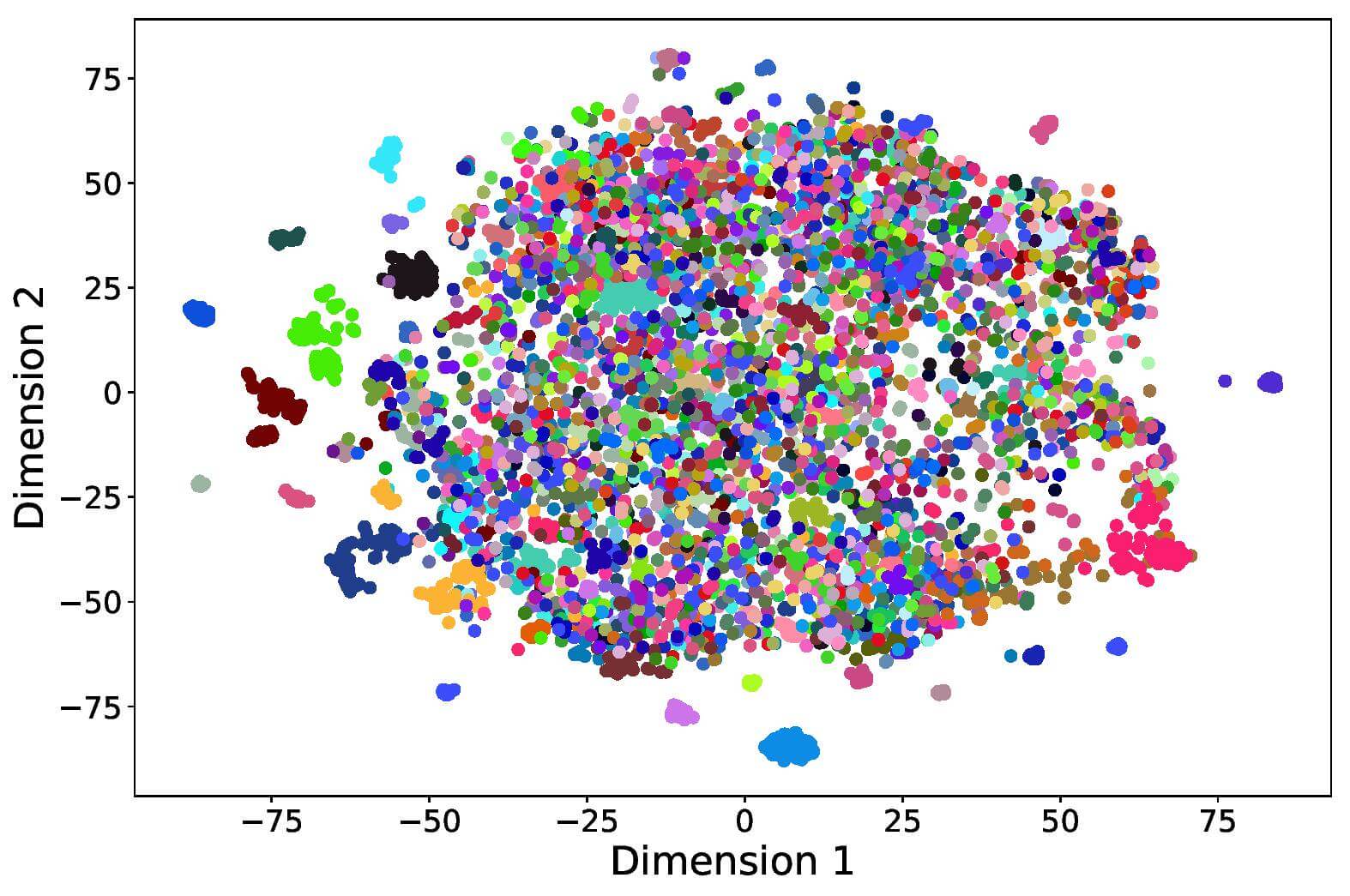}
    \subcaption{VGG16} \label{fig:vgg_websites}
  \end{minipage}
  \hfill
  \begin{minipage}{0.51\textwidth}
    \centering
    \includegraphics[width=\textwidth,height=3.2cm,keepaspectratio]{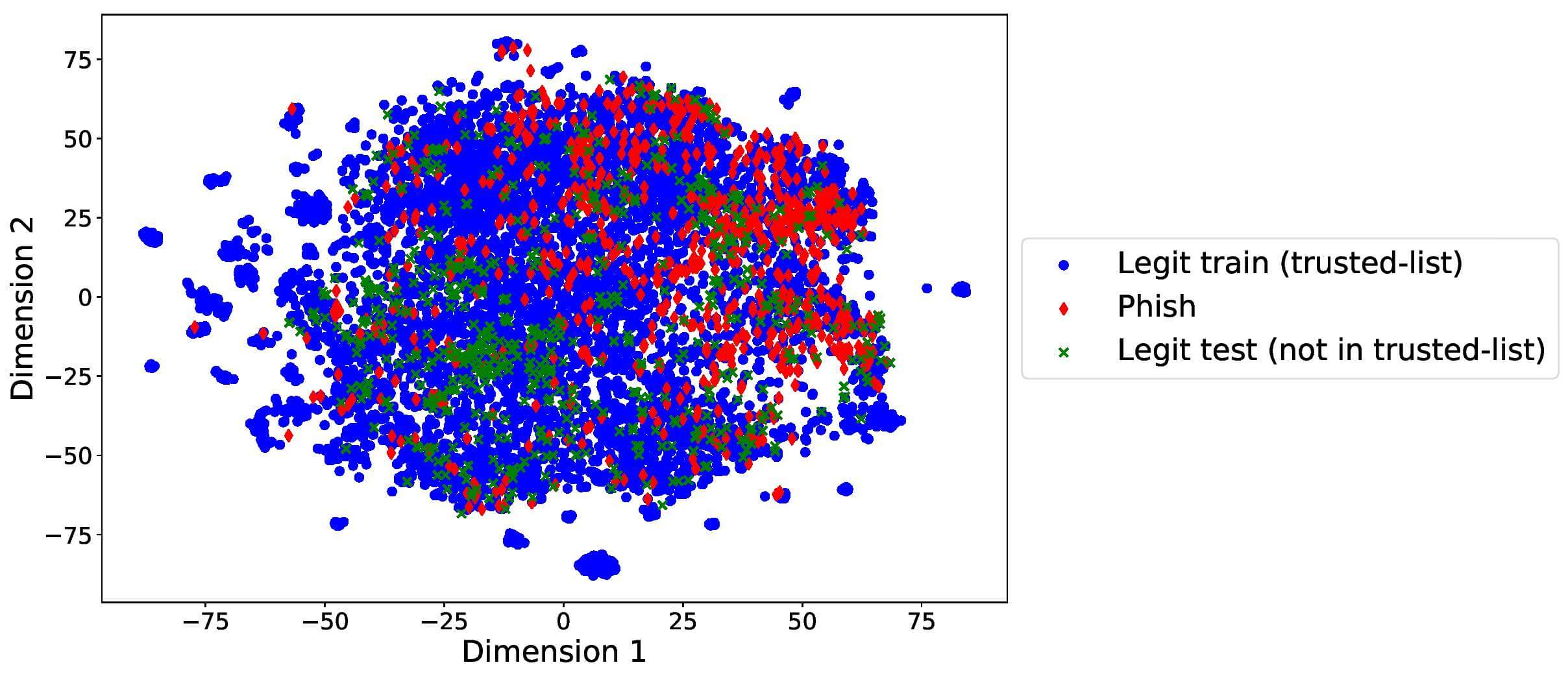}
    \subcaption{VGG16} \label{fig:vgg_sets}
  \end{minipage}
  \hfill
  \caption{t-SNE visualizations of \method{}'s embeddings compared with the pre-trained VGG16 ones as a baseline. Figures (a) and (c) show \new{the trusted} webpages color-coded by websites. Figures (b) and (d) show \new{the trusted} webpages (blue) and their phishing pages (red and orange) in comparison with legitimate test pages outside the \new{trusted-lists} (green).
  \label{fig:tsne}  }
\end{figure*}
\subsection{Comparison with Prior Work and Baselines} \label{prev_work}
Furthermore, we compared \method{} with alternative approaches that we re-implemented on the \dataset{} dataset. In recent years, deep learning and CNNs have been demonstrated to achieve a breakthrough over local and hand-crafted features (used in previous work) on many benchmarks~\cite{krizhevsky2012imagenet}. Moreover, off-the-shelf pre-trained CNNs features (even without fine-tuning) have been shown to outperform local features in many tasks~\cite{sharif2014cnn,yue2015exploiting,long2014convnets}. Therefore, we first compare \method{}'s embeddings to the embeddings of two off-the-shelf CNNs: VGG16 and ResNet50. Also, since our work is the first to utilize deep learning, the pre-trained CNNs provide a baseline for deep learning approaches. As we show in~\autoref{tab:baselines}, \method{} outperforms these two baselines with a significant performance gain. 

To provide additional evidence, we re-implemented the methods of phishing detection using SURF matching from~\cite{rao2015computer}, HOG matching from~\cite{bozkir2016use}, and ORB matching from~\cite{malisa2017detecting} which reported that ORB is more suited for the logo detection task than SIFT. Unlike previous work, our approach and dataset do not rely on page-to-page matching, thus, not all phishing pages have legitimate counterparts in the training \new{list}. This limits the applicability of methods that are based on layout segmentation and explicit block matching (such as~\cite{lam2009counteracting}). Nevertheless, HOG descriptors, which we compare to, were used to represent the page layout in~\cite{bozkir2016use}. As shown in~\autoref{tab:baselines}, the use of pre-trained CNNs (in particular VGG16) does indeed outperform the other baselines. In all of our experiments, similar to \method{} training for a fair comparison, 40\% of the phishing set was added to the \new{training list}.

This analysis demonstrates that previous image matching methods are not efficient on our dataset containing phishing pages whose contents and visual appearances were not seen in the \new{trusted-list} (as shown later in~\autoref{discuss_qual}). Additionally, it shows that pre-trained CNNs are not adequate and further optimization to find the discriminating cues, as done in \method{}, is needed.
\subsection{Embeddings Visualization}
\method{} produces a feature vector (dimensions: 512) for each screenshot that represents an encoding that resulted from minimizing the triplet loss. In this learned feature space, same-website screenshots should be in closer proximity compared with screenshots from different websites. To verify this, we used t-Distributed Stochastic Neighbor Embedding (t-SNE)~\cite{maaten2008visualizing} to reduce the dimensions of the embeddings vectors to a two-dimensional set. We show the visualization's results in~\autoref{fig:tsne} in which we compare the embeddings of \method{} with pre-trained VGG16 ones (as the best performing baseline). We first visualized the embeddings of the training \new{trusted-list}'s webpages categorized by websites as demonstrated in~\autoref{fig:white_websitess} and~\autoref{fig:vgg_websites} for \method{} and VGG16 respectively. As can be observed, the learned embeddings show higher inter-class separation between websites in the case of \method{} when compared with VGG16.  
Additionally,~\autoref{fig:white_sets} and~\autoref{fig:vgg_sets} show the training \new{trusted-list}'s pages in comparison with phishing and legitimate test ones for \method{} and VGG16 respectively. For successful phishing detection, phishing pages should have smaller distances to \new{trusted-list}'s pages than legitimate test pages, which is more satisfied in the case of \method{} than VGG16. 
\begin{table*}[!b]
\centering
\resizebox{0.92\linewidth}{!}{%
\begin{tabular}{c c|c|c|c|c|c|c} \toprule
 & \large{\textbf{Blurring}} & \large{\textbf{Darkening}} & \large{\textbf{Brightening}} & \large{\textbf{Gaussian noise}} & \large{\textbf{Salt and Pepper}} & \large{\textbf{Occlusion}} & \large{\textbf{Shift}} \\ \cline{2-8}  &&&&&\\
&\large{Sigma=1.5} & \large{Gamma=1.3} & \large{Gamma=0.8} & \large{Var=0.01} & \large{Noise=5\%} & \large{Last quarter} & \large{(-30,-30) pixels} \\
&\includegraphics[width=3.2cm,height=3cm,keepaspectratio] {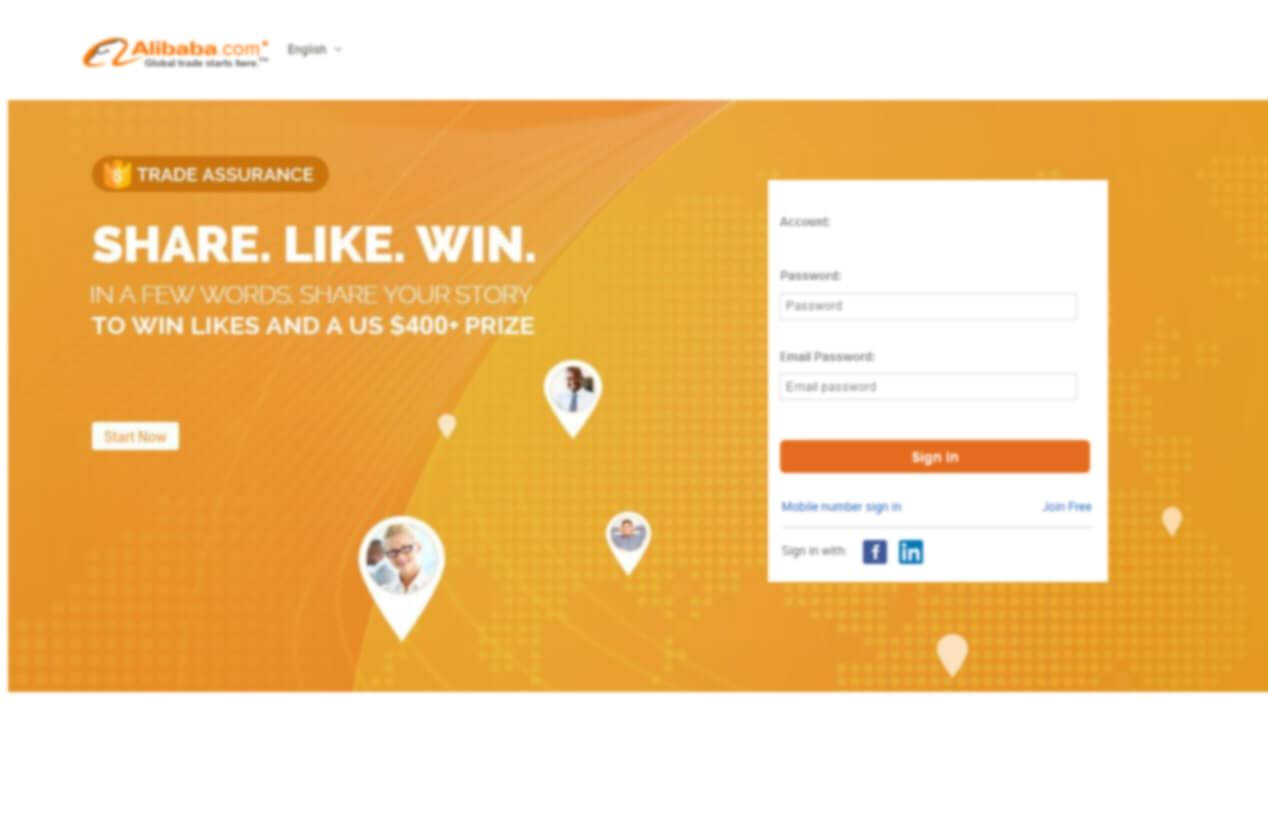} & \includegraphics[width=3.2cm,height=3cm,keepaspectratio] {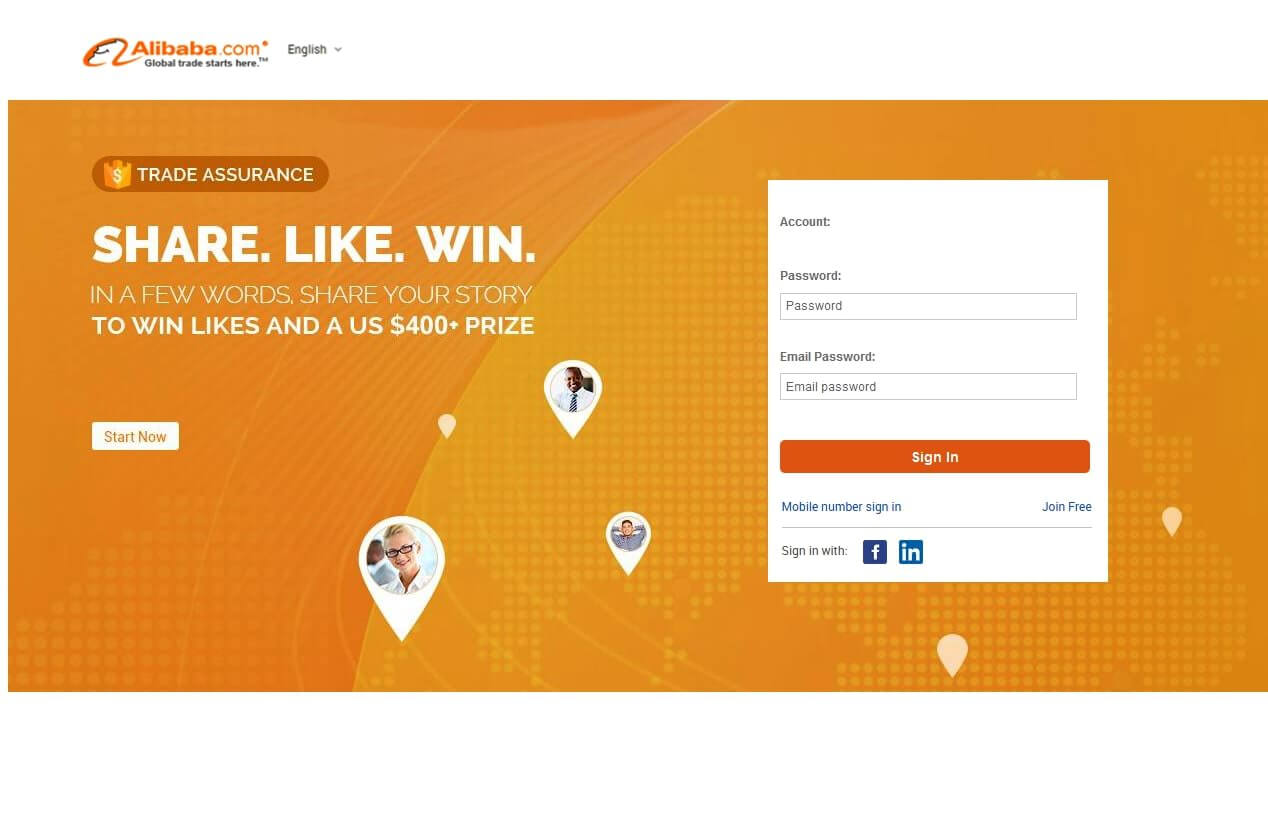} & \includegraphics[width=3.2cm,height=3cm,keepaspectratio] {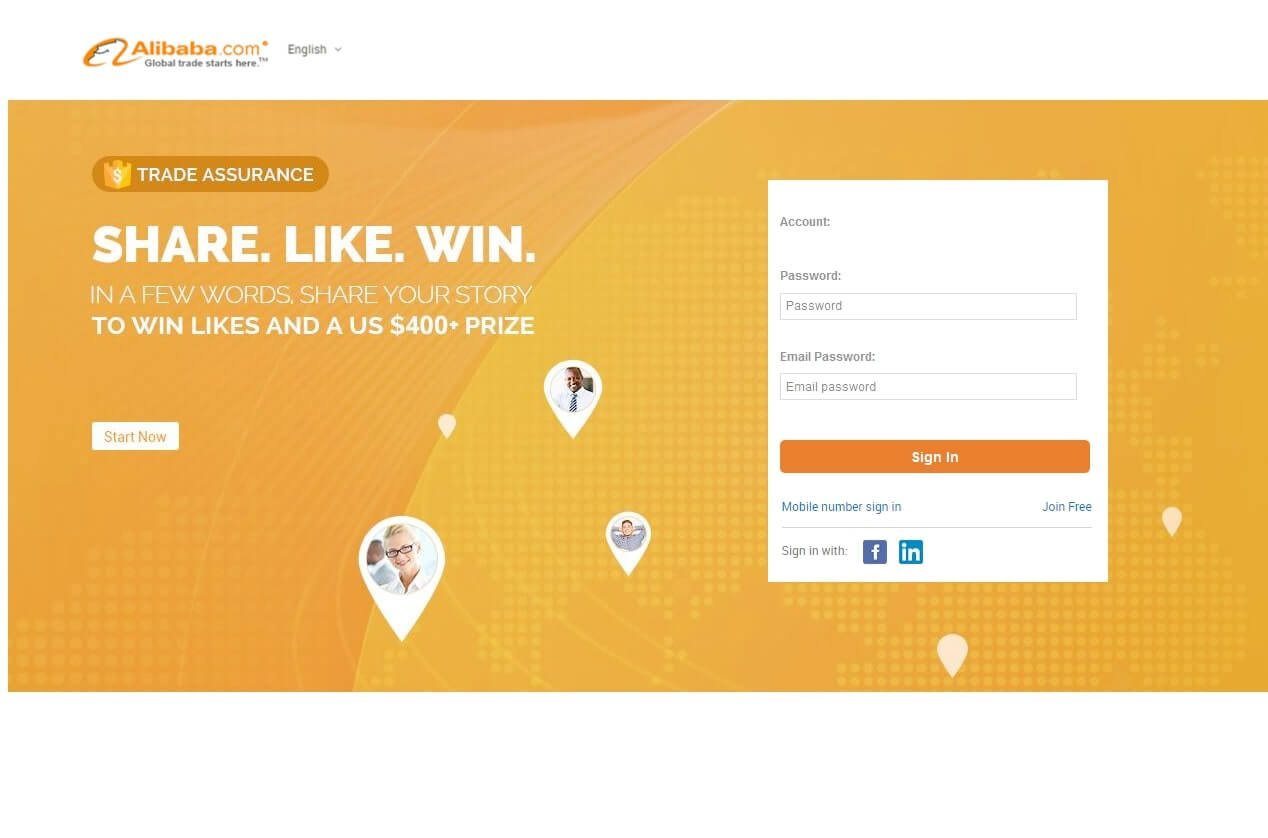} & \includegraphics[width=3.2cm,height=3cm,keepaspectratio] {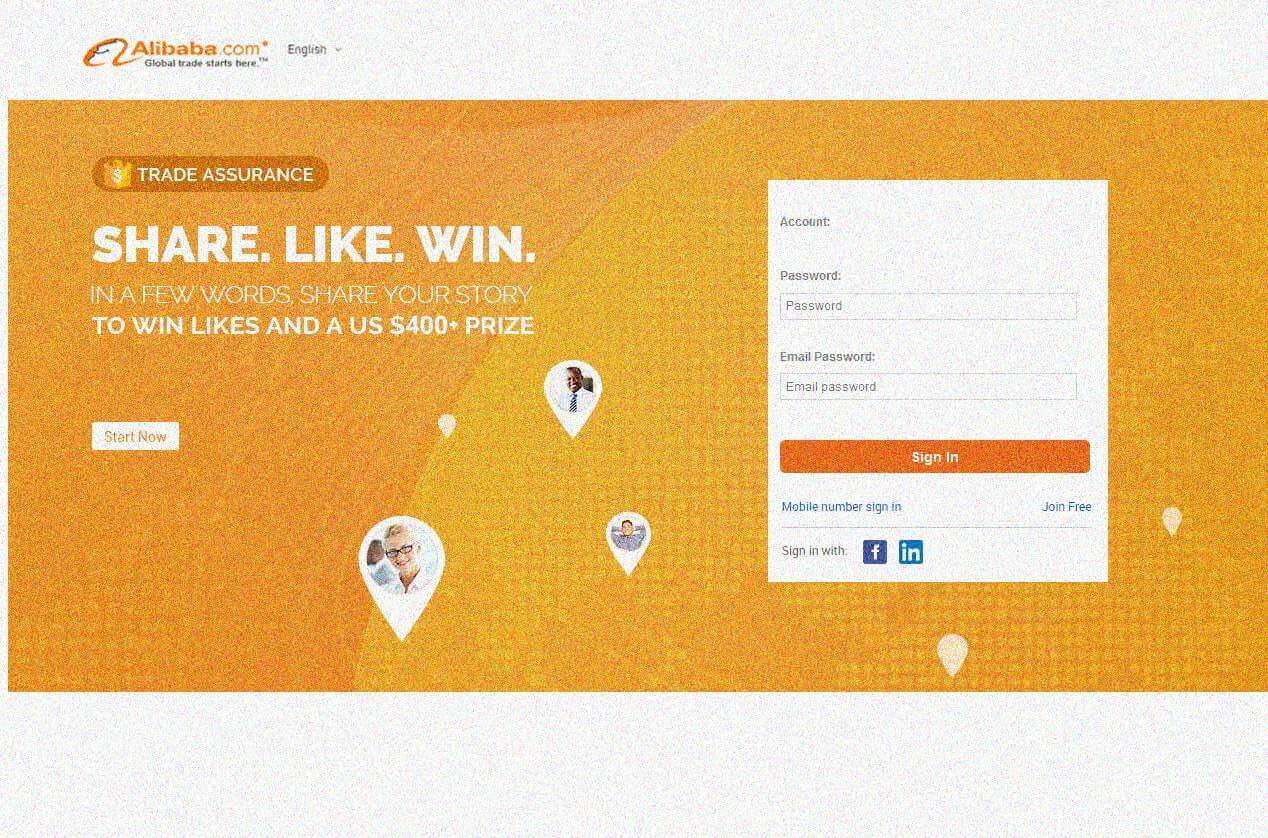} &
\includegraphics[width=3.2cm,height=3cm,keepaspectratio] {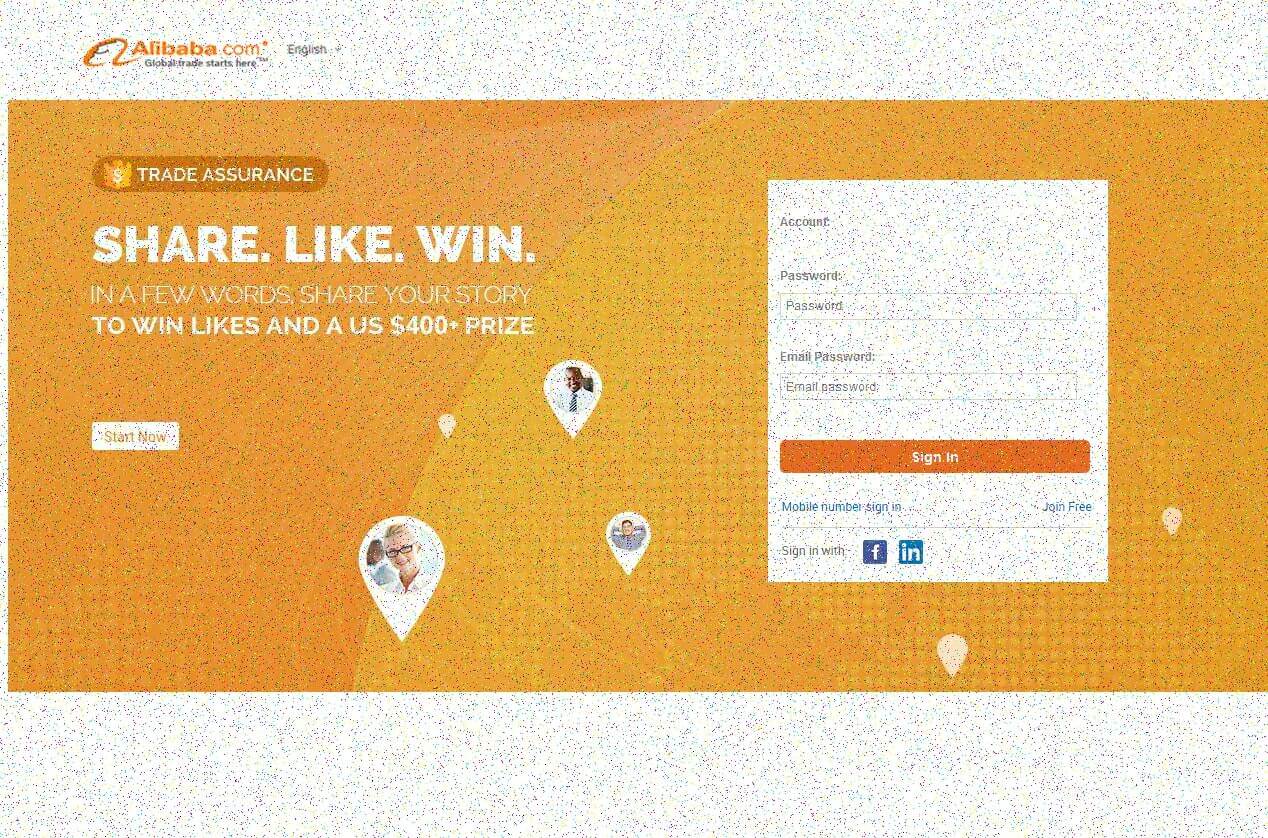} &
\includegraphics[width=3.2cm,height=3cm,keepaspectratio] {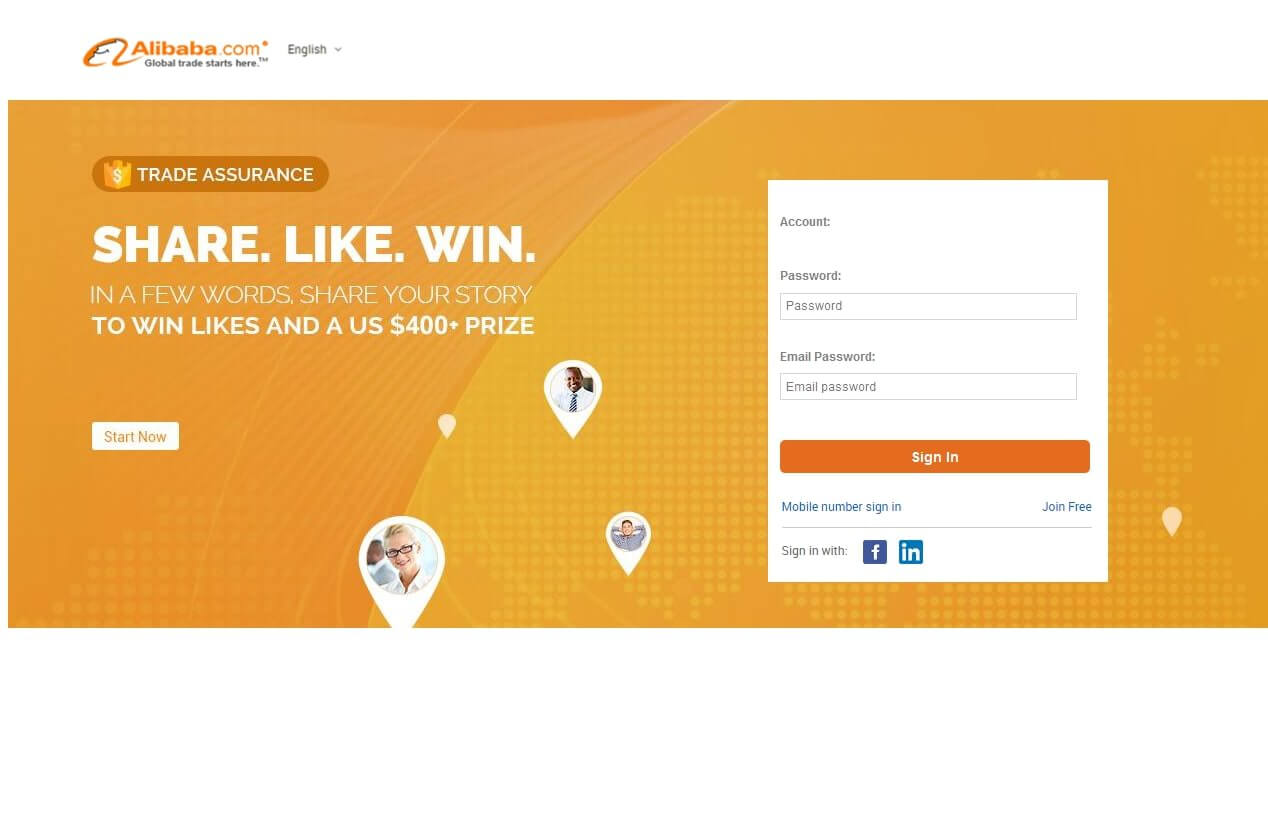} &
\includegraphics[width=3.2cm,height=3cm,keepaspectratio] {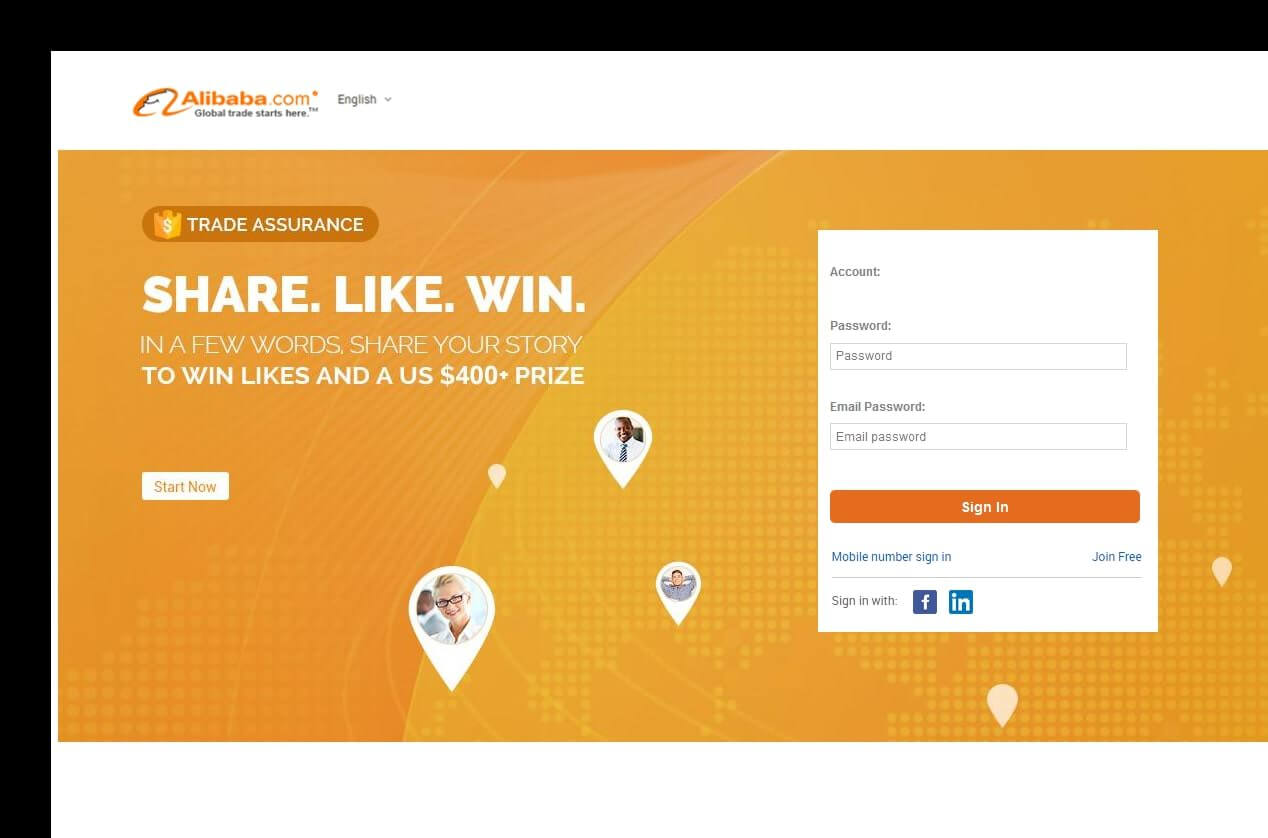} \\ 
\large{Matching drop} &\large{1.38\%} & \large{4.31\%} & \large{1.72\%} & \large{1.9\%} & \large{2.07\%} & \large{1.2\%} & \large{3.09\%} \\  
\large{ROC AUC drop} &\large{0.17\%} & \large{1.56\%} & \large{0.36\%} & \large{1.47\%} & \large{1.79\%} & \large{0.12\%} & \large{0.86\%} \\
 \cline{1-8} &&&&&\\ 
&\large{Sigma=3.5} & \large{Gamma=1.5} & \large{Gamma=0.5} & \large{Var=0.1} & \large{Noise=15\%} & \large{Second quarter} & \large{(-50,-50) pixels} \\
&\includegraphics[width=3.2cm,height=3cm,keepaspectratio] {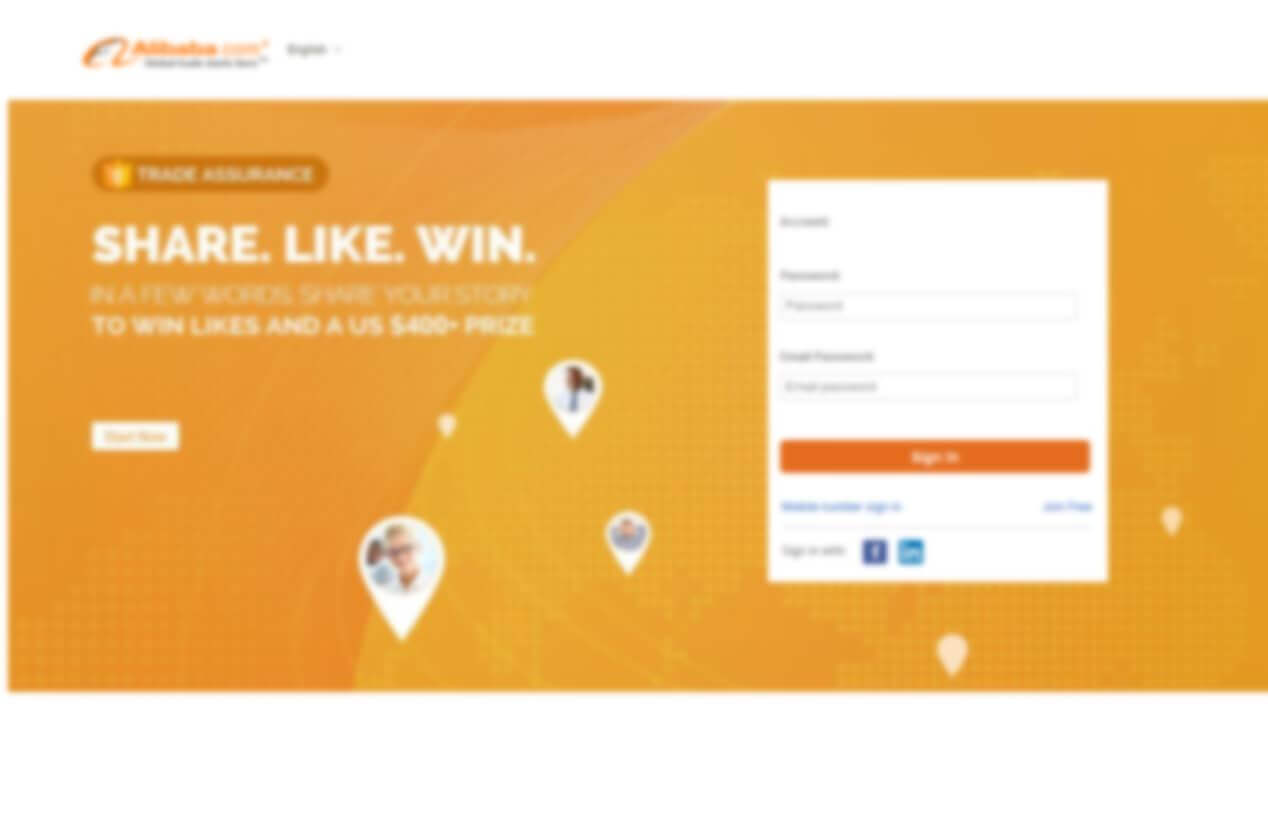} & \includegraphics[width=3.2cm,height=3cm,keepaspectratio] {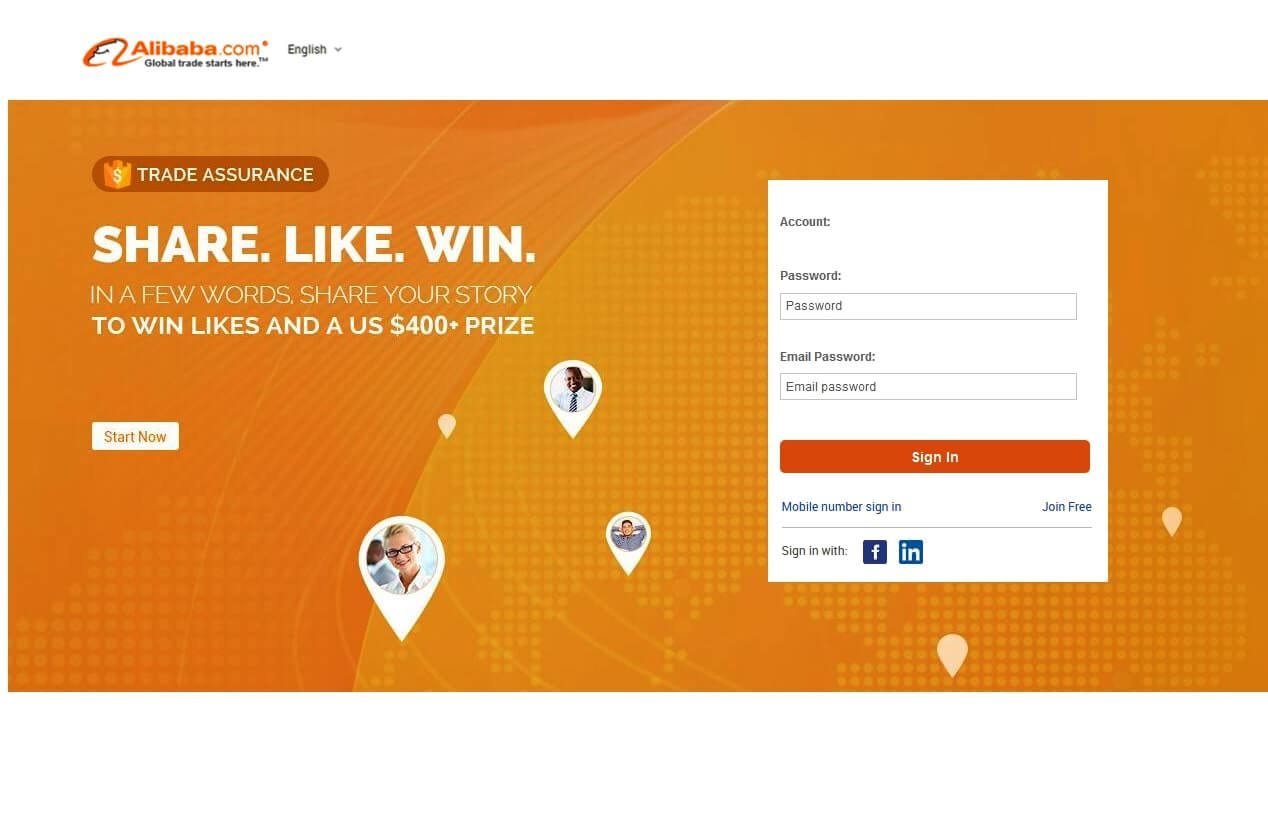} & \includegraphics[width=3.2cm,height=3cm,keepaspectratio] {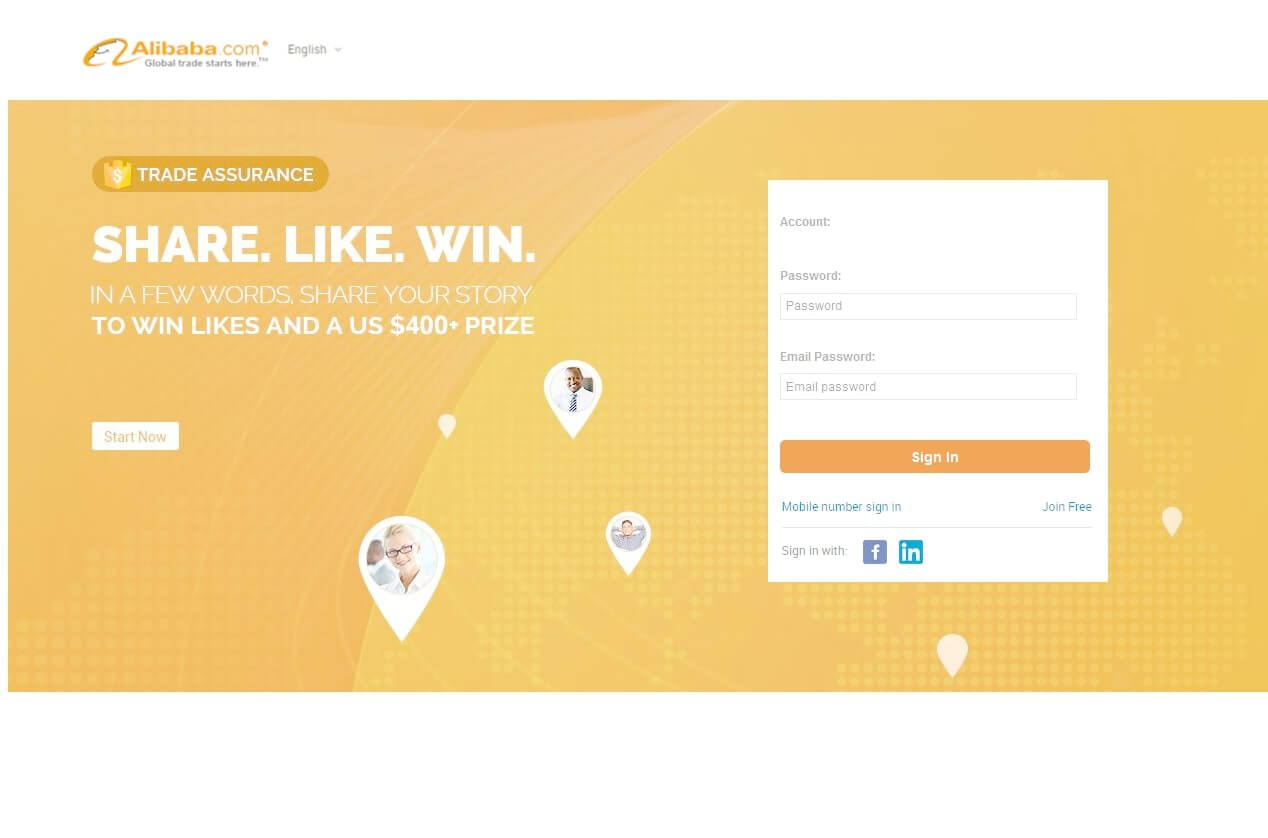} & \includegraphics[width=3.2cm,height=3cm,keepaspectratio] {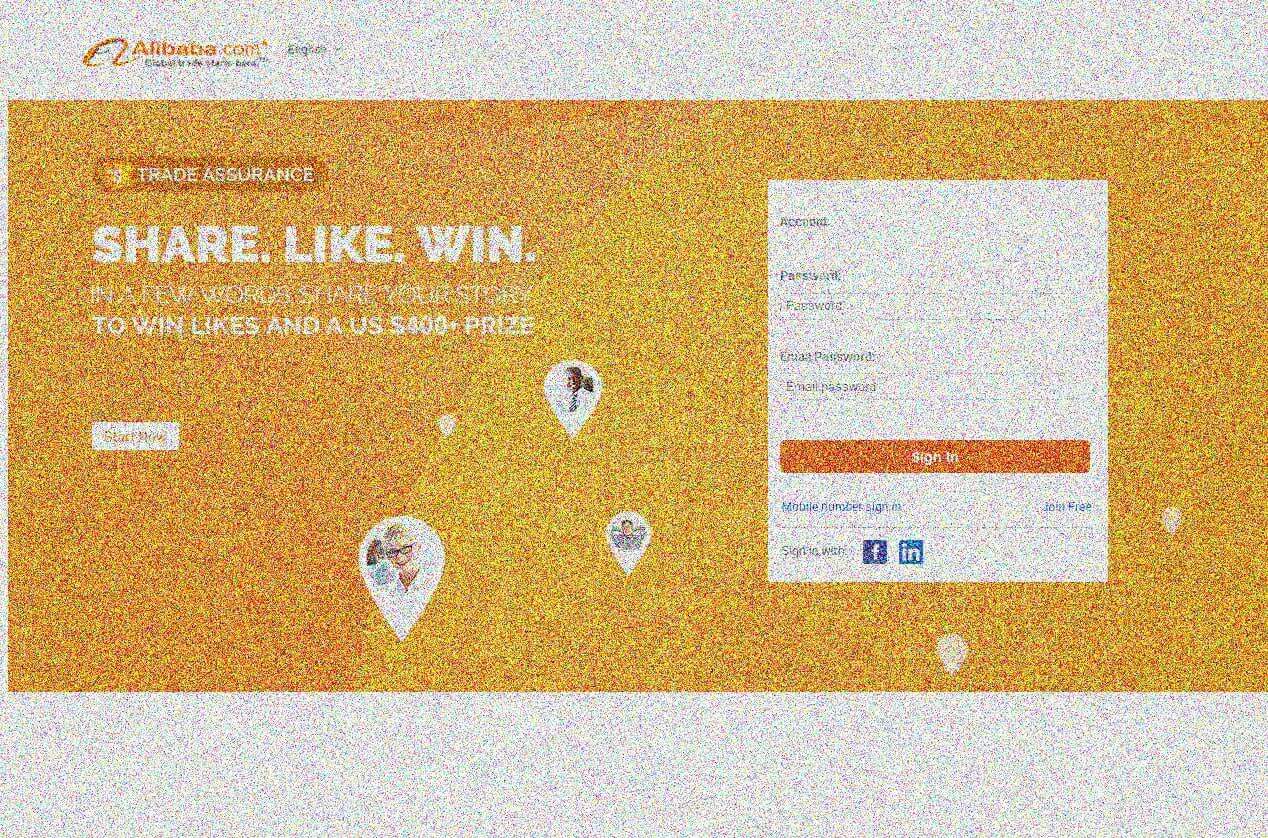} &
\includegraphics[width=3.2cm,height=3cm,keepaspectratio] {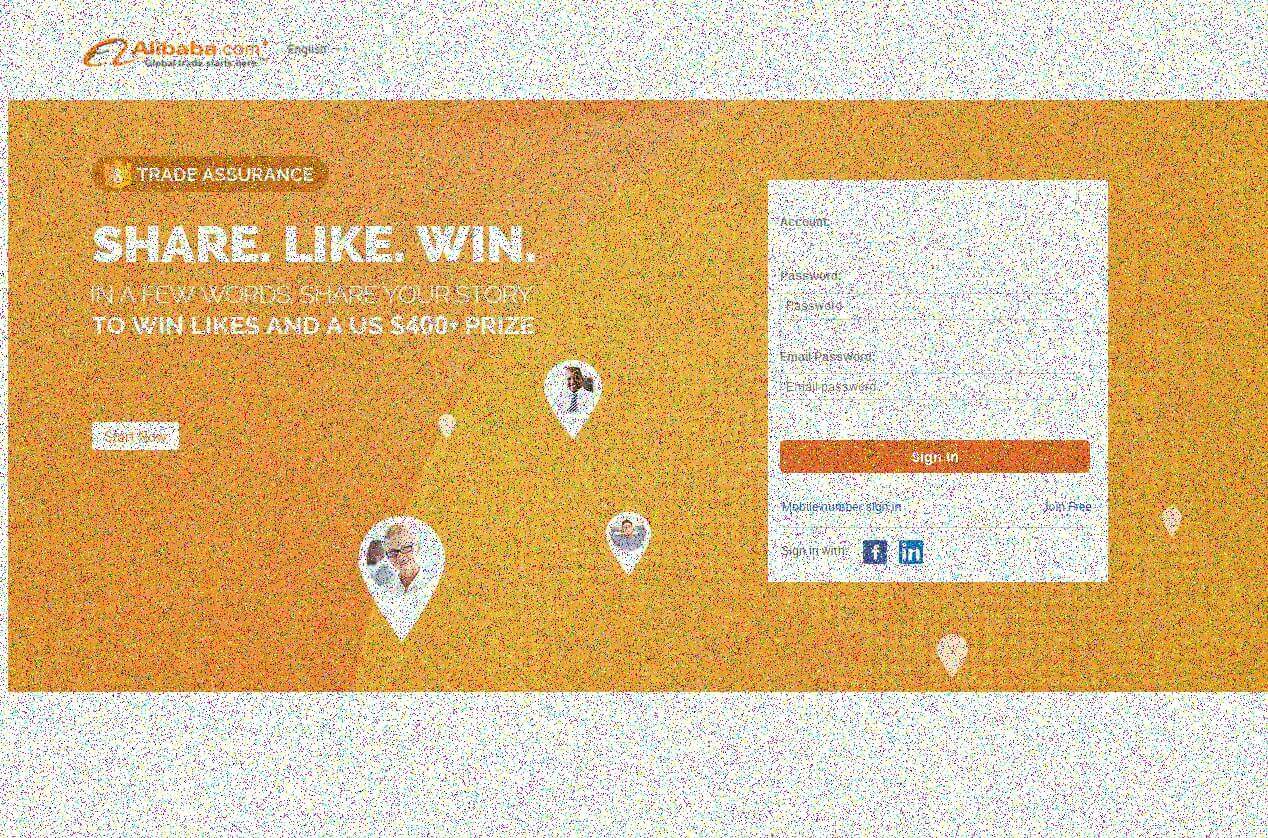} &
\includegraphics[width=3.2cm,height=3cm,keepaspectratio] {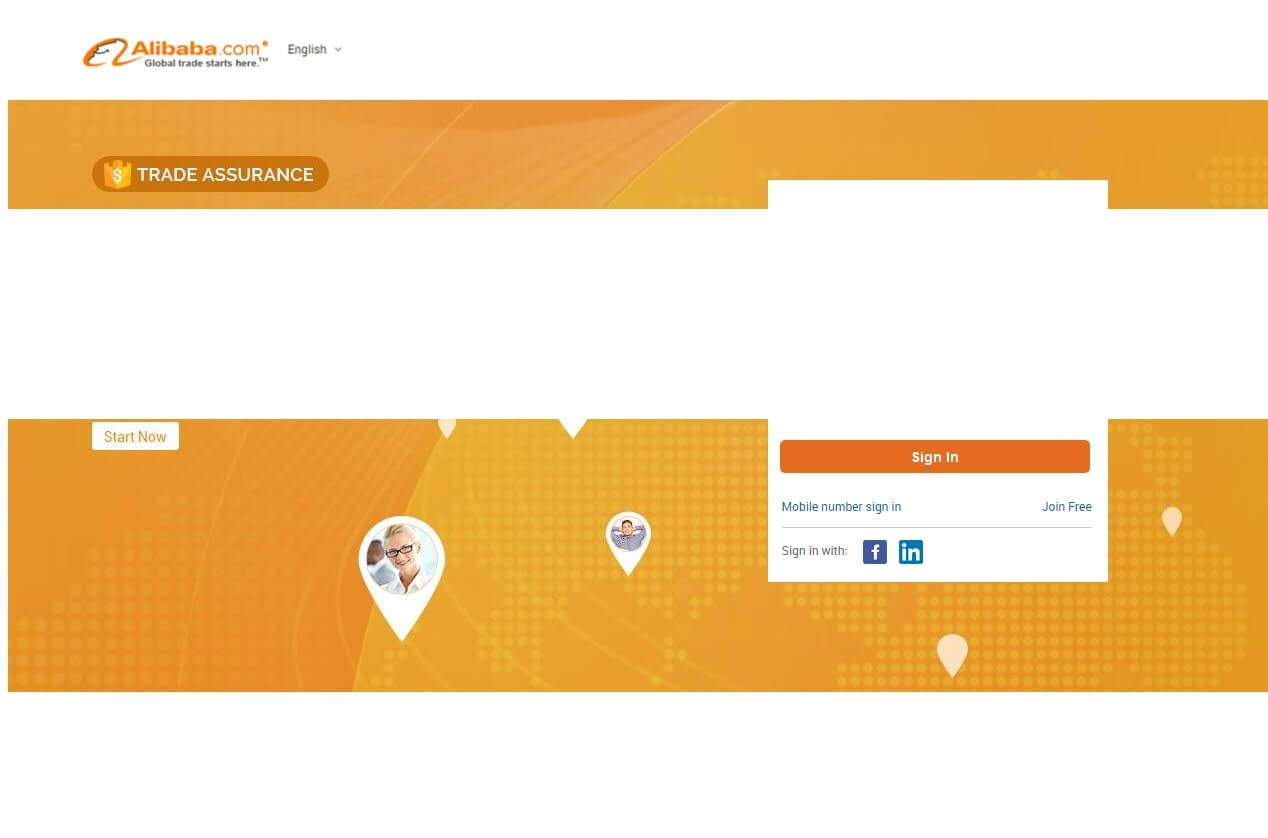} &
\includegraphics[width=3.2cm,height=3cm,keepaspectratio] {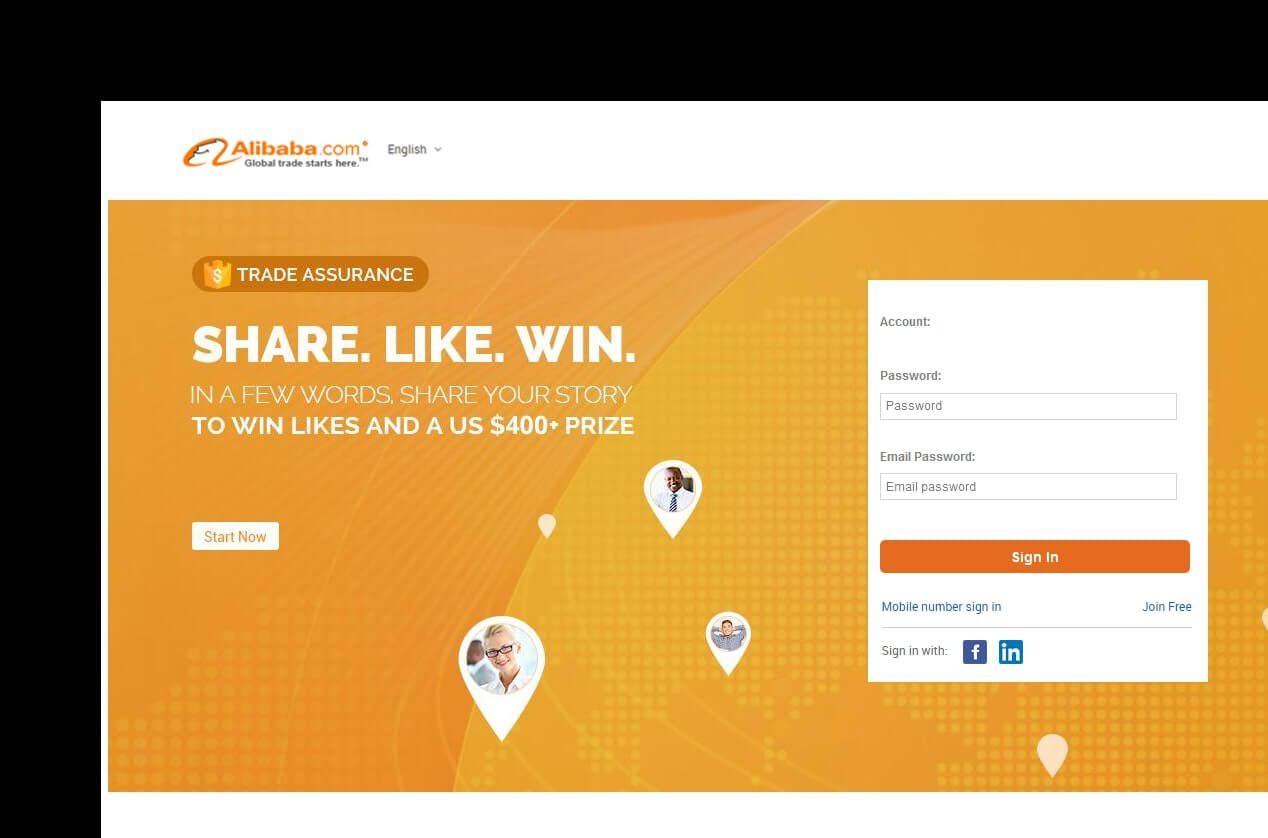} \\  
 \large{Matching drop} &\large{4.13\%} & \large{5.68\%} & \large{6.36\%} & \large{6.71\%} & \large{6.54\%} & \large{5.34\%} & \large{6.54\%} \\ 
 \large{ROC AUC drop} &\large{1.17\%} & \large{2.65\%} & \large{3.35\%} & \large{2.65\%} & \large{3.04\%} & \large{4.99\%} & \large{1.65\%} \\\bottomrule
\end{tabular}}
\caption{The studied hand-crafted perturbations applied to the phishing test set. The table shows the relative decrease in the top-1 matching accuracy and ROC AUC with respect to the performance on the original phishing set.} \label{tab:attacks1}
\end{table*}
\subsection{Distance Threshold Selection}
\begin{figure}[!b]
\centering
\begin{subfigure}{0.5\columnwidth}
  \centering
  \includegraphics[width=\textwidth]{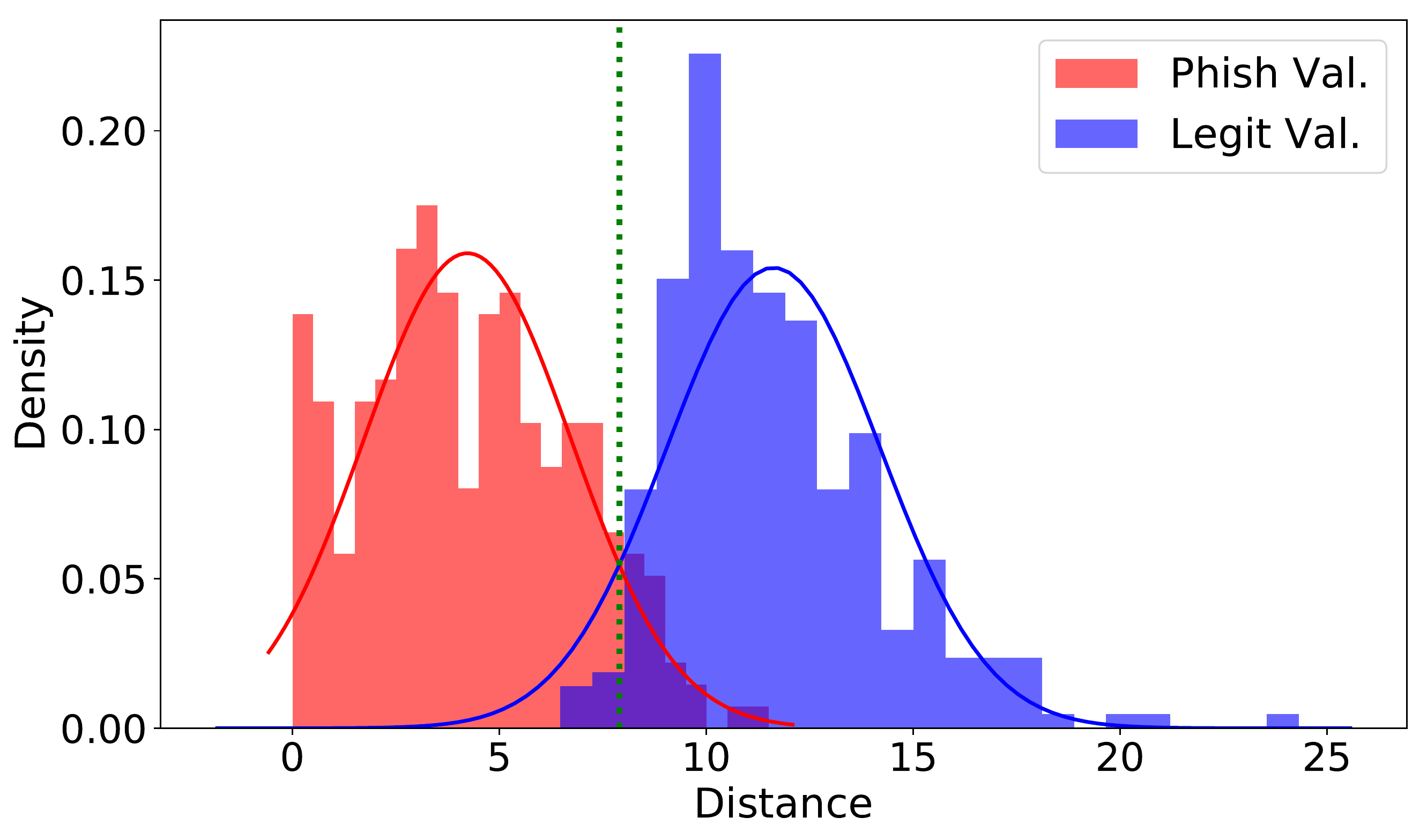}
  \caption{}
  \label{fig:validation}
\end{subfigure}%
\begin{subfigure}{0.5\columnwidth}
  \centering
  \includegraphics[width=\textwidth]{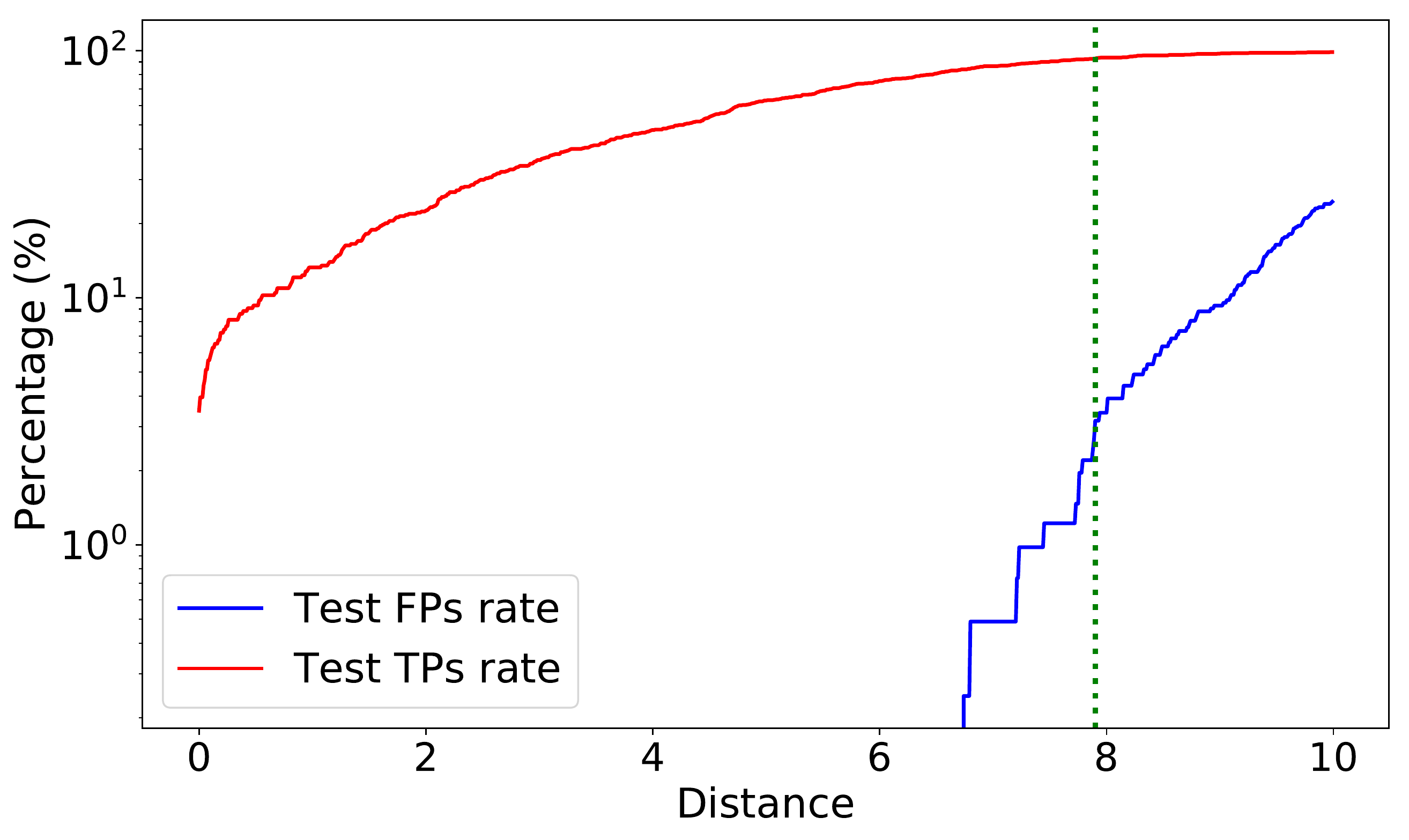}
  \caption{}
  \label{fig:distance_thresholds}
\end{subfigure}
\caption{Distance threshold selection. (a) shows a density histogram of the minimum distances between the phishing (red) and legitimate (blue) validation sets to the training \new{trusted-list}.
(b) shows the true and false positive rates of the test sets over thresholds, the vertical green line marks the threshold from (a).}
\label{fig:thresholds}
\end{figure}

To determine a suitable distance/similarity threshold for the binary classification between phishing and legitimate test sets, we split the phishing and legitimate hold-out sets to validation and test sets. We computed the minimum distances of both of them to the training \new{trusted-list}.~\autoref{fig:validation} shows the two density histograms and the fitted Gaussian Probability Density Functions (PDF) of the minimum distance for the validation sets of both classes. The vertical line (at $\approx$8) represents a threshold value with an equal error rate. Additionally,~\autoref{fig:distance_thresholds} shows the true and false positive rates of the test sets over different thresholds where the indicated threshold is the same one deduced from~\autoref{fig:validation}, which achieves $\approx$93\% true positive rate at $\approx$4\% false positive rate.

\subsection{Robustness and Security Evaluation} \label{robustness}
\begin{figure}[!t]
\centering
\begin{subfigure}{0.5\columnwidth}
  \centering
  \includegraphics[width=0.9\textwidth,height=3.2cm,keepaspectratio]{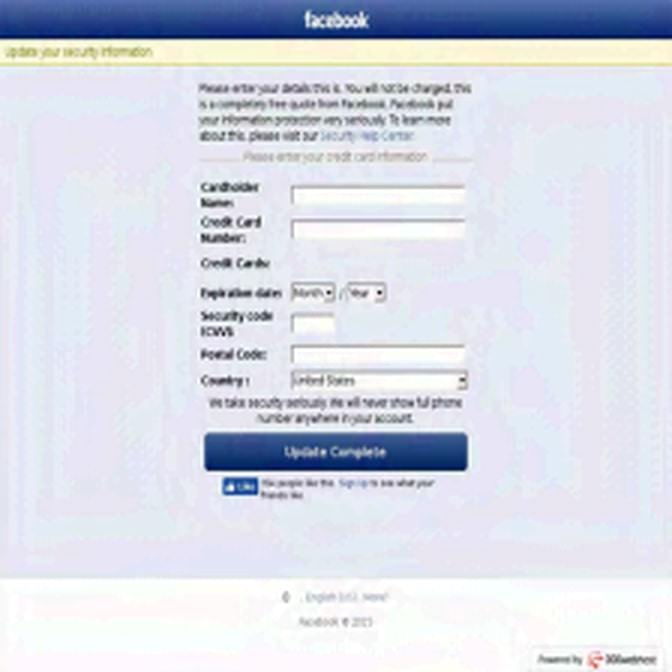}
  \caption{}
  \label{fig:epsilon1}
\end{subfigure}%
\begin{subfigure}{0.5\columnwidth}
  \centering
  \includegraphics[width=0.9\textwidth,height=3.2cm,keepaspectratio]{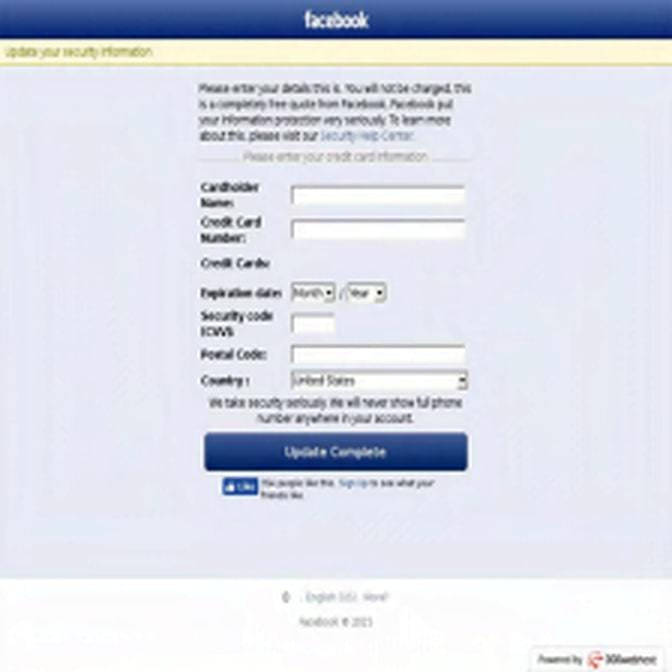}
  \caption{}
  \label{fig:epsilon2}
\end{subfigure}
\caption{Adversarial examples generated with FGSM on the triplet loss with $\epsilon = 0.01$ (a) and $\epsilon = 0.005$ (b).}
\label{fig:adv_ex}
\end{figure}

To test the robustness of \method{}, we define two models for evasion attacks. In the first one, we study how susceptible \method{} is to small changes in the input (e.g. change of color, noise, and position). In the second one, we assume a white-box attack where the adversary has full access to the target model and the dataset used in training (including the closest point to the phishing page). In both models, we assume that the attacker's goal is to violate the target model's integrity (in our case: similarity detection to the targeted website) by crafting phishing pages that show differences from their corresponding original pages that might be included in the \new{trusted-list}. However, we assume that the adversary is motivated to not introduce very perceivable degradation in the design quality for his phishing page to seem trusted and succeed in luring users.
\paragraph{Performance against hand-crafted perturbations} 
We studied 7 types of perturbations~\cite{yu19iccv} that we applied to the phishing test set (without retraining or data augmentation): blurring, brightening, darkening, Gaussian noise, salt and pepper noise, occlusion by insertion of boxes, and shifting. 
~\autoref{tab:attacks1} demonstrates an example of each of these changes along with the corresponding relative decrease in performance. Our findings revealed that the matching accuracy and the ROC area dropped slightly (by up to $\approx$4.3\% and $\approx$1.8\% respectively) for the imperceptible noise, while it dropped by up to $\approx$6.7\% and $\approx$5\% respectively for the stronger noise that we assume that it is less likely to be used. Further improvement could be achieved with data augmentation during training.
\paragraph{Adversarial perturbations}
Another direction for evasion attacks is crafting adversarial perturbations with imperceptible noise that would change the model decision when added to the input test points~\cite{kurakin2016adversarial}. There is a lot of work towards fixing the evasion problem~\cite{biggio2018wild}, however, adversarial perturbations are well-known for classification models. In contrast, \method{} is based on a metric learning approach that, at test time, is used to compute distances to the training points. We are not aware of any prior adversarial perturbation methods on similarity-based networks and therefore we propose and investigate an adaptation of the adversarial example generation methods to our problem by using the Fast Gradient Sign Method (FGSM)~\cite{goodfellow2014explaining} defined as: 
\begin{center}
$\widetilde{x} =  x + \epsilon \; \text{sign}(\nabla _x J(\theta,x,y))$ 
\end{center}
where $\widetilde{x}$ is the adversarial example, $x$ is the original example, $y$ is the example's target (0 in the triplet loss), $\theta$ denotes the model's parameters and $J$ is the cost function used in training (triplet loss in \method{}).
Adapting this to our system, we used the phishing test example as the anchor image, sampled an image from the same website as the positive image (from the training \new{trusted-list}), and an image from a different website as the negative image. We then computed the gradient with respect to the anchor image (the phishing test image) to produce the adversarial example.
We experimented with two values for the noise magnitude ($\epsilon$): 0.005 and 0.01, however, the 0.01 noise value is no longer imperceptible and causes noticeable noise in the input (as shown in~\autoref{fig:adv_ex}). We also examined different triplet sampling approaches when generating the adversarial examples, in the first one, we select the positive image randomly from the website's images. However, since the matching decision is based on the closest distance, in the second approach, we select the closest point as the positive. We demonstrate our results in~\autoref{tab:adv} where we show the relative decrease in the top-1 matching accuracy and the ROC AUC for each case averaged over 5 trials as we randomly sample triplets for each example. Our results showed that the matching accuracy and the AUC dropped by $\approx$10.5\% and $\approx$6.5\% for the 0.005 noise and by $\approx$22.8\% and $\approx$12.4\% for the higher 0.01 noise. Also, targeting the closest example was similar to sampling a random positive image. In addition, we tested an iterative approach of adding a smaller magnitude of noise to the closest point at each step (0.002 noise magnitude for 5 steps) which was comparable to adding noise with a larger magnitude (0.01) at only one step.

\begin{table} [!b]
\centering
\resizebox{0.9\linewidth}{!}{%
\begin{tabular}{c|ll|ll}
\toprule
Model &Epsilon ($\epsilon$) & Sampling & Matching drop & ROC AUC drop \\  \midrule 
\multirow{4}{*}[0.1cm]{\small{Original}} &0.005 & random & 10.5\% & 6.47\%\\ 
&0.005 & closest point & 10.11\% & 6.07\% \\
&0.01 & random & 22.81\% & 12.35\% \\ 
&0.002 & iterative & 20.8\% & 12.05\% \\ \hline
\multirow{2}{*}[-0.1cm]{\small{Retrained}} & 0.005 & random & \textbf{2.54\%} & \textbf{0.07\%} \\
& 0.01 & random & 9.78\% & 3.61\%\\\bottomrule
\end{tabular}}
\caption{The relative performance decrease (with respect to the original test set) of the FGSM adversarial examples.} \label{tab:adv}
\end{table}
We then performed adversarial training by fine-tuning the trained \method{} for 3000 mini-batches. In each mini-batch, half of the triplets were adversarial examples generated with FGSM with an epsilon value that is randomly generated from a range of 0.003 and 0.01. After training, we again applied FGSM on the phishing test set using the tuned model. As shown in the last two columns of~\autoref{tab:adv}, the performance improved to reach a comparable performance to the original set in the case of the 0.005 noise. These results demonstrate that \method{}, after retraining, is robust against adversarial attacks with slightly added noise.
\paragraph{Evaluating different browsers}
We studied the effect of the changes caused by other browsers than the one we used to build the dataset (Firefox) as an example of one of the factors that could be different when deploying the system. Thus, we created a subset of 50 URLs from 14 websites, and we used Firefox, Opera, Google Chrome, Microsoft Edge, and Vivaldi browsers to take screenshots of these pages of which we computed the \method{}'s embeddings. In~\autoref{tab:browsers}, we quantify the browsers' changes by comparing the L2 differences between Firefox's embeddings (to match the dataset) and other browsers' ones, which we found smaller by at least ${\approx}6.6x$ than the differences caused by the slight hand-crafted perturbations (applied on Firefox screenshots) we previously showed in the first row of ~\autoref{tab:attacks1} and demonstrated that they already had a small effect on the performance. Additionally, some of these browser differences were due to advertisement or color differences which are already included in the constructed dataset (see~\autoref{appendix_dataset}). 
\begin{table} [!b]
\centering
\resizebox{0.83\linewidth}{!}{%
\begin{tabular}{l|llll}
\toprule
\large{Browser} & \large{Chrome} & \large{Edge} & \large{Opera} & \large{Vivaldi} \\   \midrule
& \large{0.278\rpm0.54} & \large{0.23\rpm0.75} & \large{0.271\rpm0.21} & \large{0.41\rpm0.57} \\ \hline \hline
\large{Noise} & \large{Blurring} & \large{Gaussian} & \large{Salt and Pepper} & \large{Shift} \\ \midrule
& \large{4.92\rpm2.71} & \large{2.73\rpm1.02} & \large{6.80\rpm2.24} & \large{5.43\rpm2.36} \\\bottomrule
\end{tabular}}
\caption{The L2 difference between Firefox screenshots' embeddings and other browsers' ones, compared to the L2 difference due to the studied slight perturbations.} \label{tab:browsers}
\end{table}

\subsection{Testing with New Crawled Data} \label{new_data}

\paragraph{Zero-day pages}
To provide additional evidence for the efficacy of \method{} in detecting zero-day pages, we crawled recent 955 PhishTank pages targeting the \new{trusted-list} (examples in~\autoref{appendix_dataset}). \new{These are new pages that were created and captured after dataset collection, training, and evaluating the model with all previous experiments, and therefore, they are future pages with temporal separation with respect to the model.} Additionally, We used a different browser, machine, and screen size from the ones used to collect the dataset to further test against possible variations. We then tested the trained model with this new set (without retraining), and 93.25\% were correctly matched (top-5: 96\%), compared to 81\% (top-5: 88\%) on the harder and more dissimilar dataset's phishing pages (see~\autoref{appendix_results} for matching examples). \new{Additionally, as a baseline, the VGG-16 matching accuracy of this new set is 65.8\%.} 

\paragraph{Alexa top-10K}
\new{To further test the false positives, we crawled the Alexa top 10K websites (excluding \new{the trust-list's} domains) to use as a benign test set. Using the same trained model, the ROC AUC of classifying this new benign set against the original phishing set is 0.974, while the partial ROC AUC at 1\% false positives is 0.0079 (compared to 0.987 and 0.0087, respectively, on the smaller benign subset). Similarly, the corresponding VGG-16 ROC AUC is 0.781.}

\section{Discussion}
\begin{table}[!b]
\centering
\resizebox{0.9\linewidth}{!}{%
\begin{tabular}{l|c|c|c}
\rot{\large{Phishing test}} & \includegraphics[width=3cm,height=3cm,keepaspectratio] {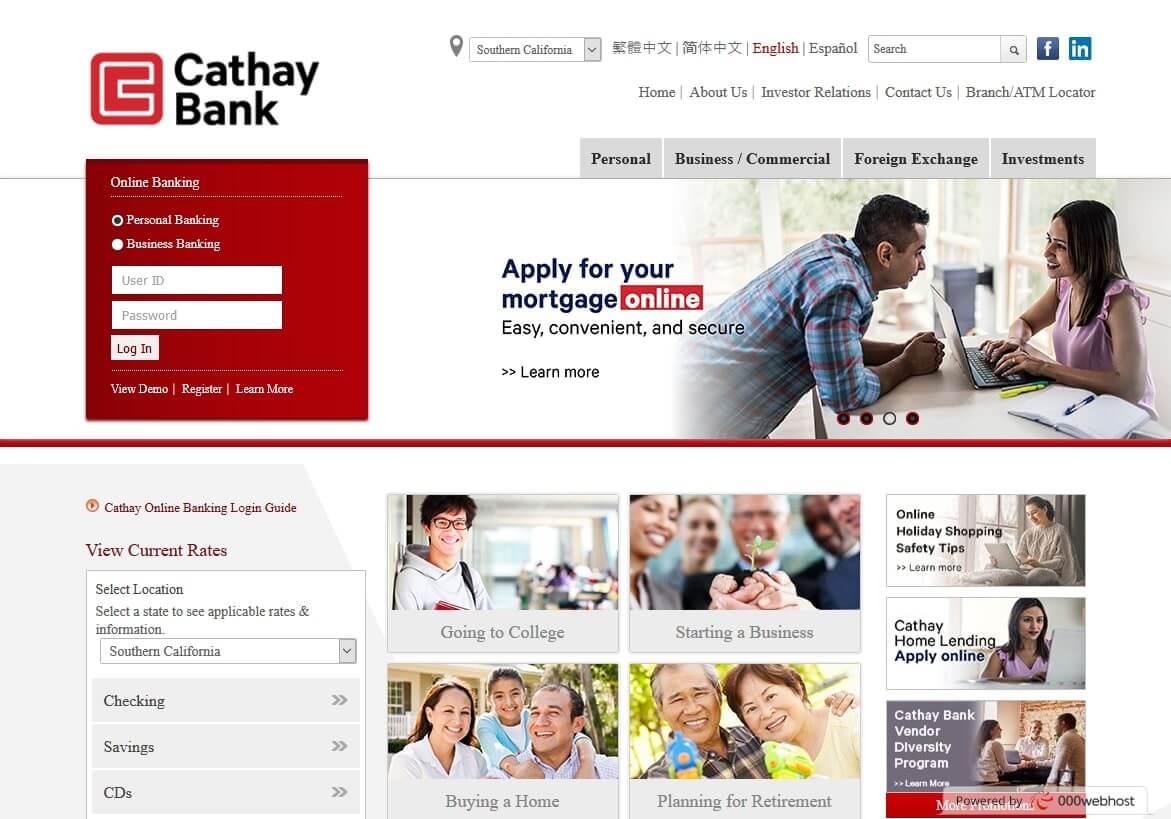} &
\includegraphics[width=3cm,height=3cm,keepaspectratio] {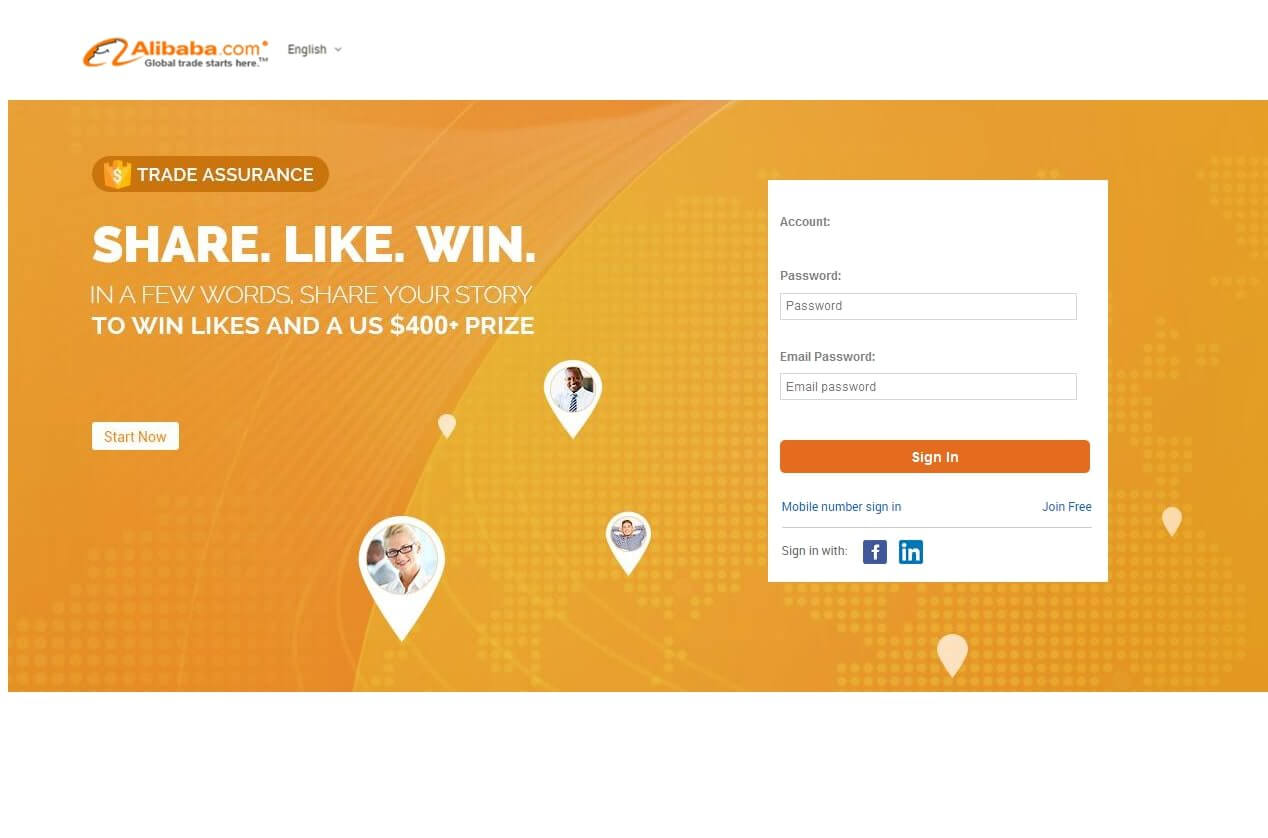} &
\includegraphics[width=3cm,height=3cm,keepaspectratio] {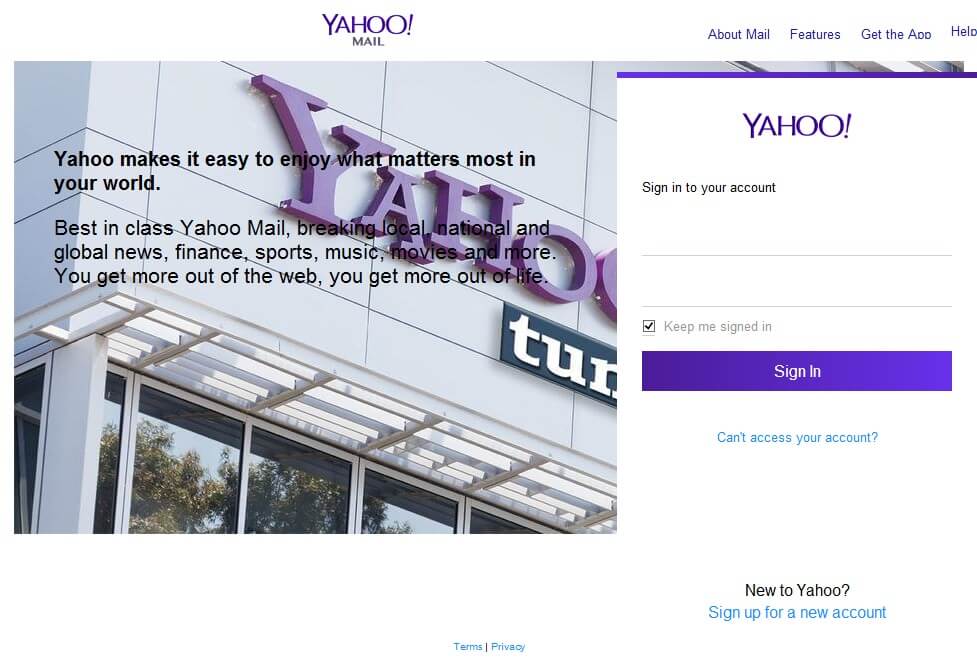} \\ 
&&\\
\rot{\large{Closest match}} & \includegraphics[width=3cm,height=3cm,keepaspectratio] {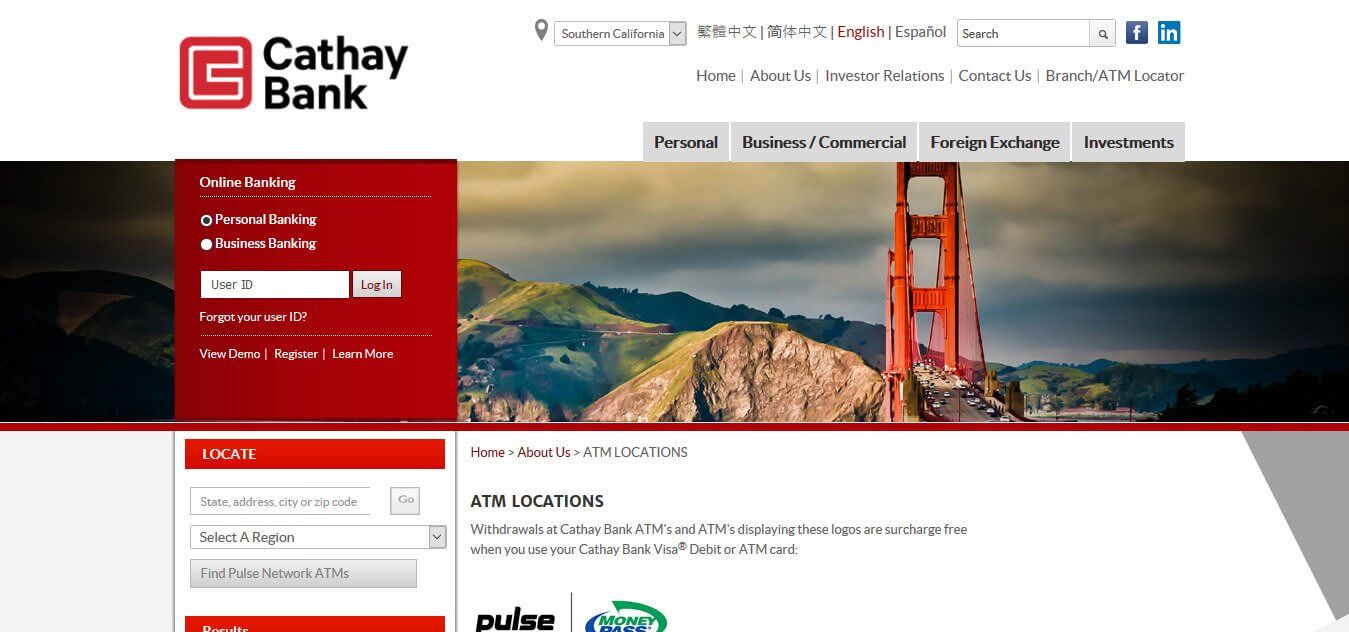} &
\includegraphics[width=3cm,height=3cm,keepaspectratio] {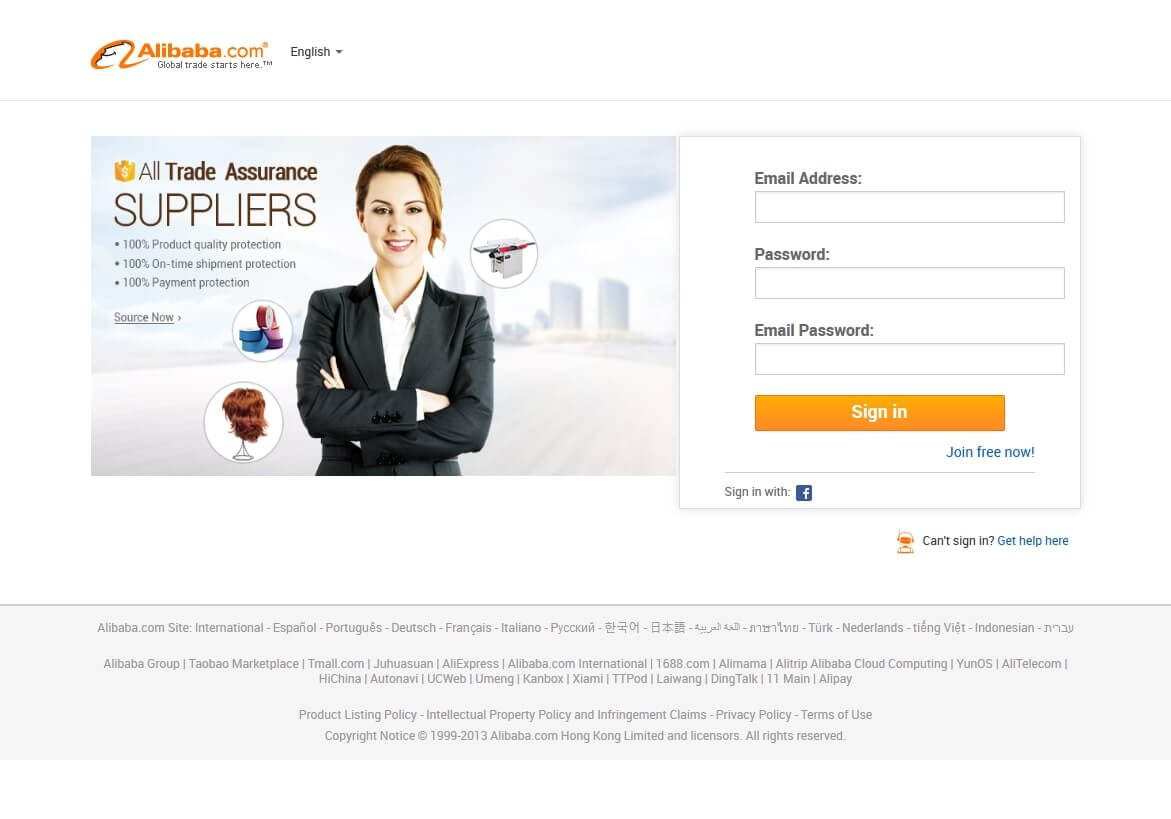} & \includegraphics[width=3cm,height=3cm,keepaspectratio] {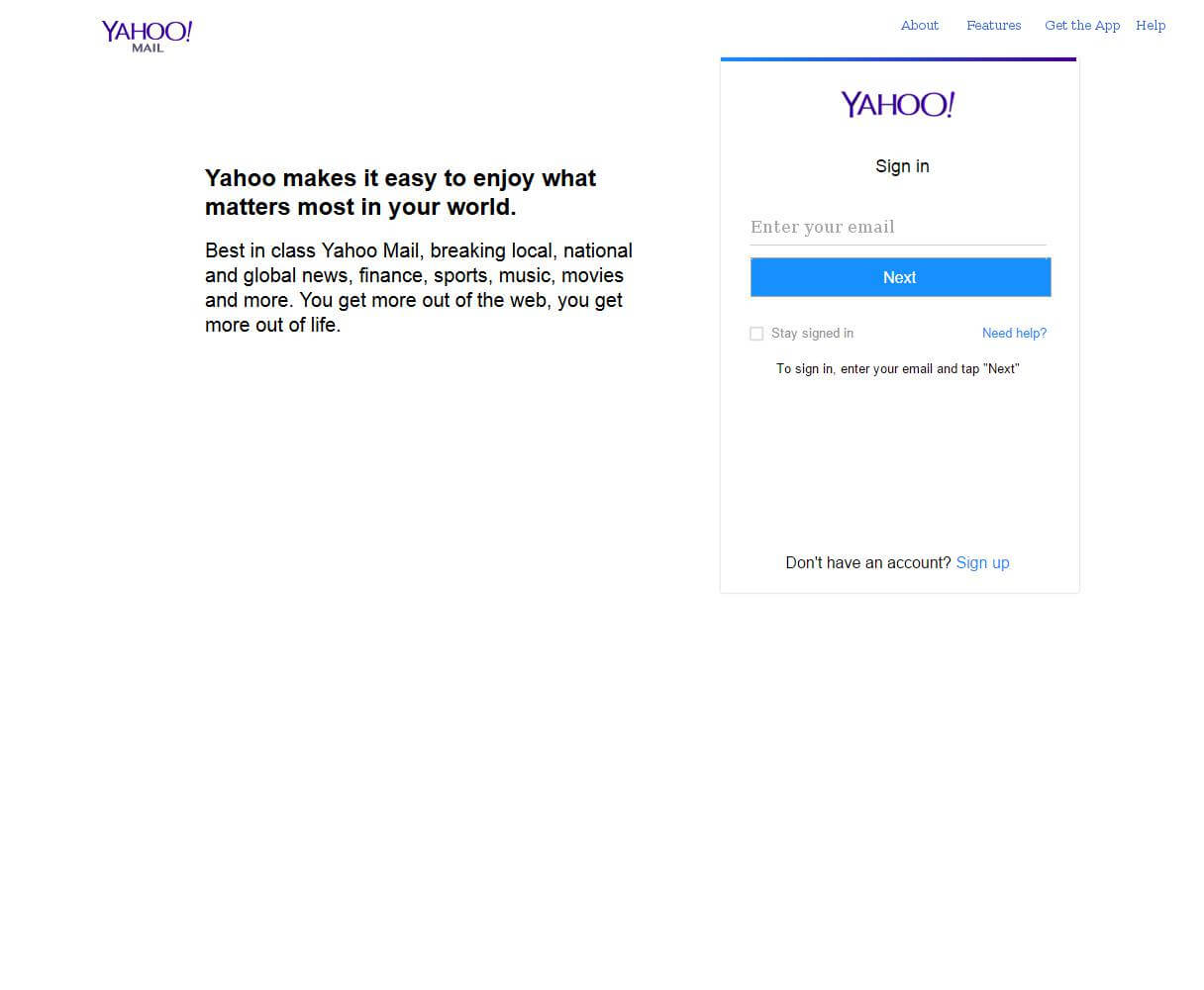}\\
\end{tabular}}
\captionof{figure}{Test phishing pages (first row) that were correctly matched to the targeted websites (closest match from the training set in the second row) with the closest pages having a relatively similar layout but different colors and content.} \label{tab:easy}
\end{table}
\begin{table*}[!b]
\centering
\resizebox{0.95\linewidth}{!}{%
\begin{tabular}{l|c|c|c|c|c|c|c}
\rot{\large{Phishing test}} & 
\includegraphics[width=3cm,height=3cm,keepaspectratio] {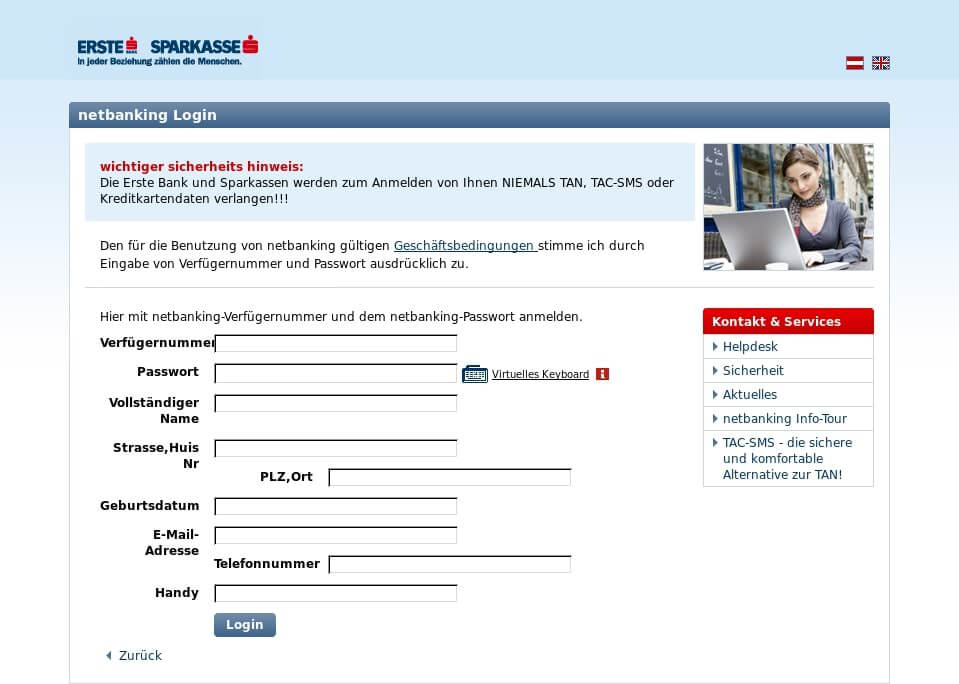} &
\includegraphics[width=3cm,height=3cm,keepaspectratio] {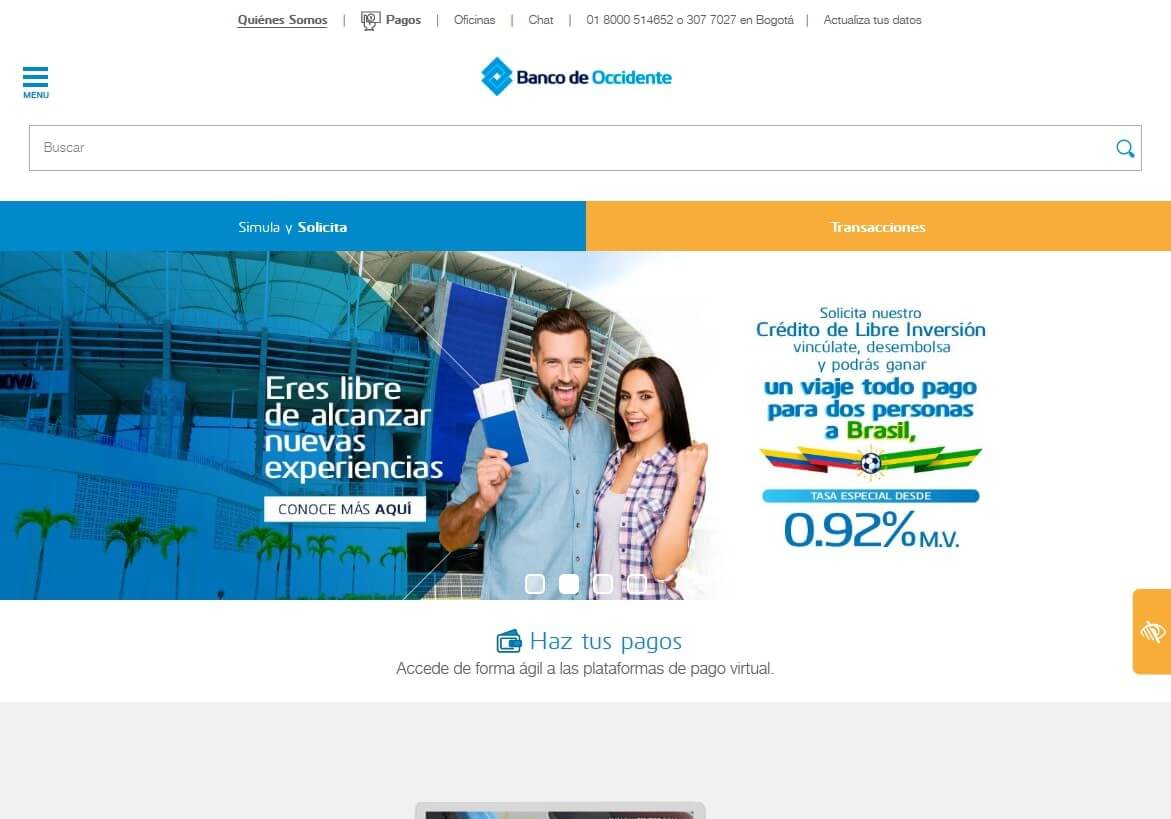} &
\includegraphics[width=3cm,height=3cm,keepaspectratio] {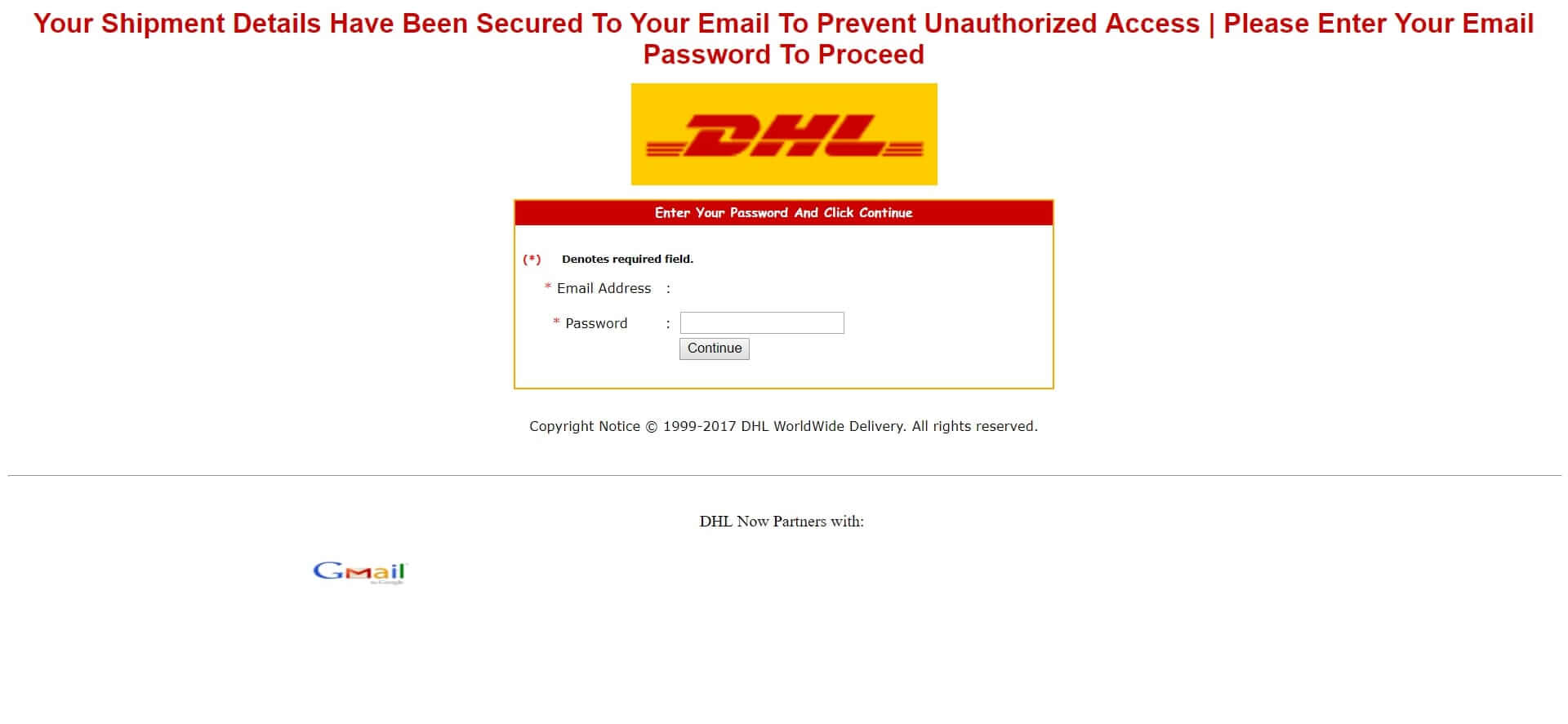} &
\includegraphics[width=3cm,height=3cm,keepaspectratio] {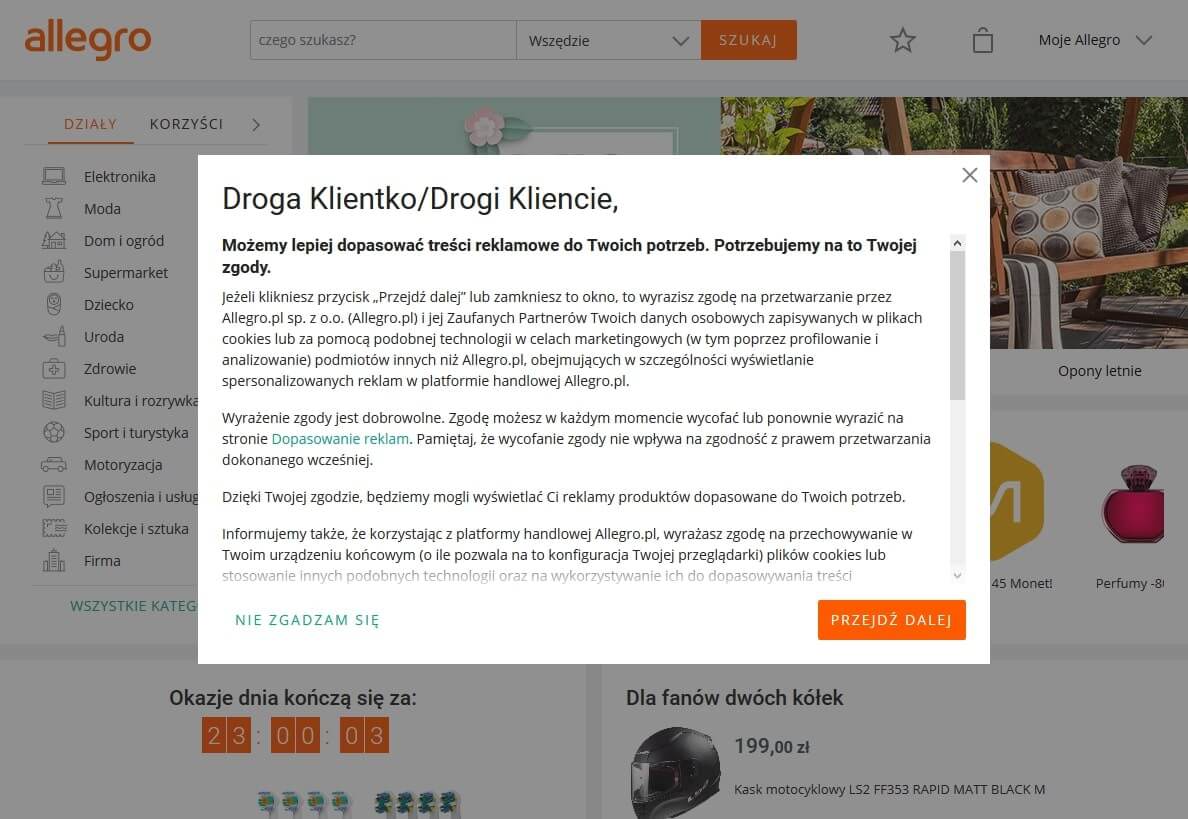} & \includegraphics[width=3cm,height=3cm,keepaspectratio] {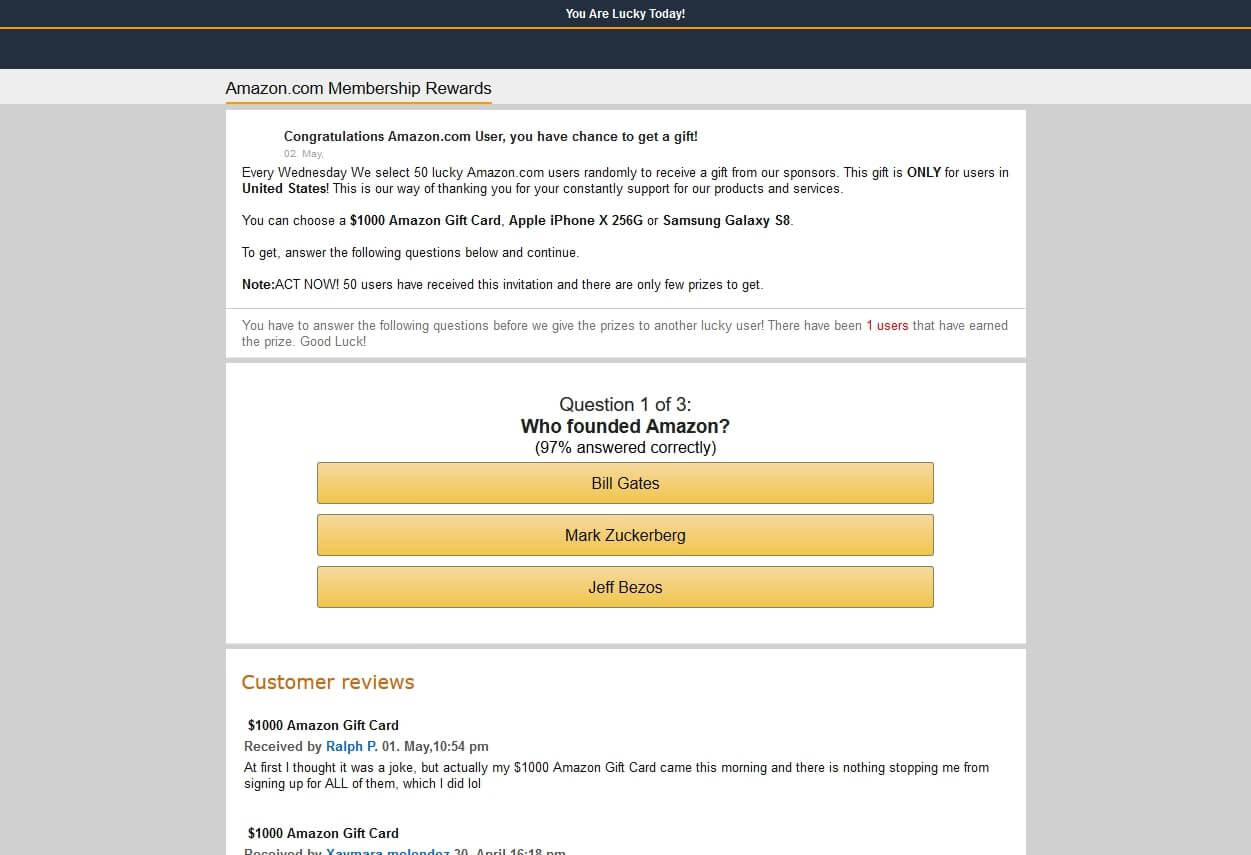} & \includegraphics[width=3cm,height=3cm,keepaspectratio] {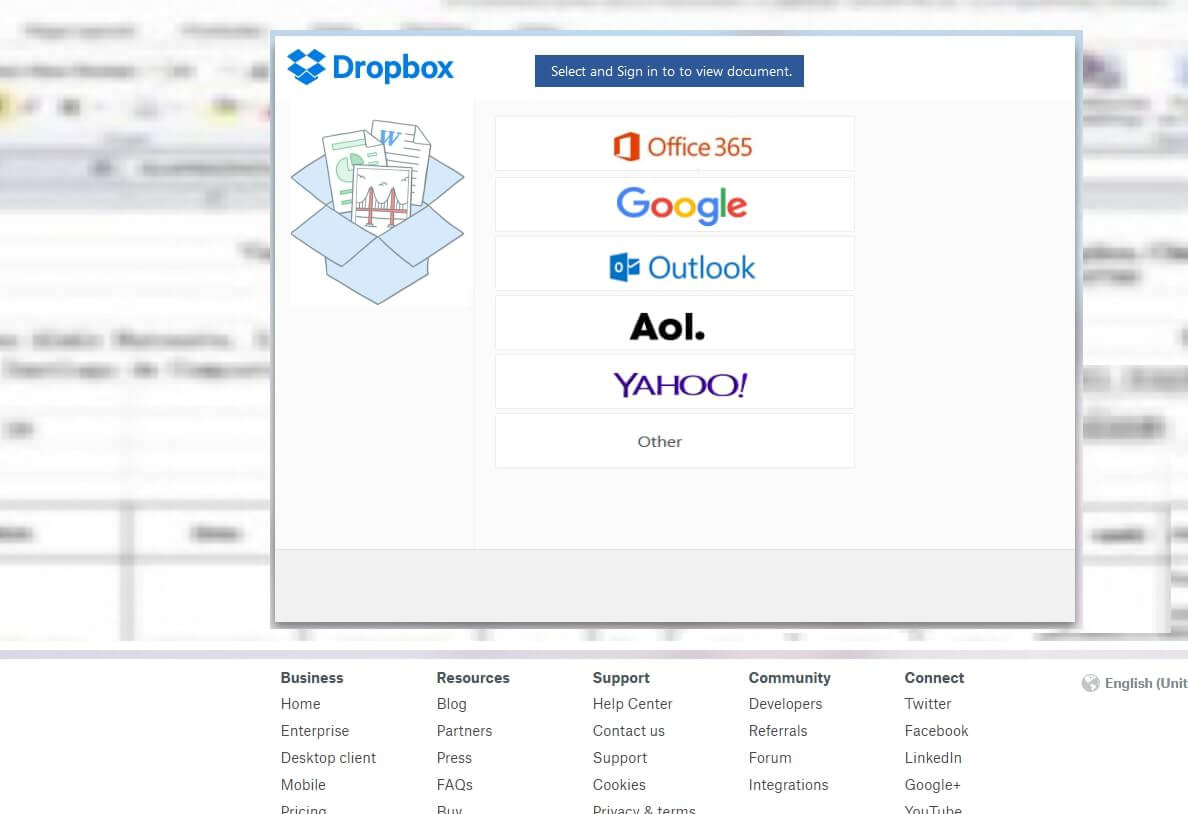} & \includegraphics[width=3cm,height=3cm,keepaspectratio] {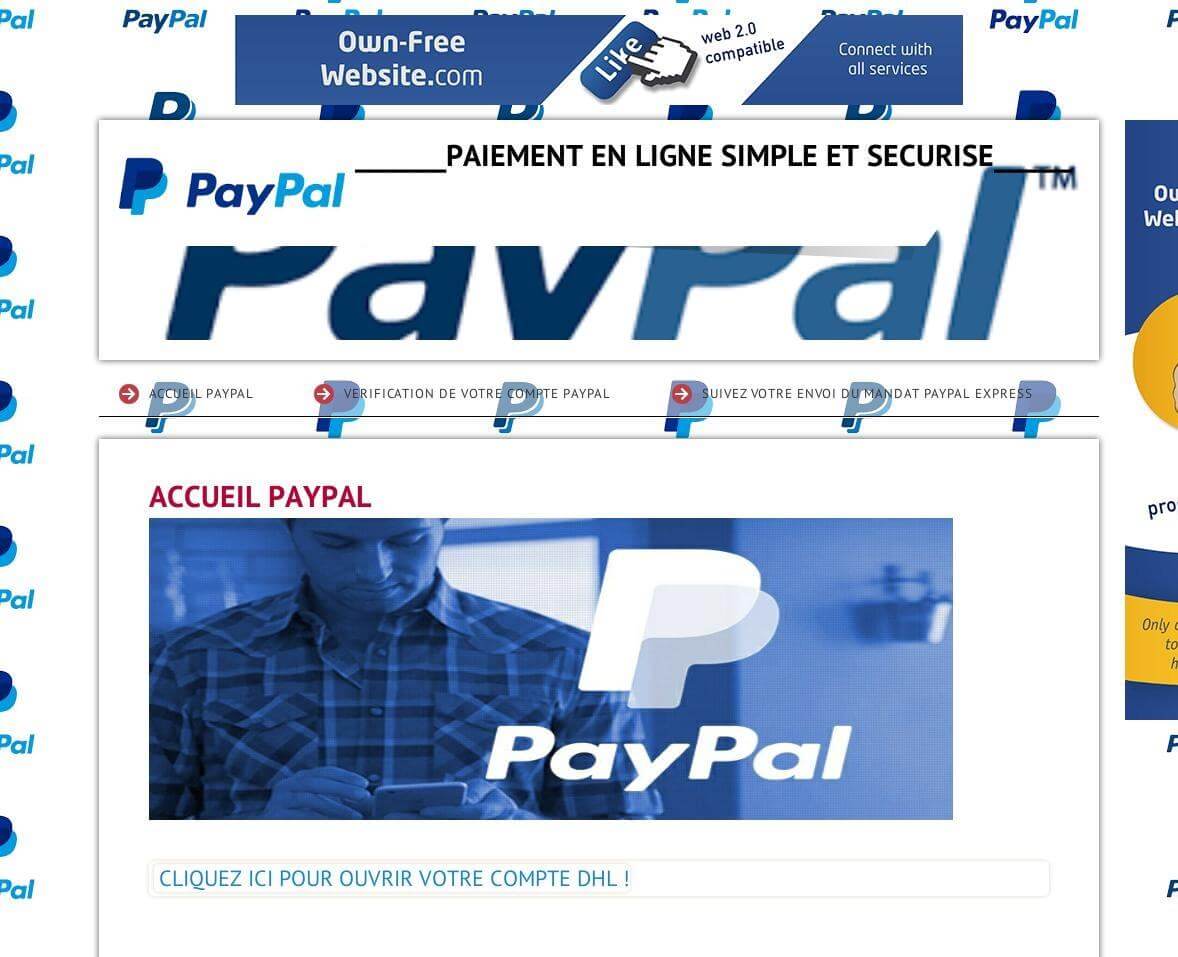} \\ 
&&&&&&\\
\rot{\large{Closest match}} &
\includegraphics[width=3cm,height=3cm,keepaspectratio] {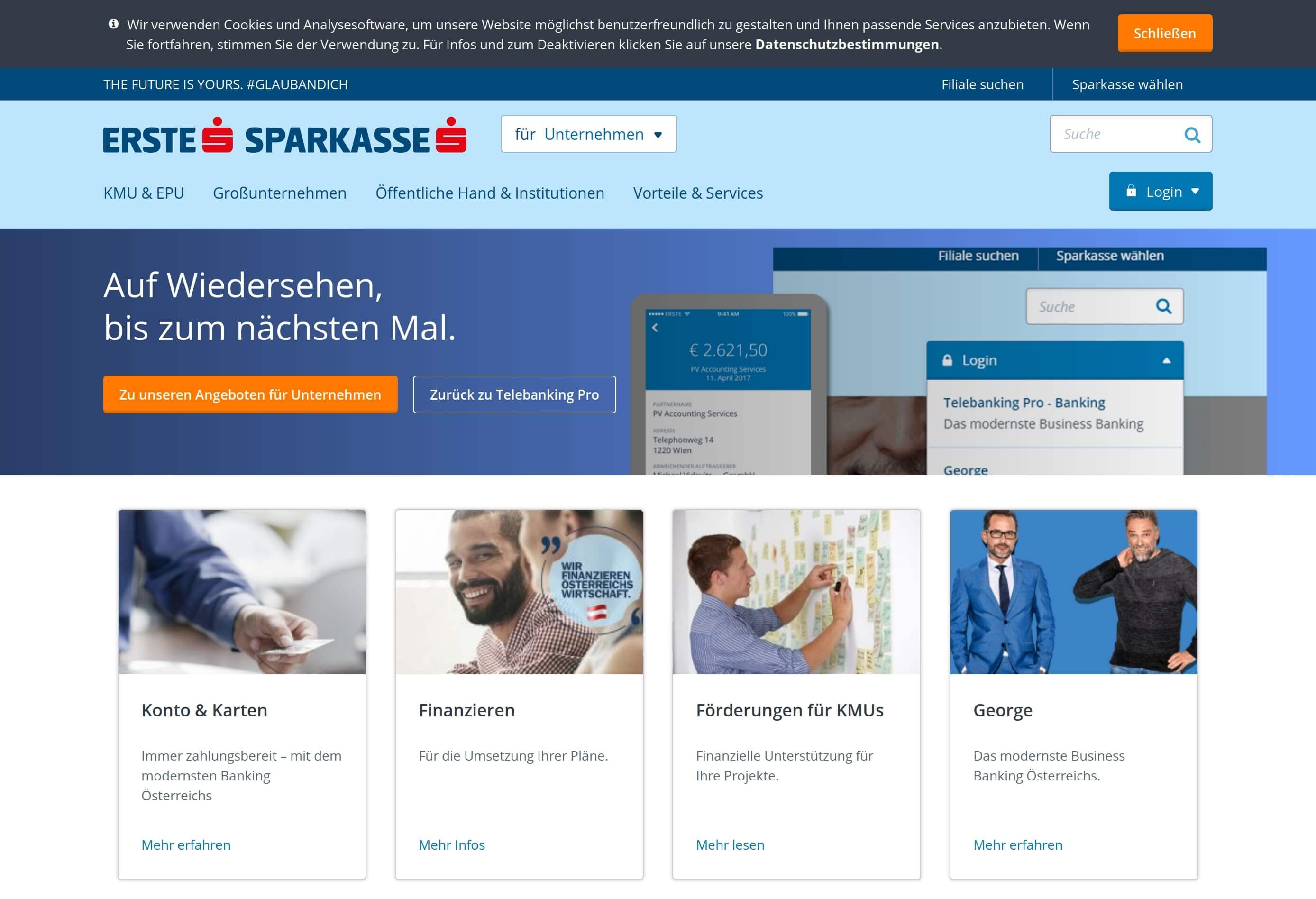} &
\includegraphics[width=3cm,height=3cm,keepaspectratio] {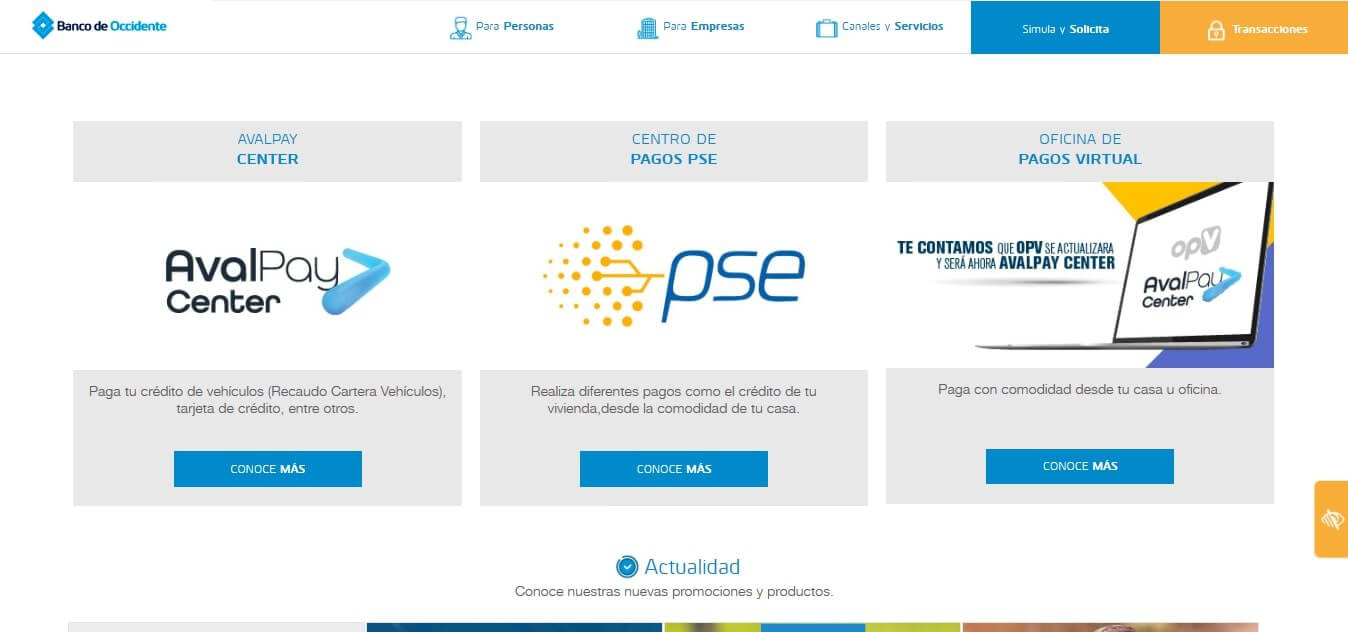} &
\includegraphics[width=3cm,height=3cm,keepaspectratio] {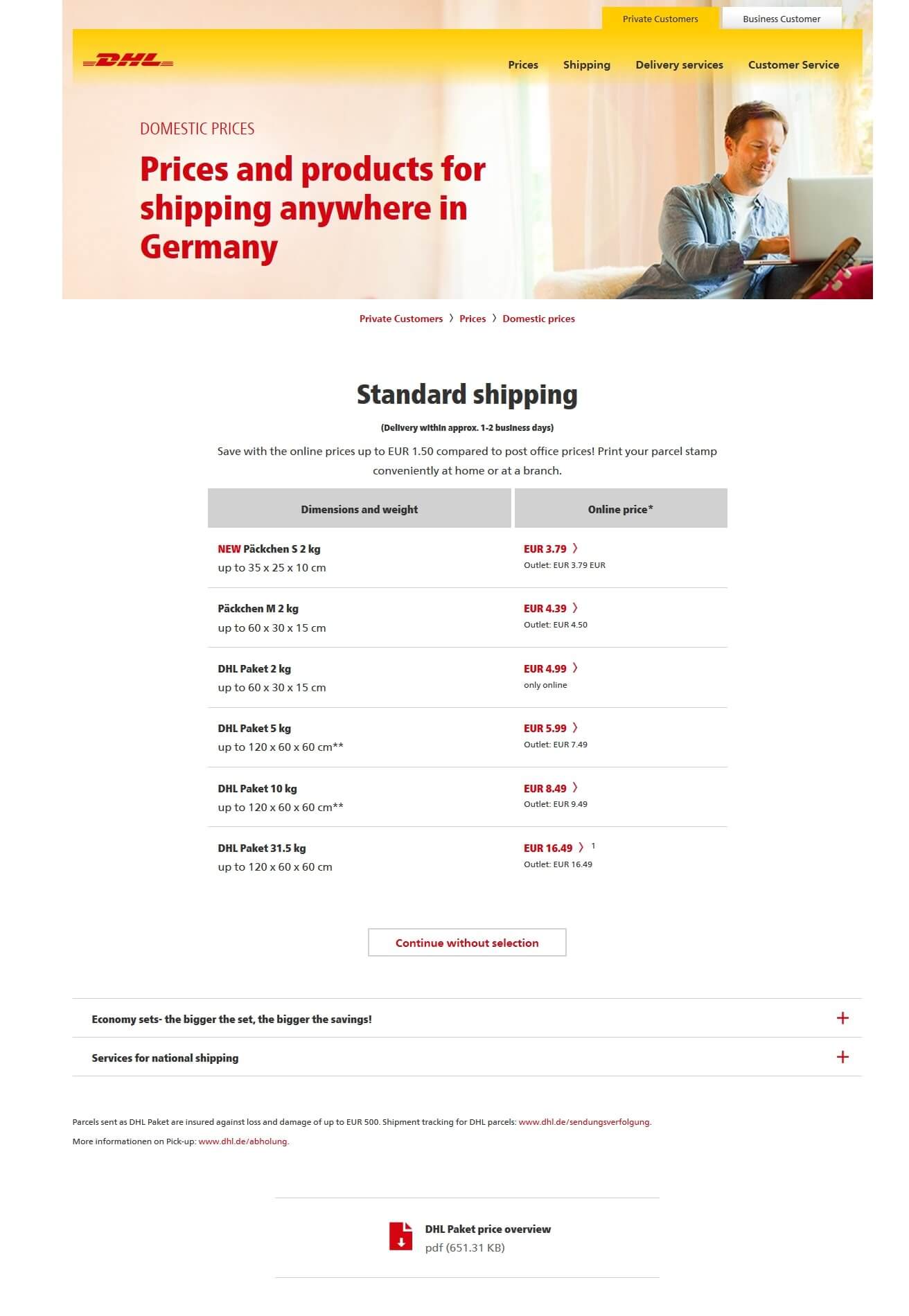} &
\includegraphics[width=3.5cm,height=3cm,keepaspectratio] {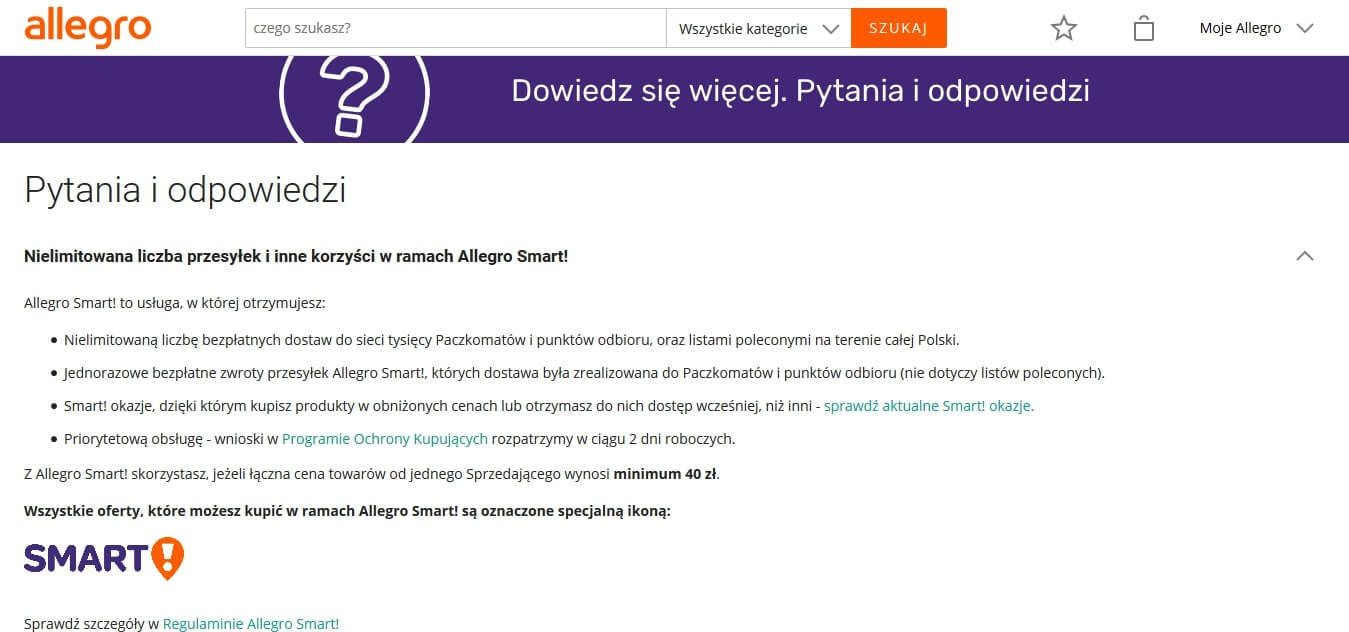} & \includegraphics[width=3cm,height=3cm,keepaspectratio] {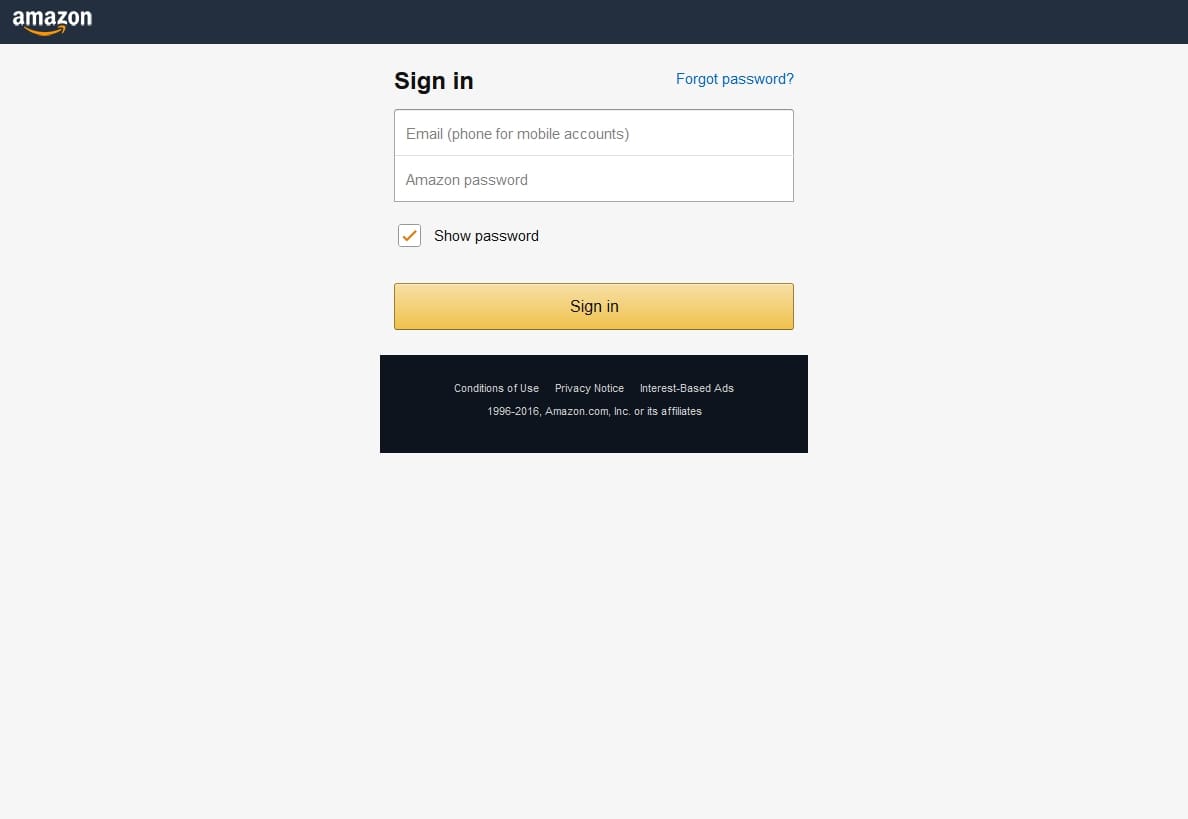} & \includegraphics[width=3cm,height=3cm,keepaspectratio] {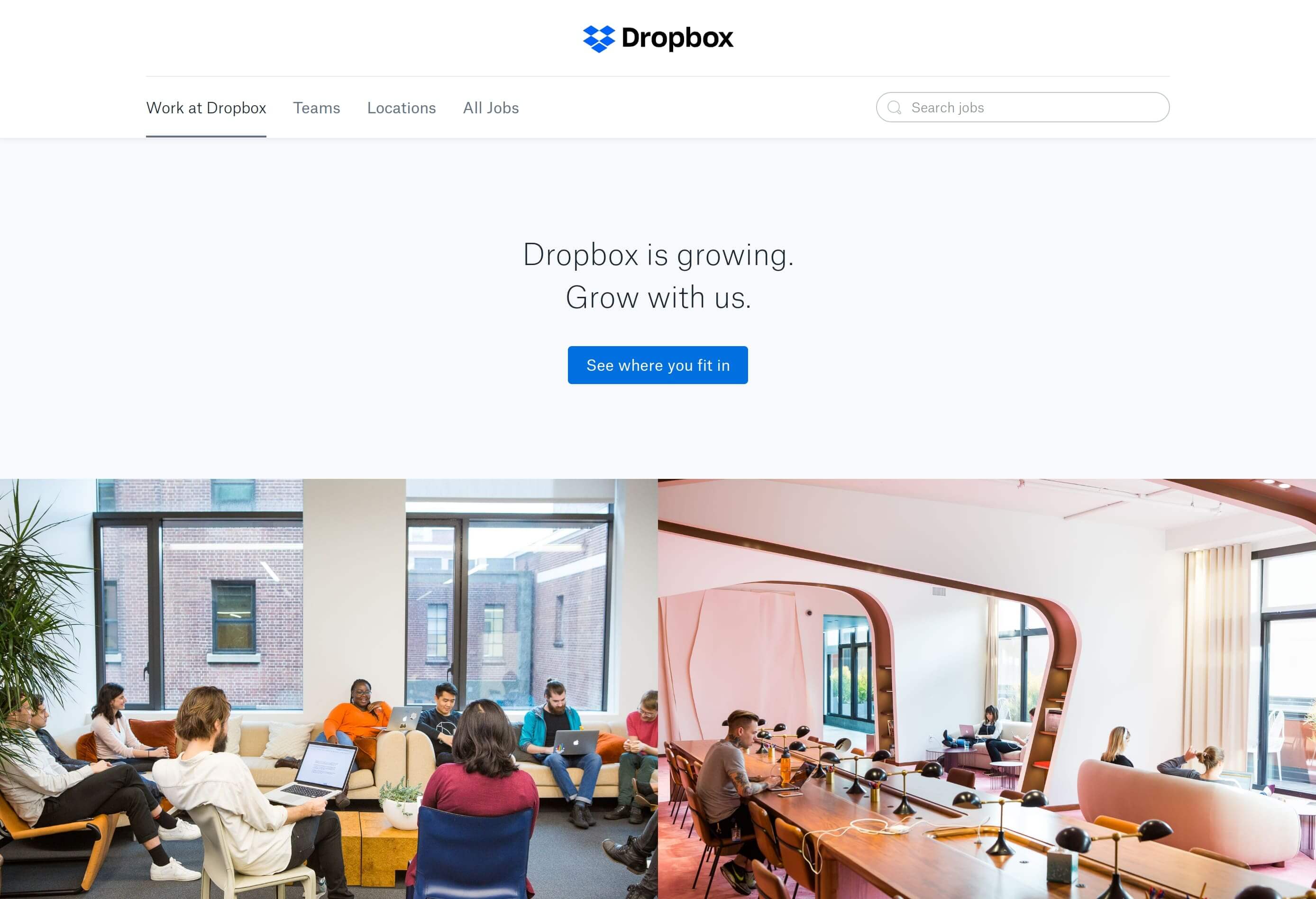} & \includegraphics[width=3cm,height=3cm,keepaspectratio] {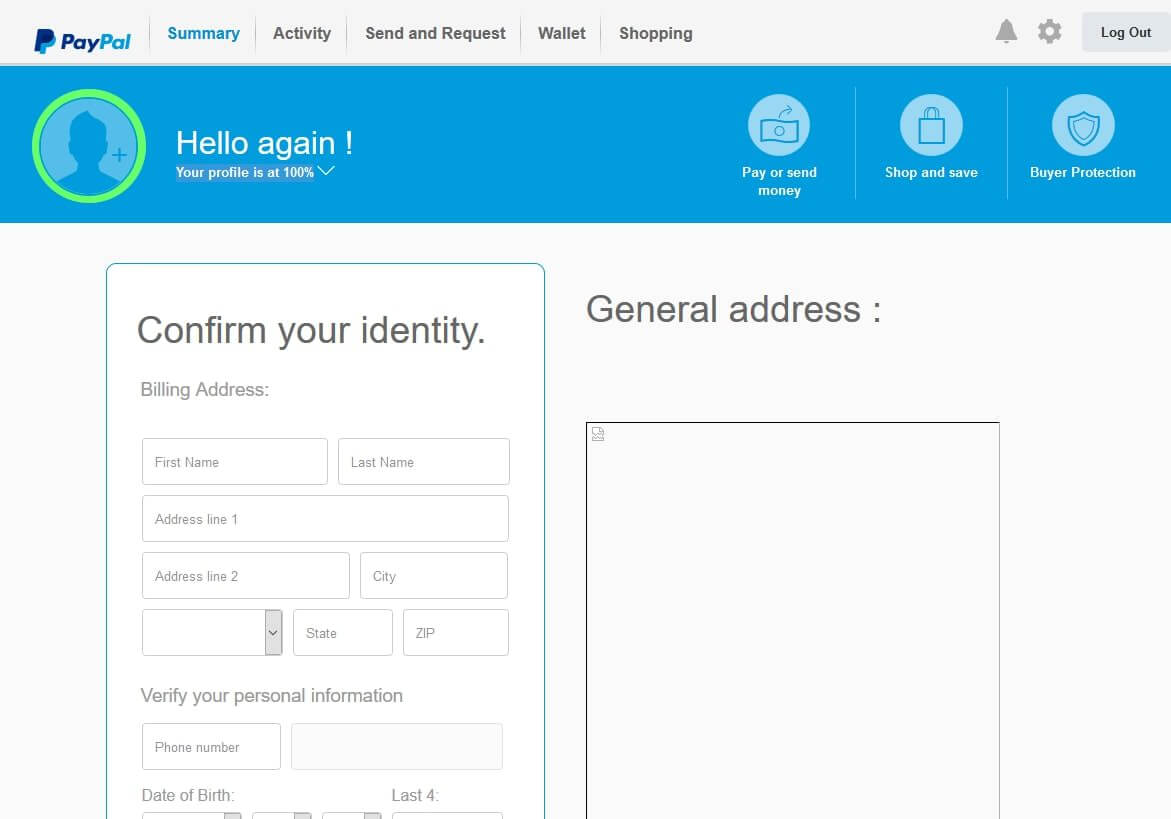} \\
\end{tabular}}
\captionof{figure}{Examples of test phishing webpages that were correctly matched to the targeted websites despite having large differences in layout and content.}\label{tab:hard}
\end{table*}
We discuss the implications of the efficacy of \method{} by showcasing phishing examples that were correctly detected, and failure modes with both false positive and false negative examples.
\subsection{Evaluating Successful Cases} \label{discuss_qual}
We categorize the successfully classified phishing pages into three main categories. The first one is the easily classified ones consisting of exact or very close copying of a corresponding legitimate webpage from the training \new{trusted-list}. However, our model still showed robustness to small variations such as the text language of login forms (which shows an advantage over text-similarity methods), small advertisements' images changes, the addition or removal of elements in the page, and changes in their locations. We observed that these pages have approximately a minimum distance in the range of 0-2 to the training set (as shown in the distances' histogram in~\autoref{fig:thresholds}) and constitute around 25\% of the correct matches. The second category, which is relatively harder than the first one, is the phishing webpages that look similar in style (e.g. location of elements and layout of the page) to training pages, however, they are highly different in content (e.g. images, colors, and text). We show examples of this second category in Figure~\ref{tab:easy}. Similarly, these pages correspond approximately to the distance range of 2-4 in~\autoref{fig:thresholds} and constitute around 35\% of the correct matches.

Finally, the hardest category is the phishing pages showing disparities in design when compared to the training examples as shown in Figure~\ref{tab:hard}. These pages had distances to the training set which were higher than 4 and increased according to their differences and they constitute around 40\% of the correct matches. For example, the first three columns show a match between pages with different designs and elements' locations. Also, the fourth phishing page has a pop-up window that partially occludes information and changes the page's colors. The fifth phishing page is challenging as it does not show the company logo, yet it was correctly matched to the targeted website due to having other similar features. This suggests that \method{} captures the look and feel of websites, which makes it have an advantage over previous matching methods that relied only on logo matching such as~\cite{afroz2011phishzoo,dunlop2010goldphish}. 
The last two pages are highly dissimilar to the matched page except for having the same logo and other similar colors. Even though these examples could arguably be easily recognized as phishing pages by users, they are more challenging to be detected based on similarity and therefore they were excluded in previous studies such as~\cite{mao2017phishing}, however, we included them for completeness.
This analysis shows the ability of \method{} to detect the similarity of phishing pages that are partially copied or created with poor quality in addition to unseen phishing pages with no counterparts in the training \new{trusted-list}, which all are possible attempts to evade detection in addition to the ones we previously discussed. We also show in~\autoref{appendix_results} phishing examples targeting different websites that have highly similar colors but they were correctly distinguished from each other.

\begin{table*}[!t]
\centering
\resizebox{0.97\linewidth}{!}{%
\begin{tabular}{l|c|c|c|c|c|c|c}
\rot{\large{Phishing test}} & \includegraphics[width=3cm,height=3cm,keepaspectratio] {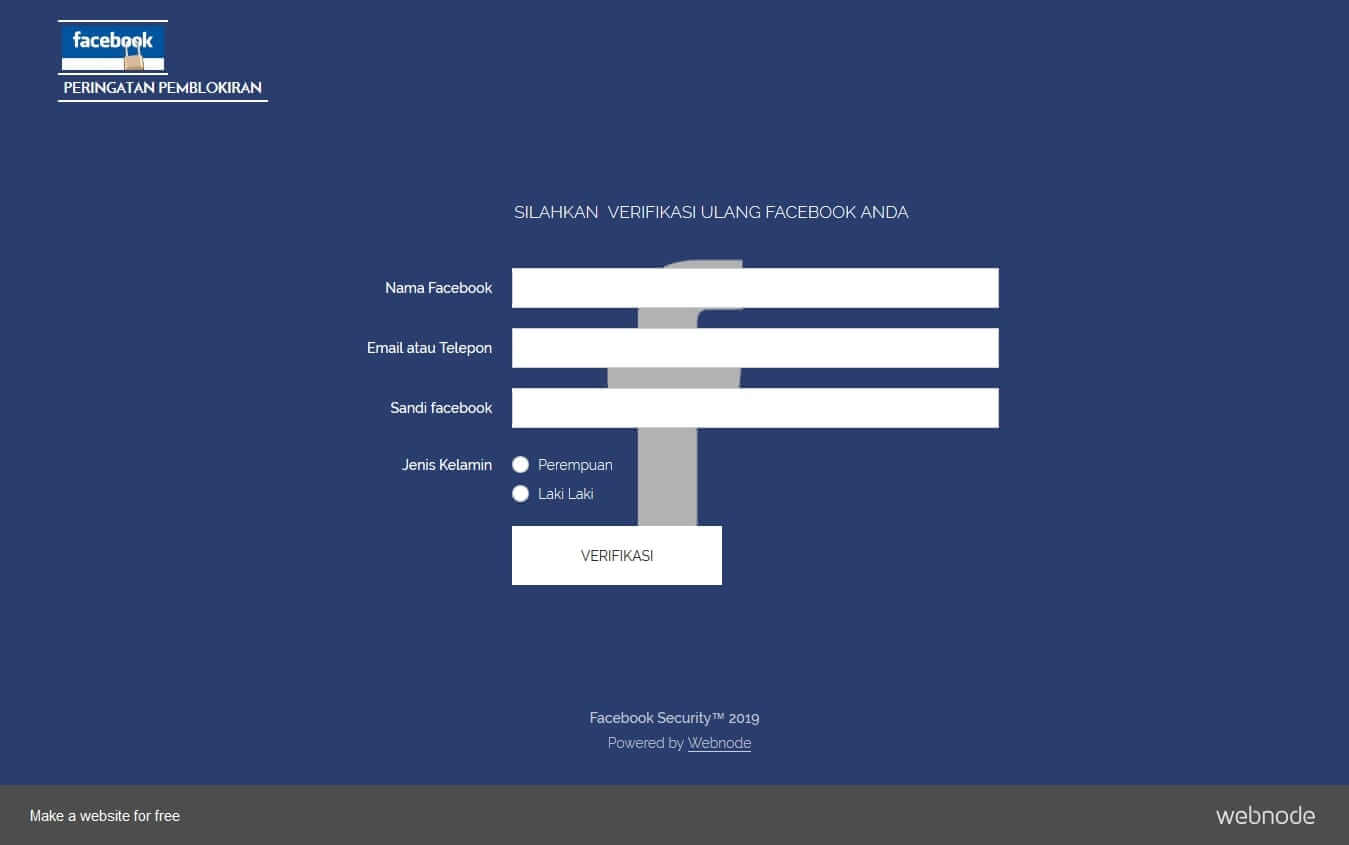} & \includegraphics[width=3cm,height=3cm,keepaspectratio] {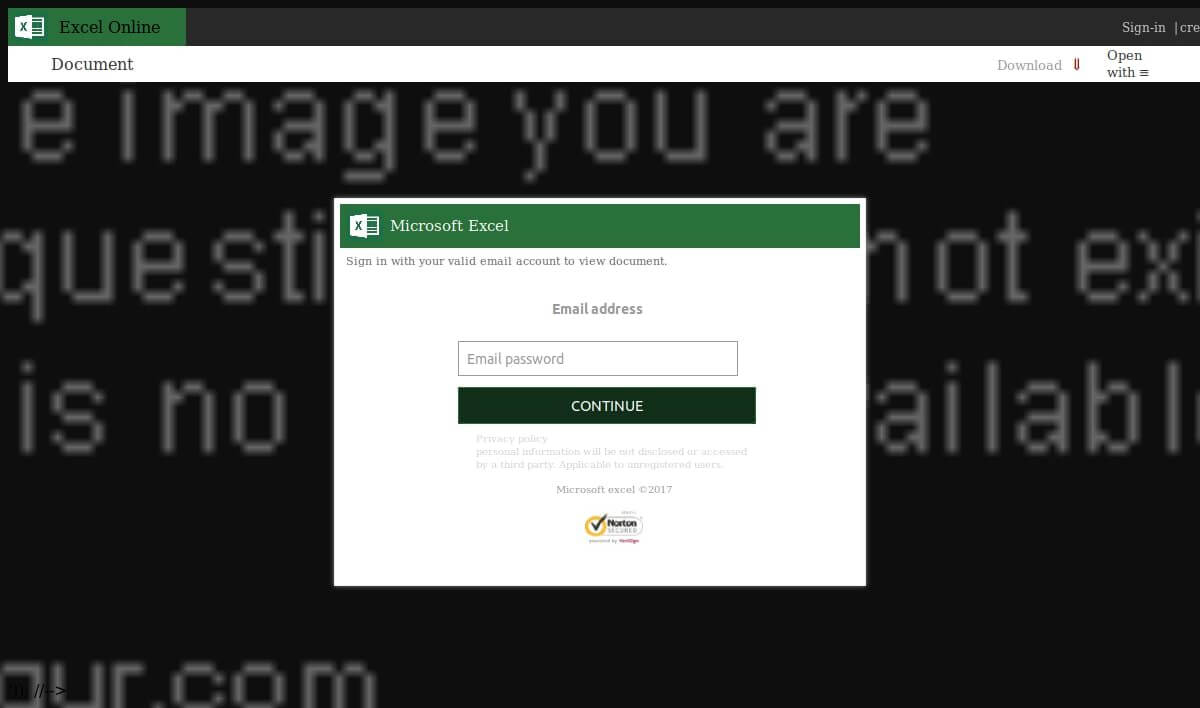} &
\includegraphics[width=3cm,height=3cm,keepaspectratio] {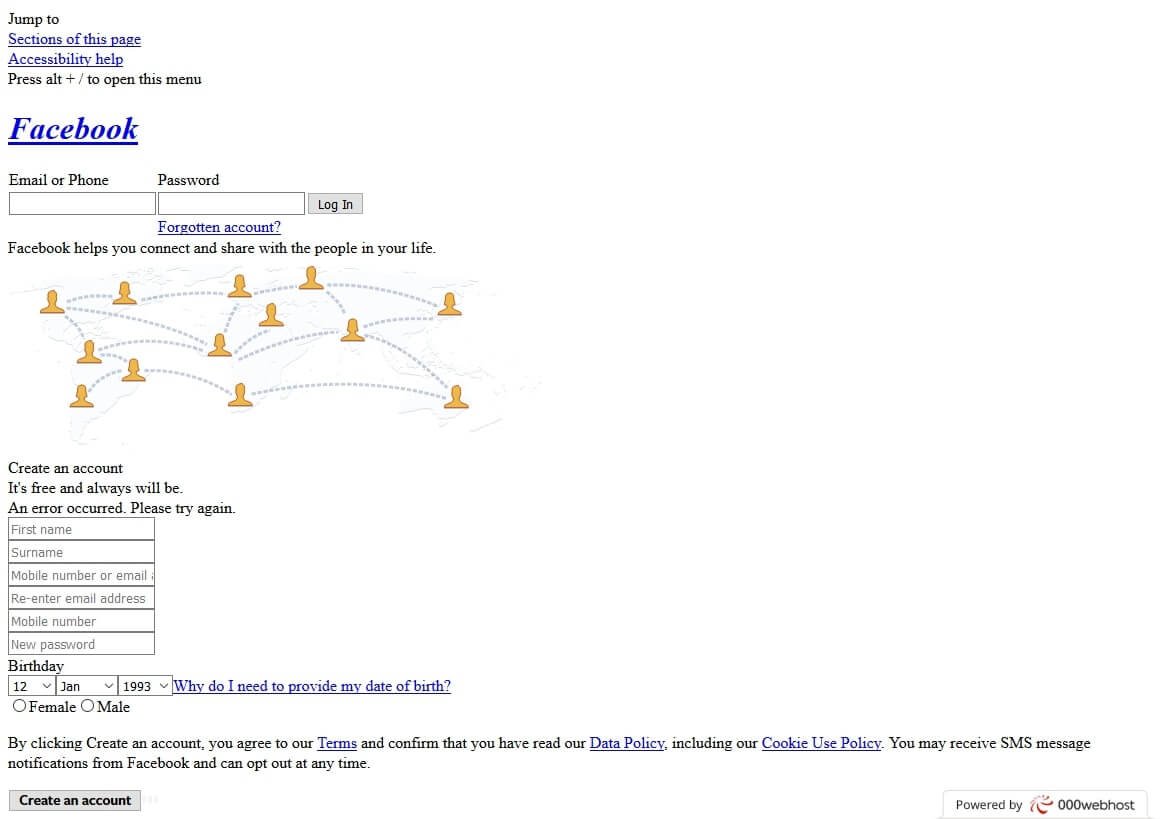} &
\includegraphics[width=3cm,height=3cm,keepaspectratio] {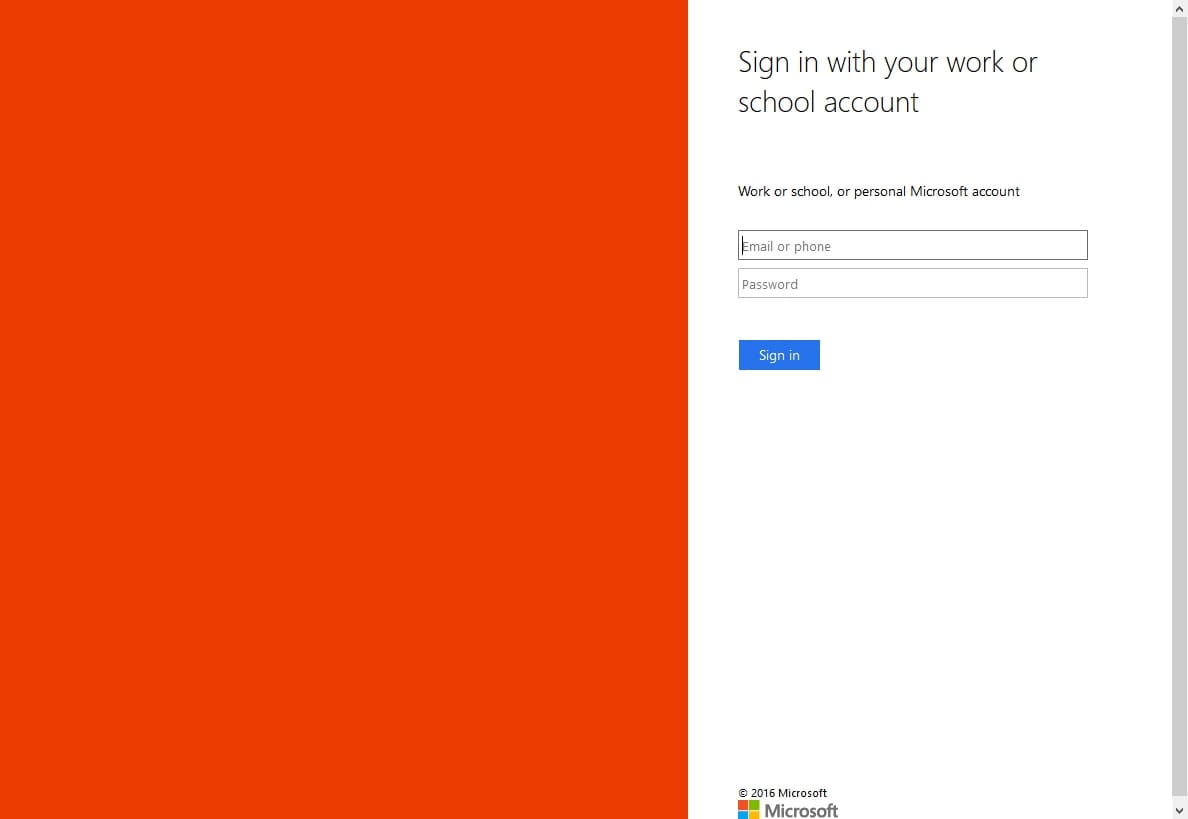} & \includegraphics[width=3cm,height=3cm,keepaspectratio] {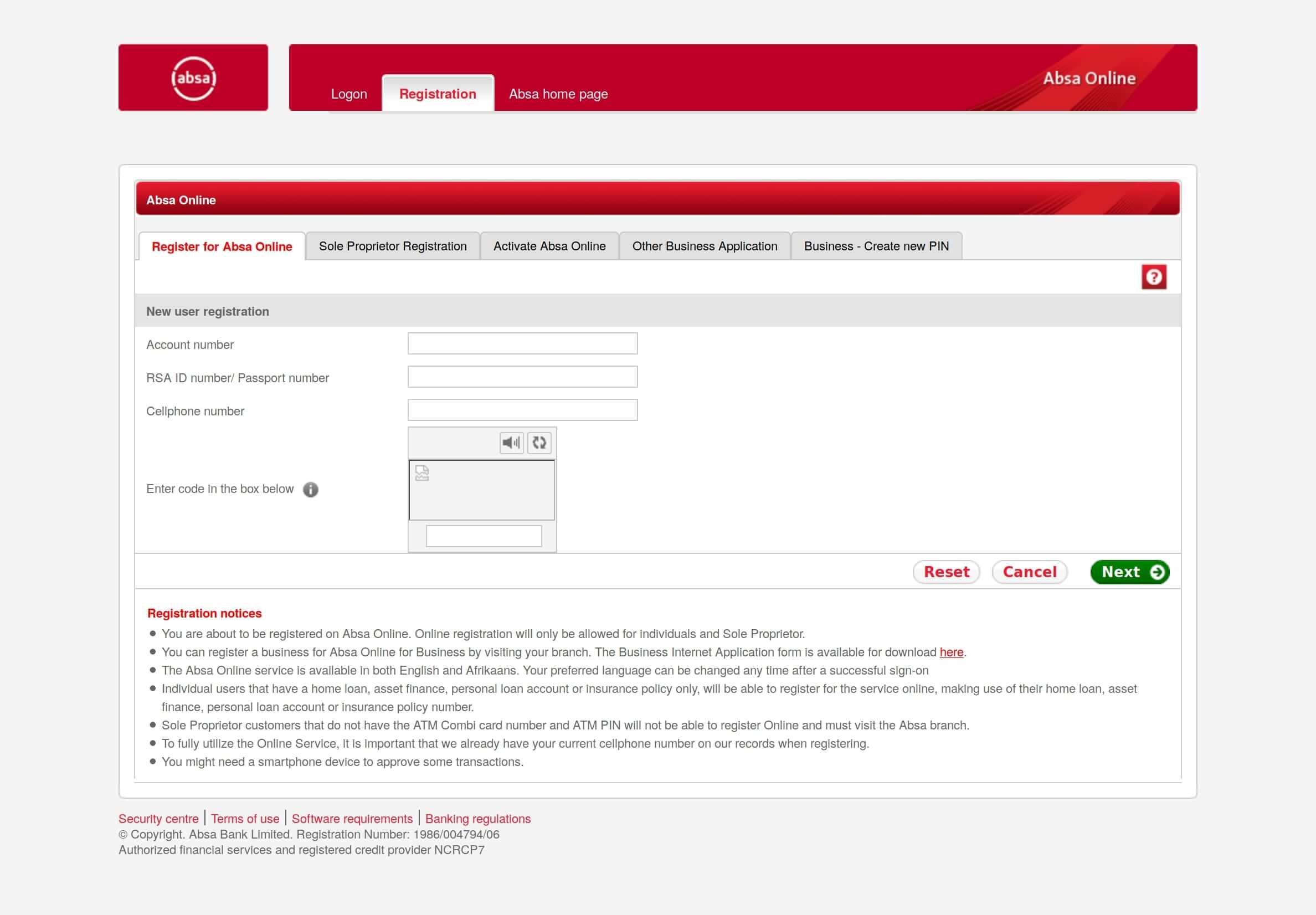} & \includegraphics[width=3cm,height=3cm,keepaspectratio] {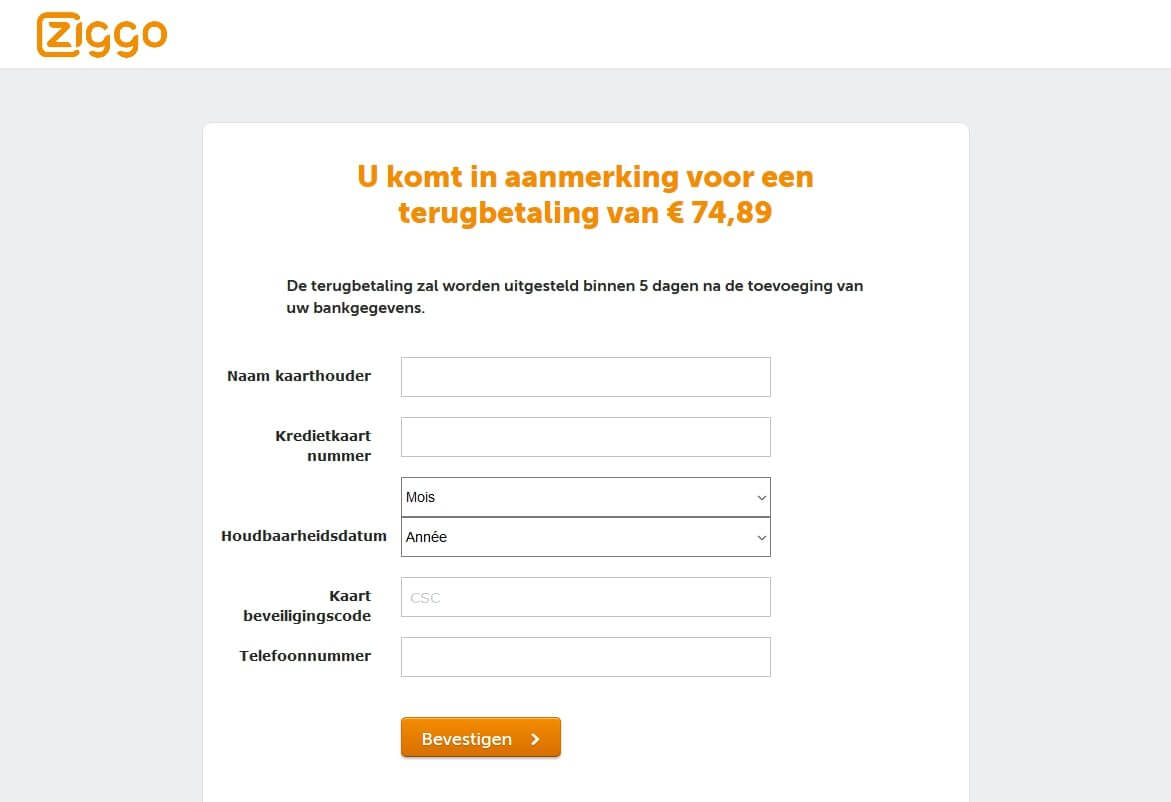} &
\includegraphics[width=3cm,height=3cm,keepaspectratio] {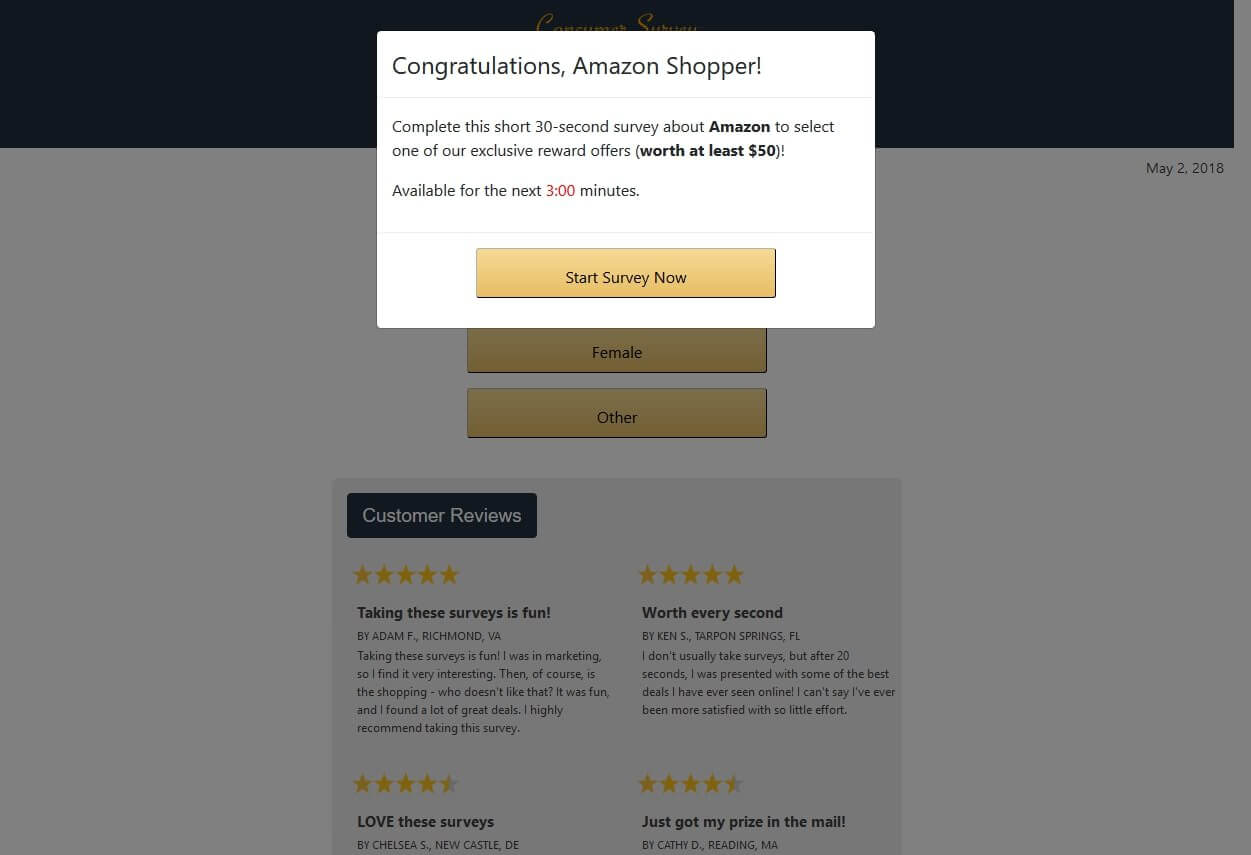} \\ 
&&&&&&\\
\rot{\large{Closest match}} & \includegraphics[width=3cm,height=3cm,keepaspectratio] {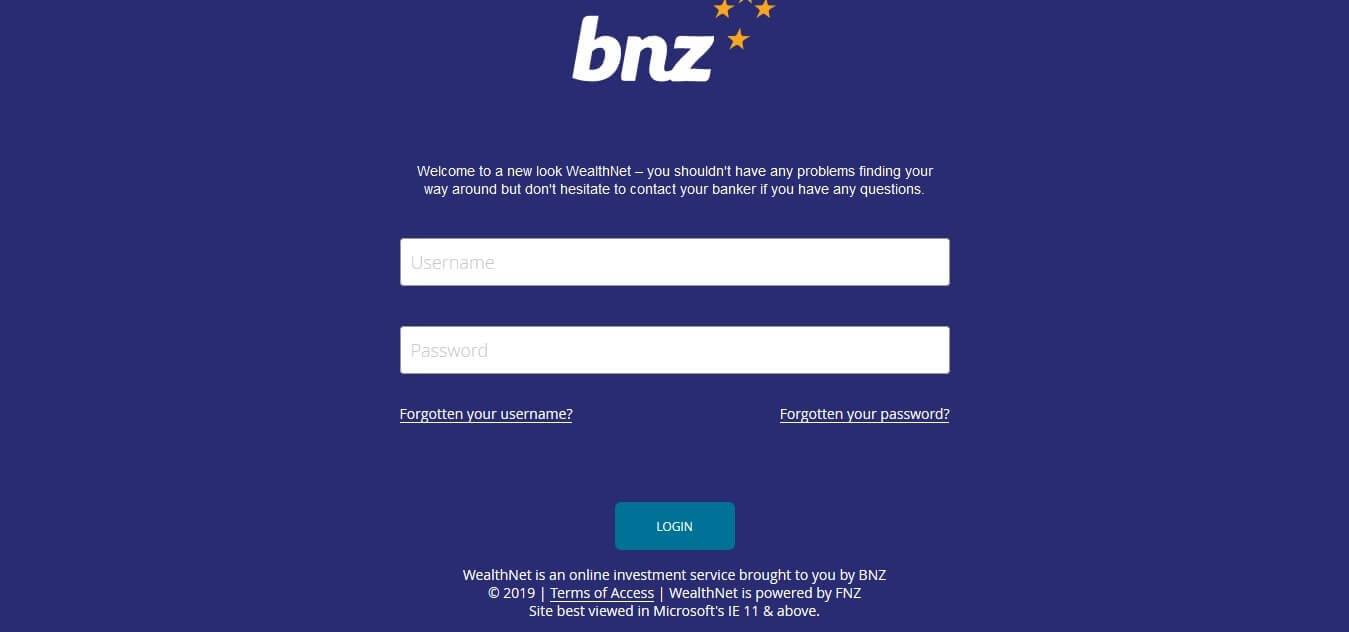} & \includegraphics[width=3cm,height=3cm,keepaspectratio] {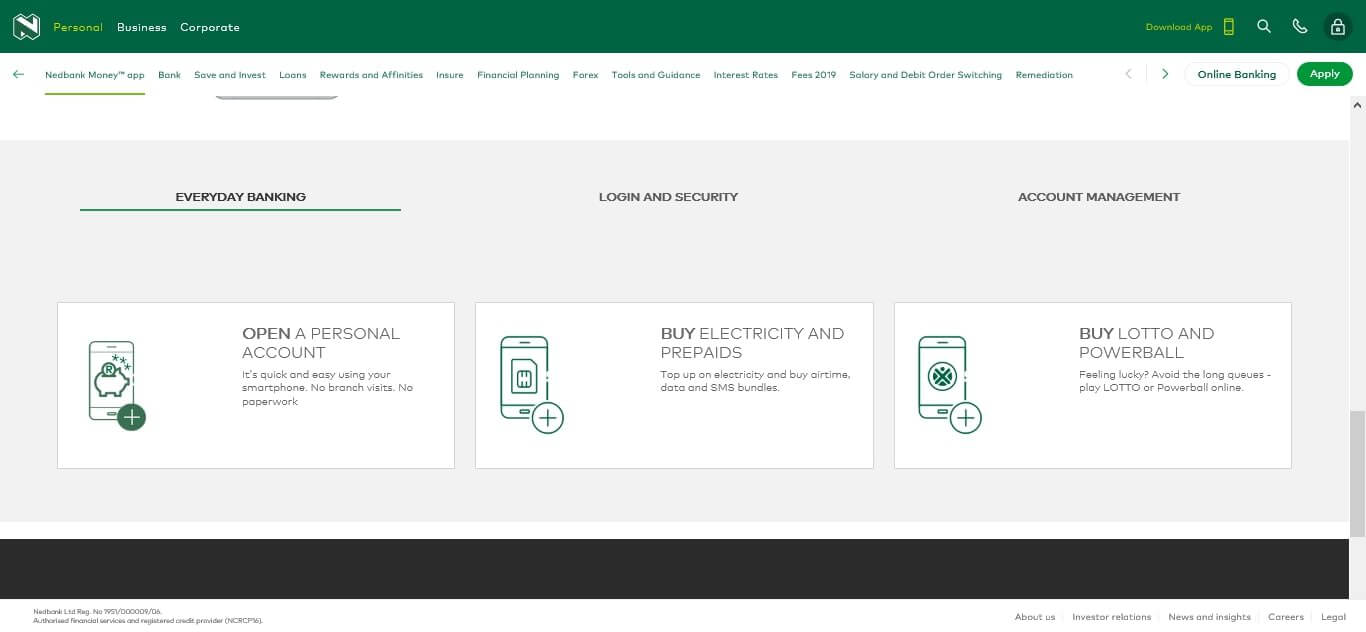} &
\includegraphics[width=3cm,height=3cm,keepaspectratio] {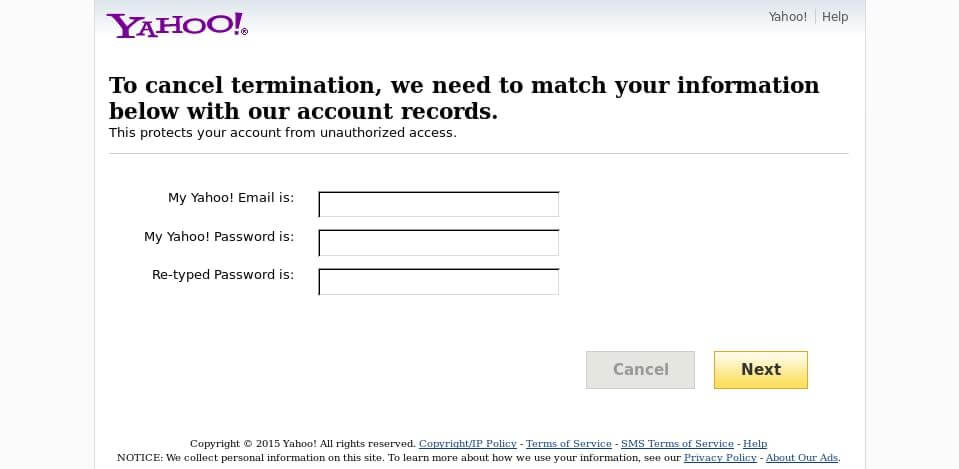} &
\includegraphics[width=3cm,height=3cm,keepaspectratio] {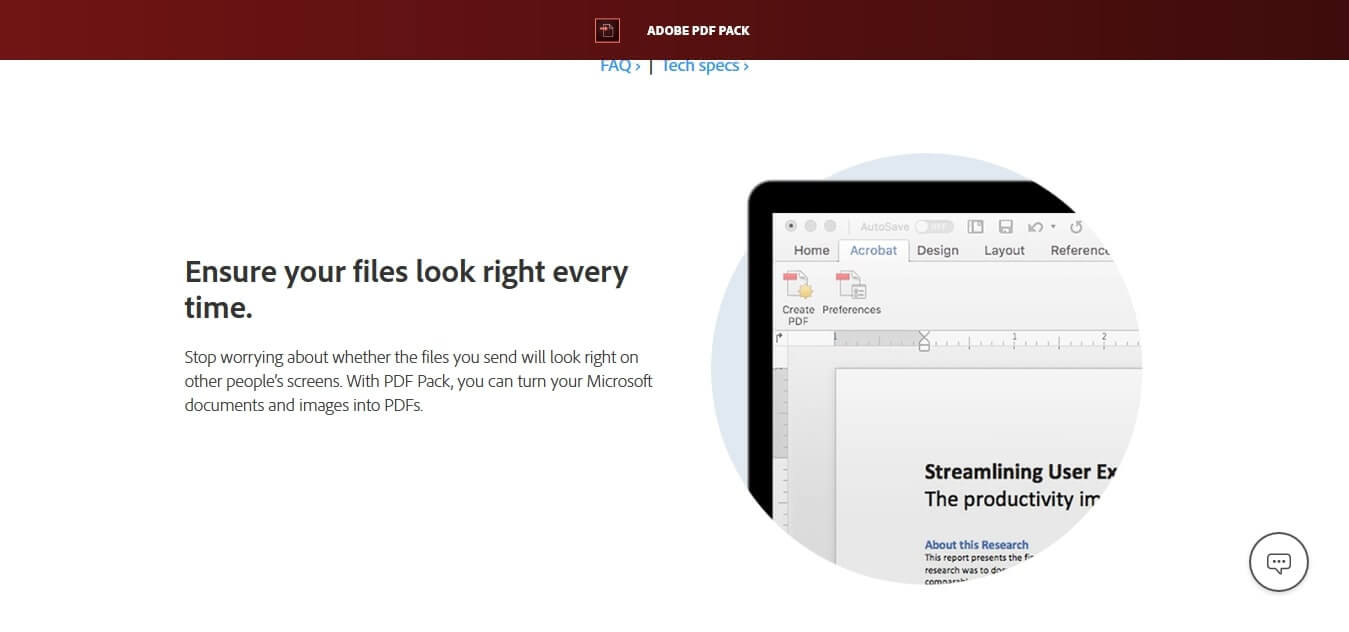} & \includegraphics[width=3cm,height=3cm,keepaspectratio] {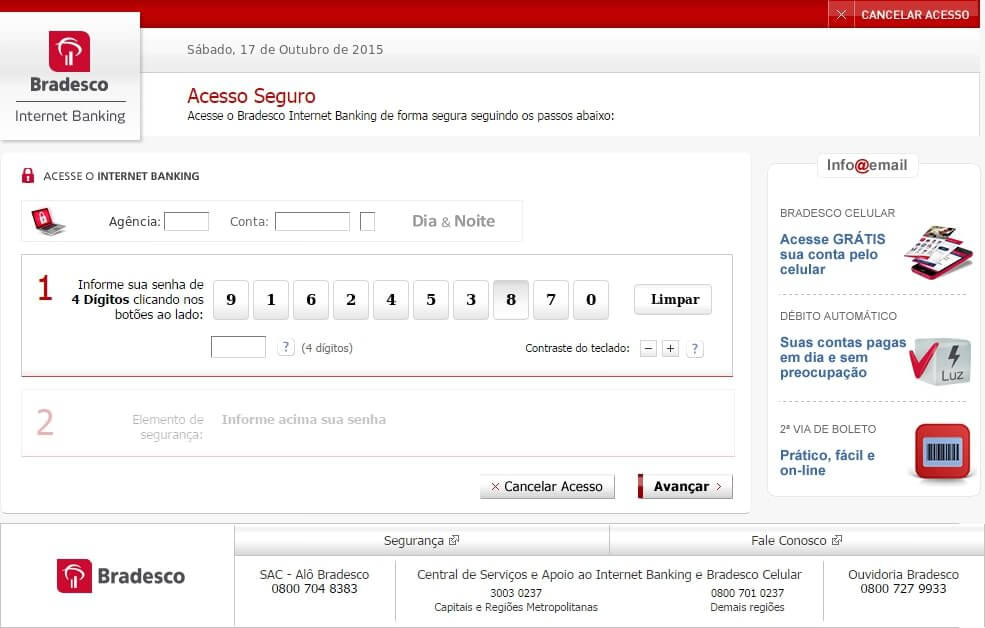}  & \includegraphics[width=3cm,height=3cm,keepaspectratio] {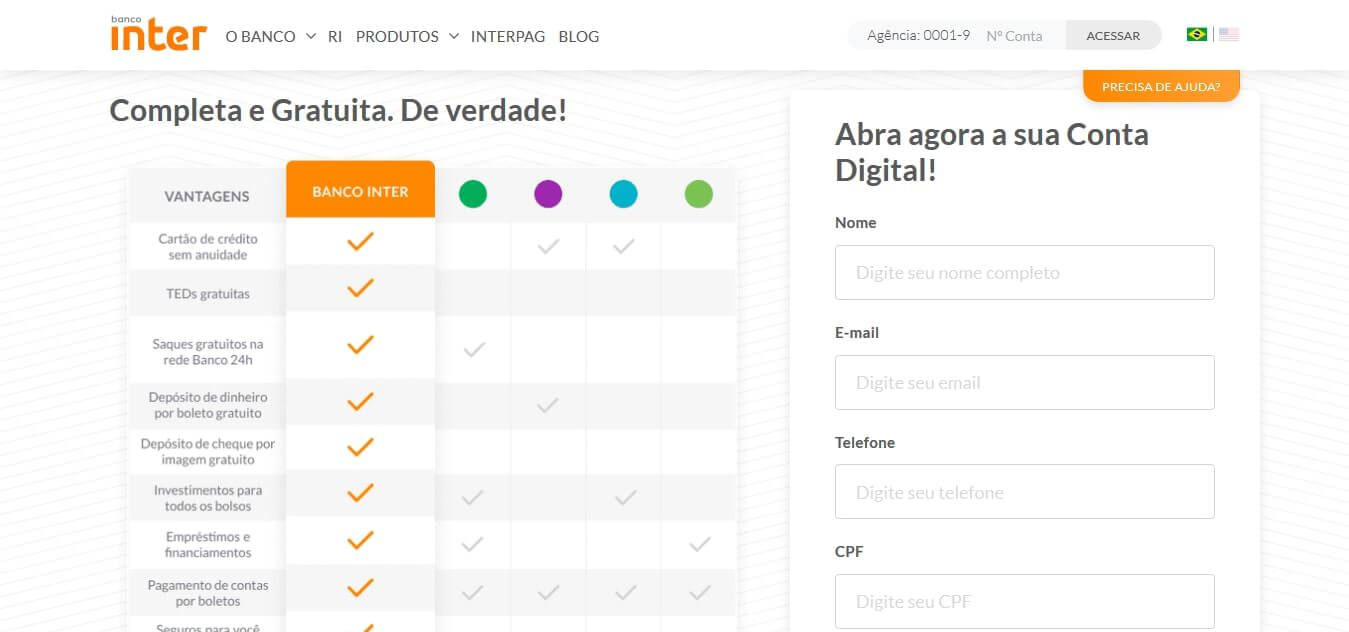} & 
\includegraphics[width=3cm,height=3cm,keepaspectratio] {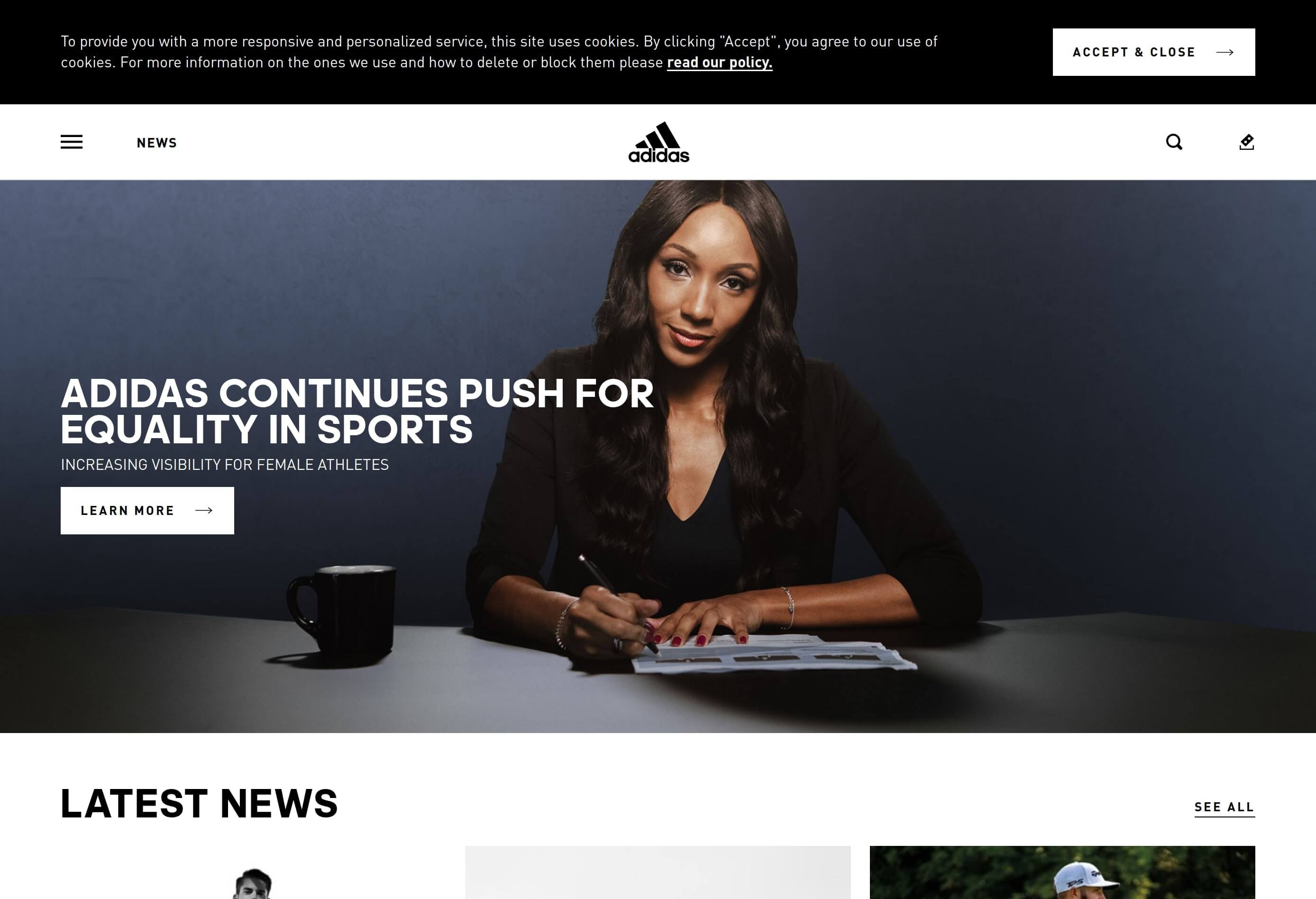} \\
\end{tabular}}
\captionof{figure}{Examples of test phishing webpages that were matched to the wrong website from the training set.} \label{tab:wrong}
\end{table*} 

Since these successful matches suggest that the logo of a page plays an important factor in the matching decision, possible false matches could happen if a benign page contains another website logo. To evaluate this, we collected a benign subset of 125 pages (see~\autoref{appendix_dataset}) that contain the logos of one or more of 9 \new{trusted} websites. These pages are articles about a website, or login pages with other websites' sign-in or sharing options. However, only 3.07\% of these pages were matched to the website whose logo appears in the screenshot which indicates that the learnt profiles incorporate more visual cues than logos only.

\subsection{Evaluating Failure Modes} \label{failures}
We also analysed the failure modes of the model including wrong websites matches and false positives. We found that the highest mismatches are for phishing examples belonging to Facebook, Dropbox, Microsoft one drive, Microsoft Office, and Adobe. We found that these websites have many phishing pages with dissimilar appearances (and poor designs) compared to the targeted websites, such as the first three phishing pages targeting Facebook and Microsoft Excel in Figure~\ref{tab:wrong} (see also~\autoref{appendix_dataset} for more examples). On the other hand, phishing pages targeting banks had higher quality in copying and appeared plausible and similar to the targeted websites making them have fewer mismatches (see~\autoref{appendix_results} for a histogram of wrong matches). To analyse how successful these dissimilar pages in fooling users, we conducted an online study where users were shown dissimilar and relatively similar phishing pages and were asked to evaluate how trustworthy they seem based only on their appearance. Only 3.02\% said that they would trust the dissimilar examples as opposed to 65.3\% in the case of the relatively similar ones (see~\autoref{appendix_dataset} for examples used in the study). 

We also found some phishing pages that used outdated designs or earlier versions of certain login forms such as the fourth example in Figure~\ref{tab:wrong} (that is now changed entirely in Microsoft website) and were, therefore, matched to a wrong website. This could be improved by including earlier versions of websites in the training data. Moreover, the last three examples in Figure~\ref{tab:wrong} show some of the main limitations. Since our \new{training trust-list} contains a large number of screenshots per website, we have many distractors of potentially similar pages to the query screenshot, such as the fifth and sixth examples in Figure~\ref{tab:wrong} that were matched to similar screenshots from different websites. We also found that some phishing pages have pop-up windows that completely covered the logo and the page's colors and structure, and were then matched to pages with darker colors such as the last example in Figure~\ref{tab:wrong}. The wrong matches had generally higher distances than the correct matches which could make them falsely classified as legitimate examples.
 
We also show false positive examples (benign test pages) that had high similarity to pages from the training set in Figure~\ref{img:false_pos} and would be falsely classified as phishing pages based on the threshold in~\autoref{fig:thresholds}. We observed that pages with forms were harder to identify as dissimilar to other pages with forms in the \new{trust-list} especially when having similar colors and layout, since they contain few distinguishable and salient elements and they are otherwise similar. We believe that using the screenshot's text (possibly extracted by OCR), or more incorporation of the logo features along with other visual cues by region-based convolution~\cite{ren2015faster} could be future possible model optimization directions to help reduce the false positives and also improve the matching of hard examples. Additionally, tackling the phishing problem has many orthogonal aspects; while we focus on visual similarity to detect zero-day \new{and unseen} pages and achieve a significant leap in performance, our approach could still be used along with other allow-listing of trusted domains to further reduce the false positives.
\begin{table}[!b]
\centering
\resizebox{\linewidth}{!}{%
\begin{tabular}{l|c|c|c|c}
\rot{\small{Legitimate test}} & \includegraphics[width=2cm,height=3.5cm,keepaspectratio] {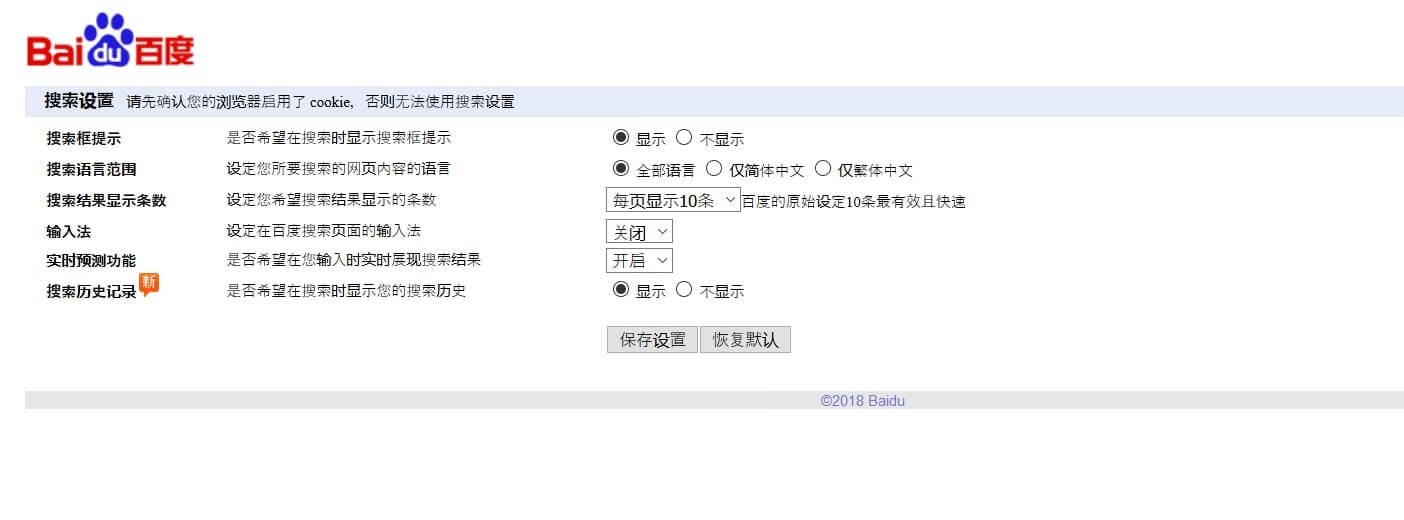} & \includegraphics[width=2.5cm,height=2.5cm,keepaspectratio] {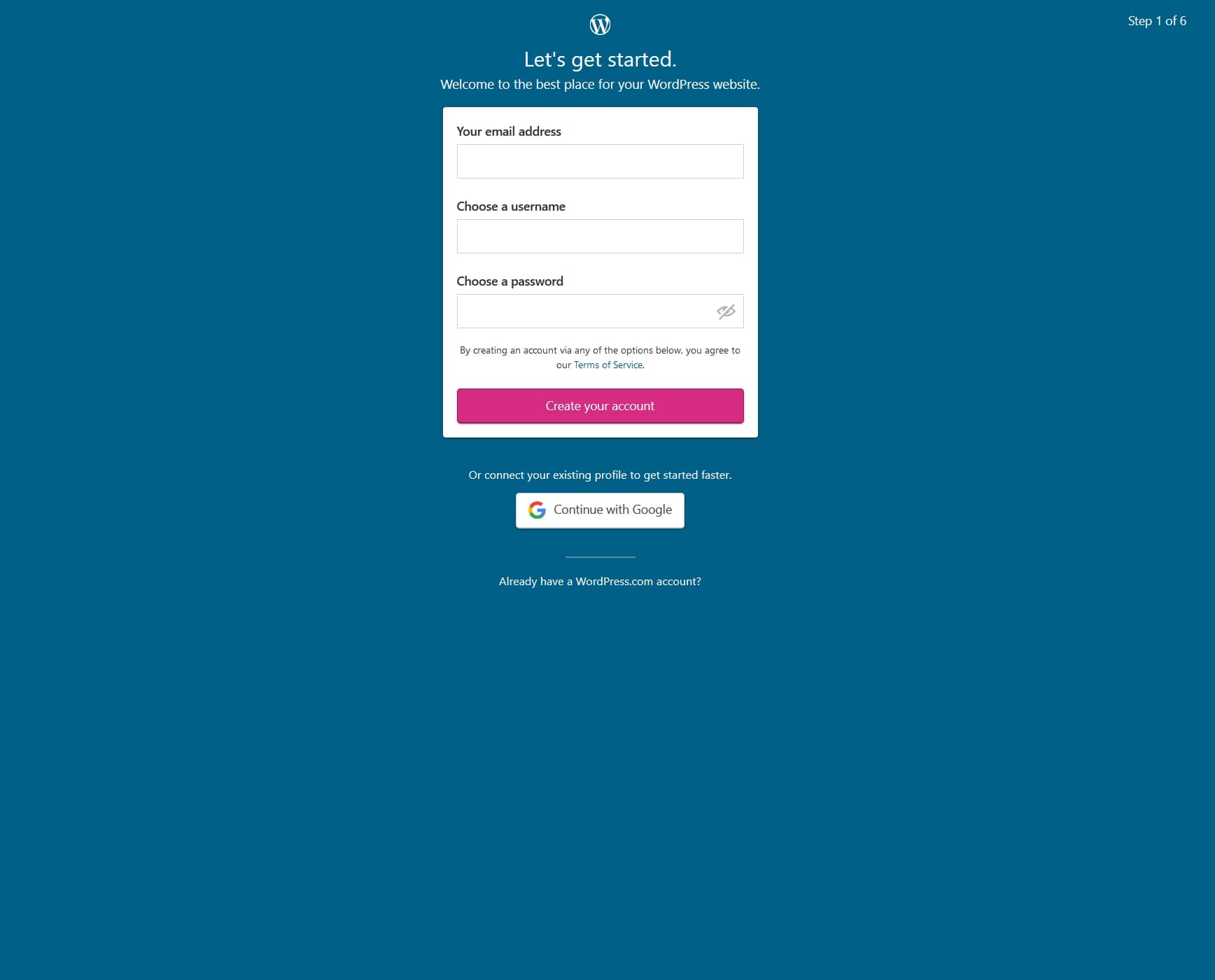}  & 
\includegraphics[width=2.5cm,height=2.5cm,keepaspectratio] {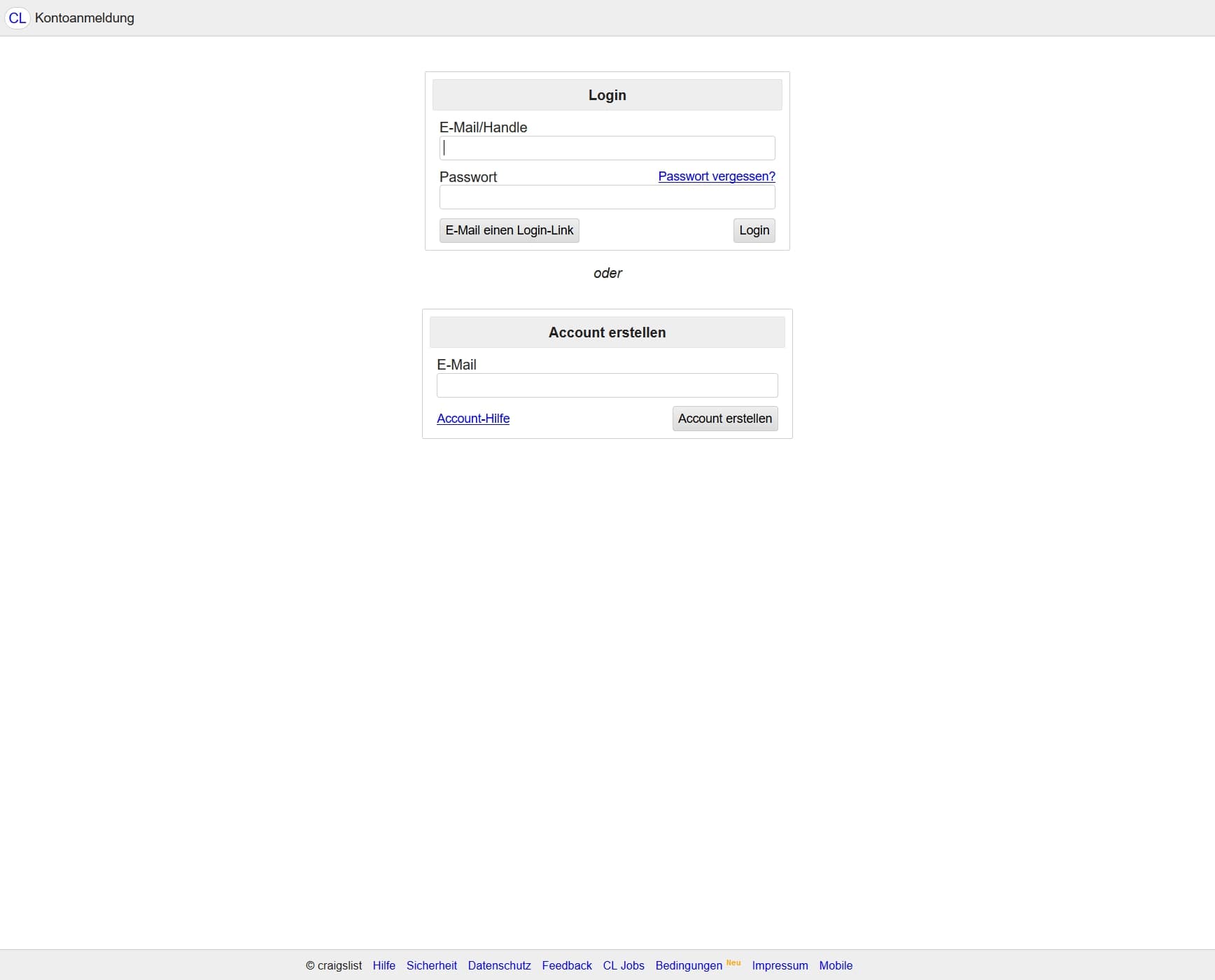} &
\includegraphics[width=2.5cm,height=2.5cm,keepaspectratio] {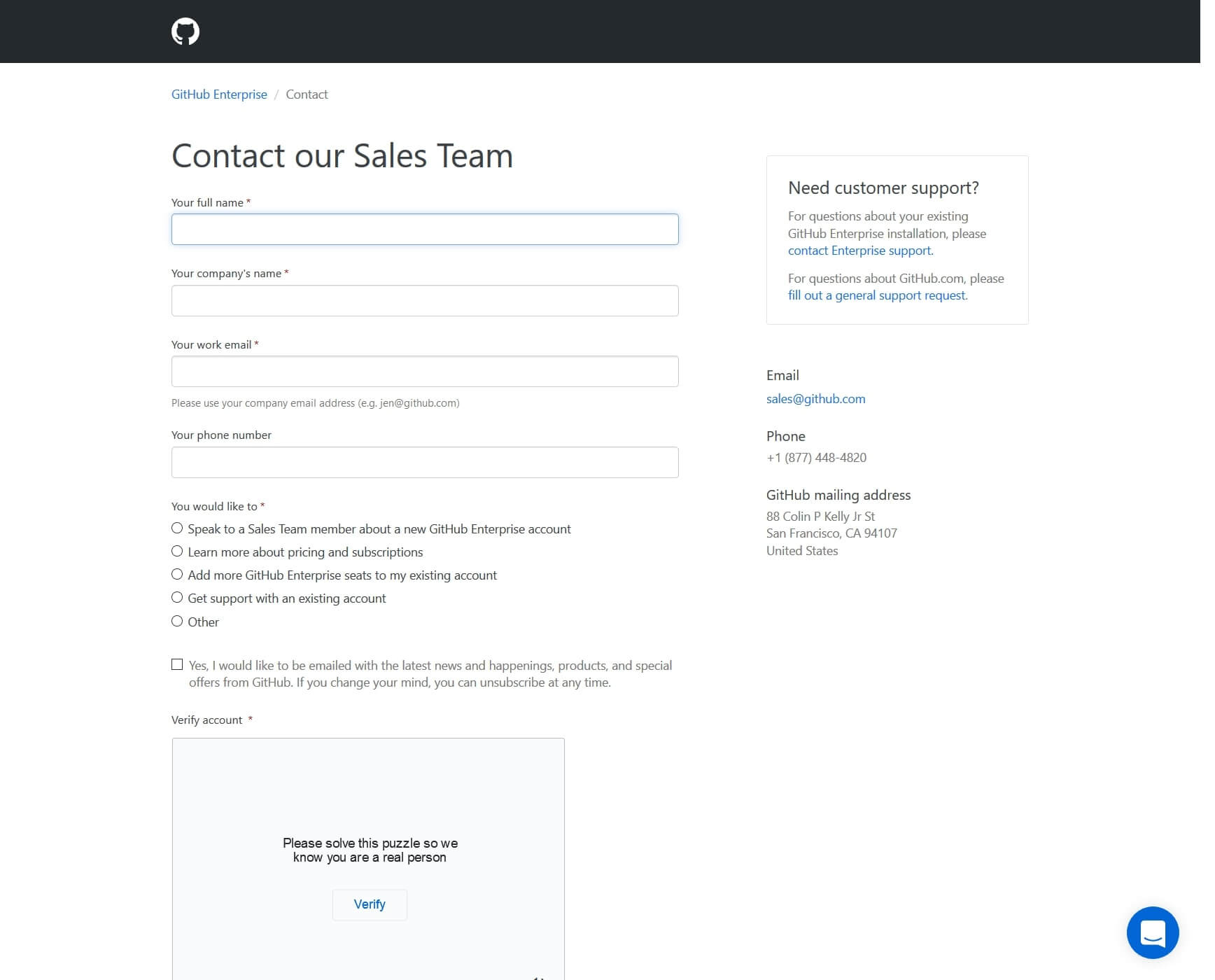} \\ 
&&&&\\
\rot{\small{Closest match}} & \includegraphics[width=2cm,height=3.5cm,keepaspectratio] {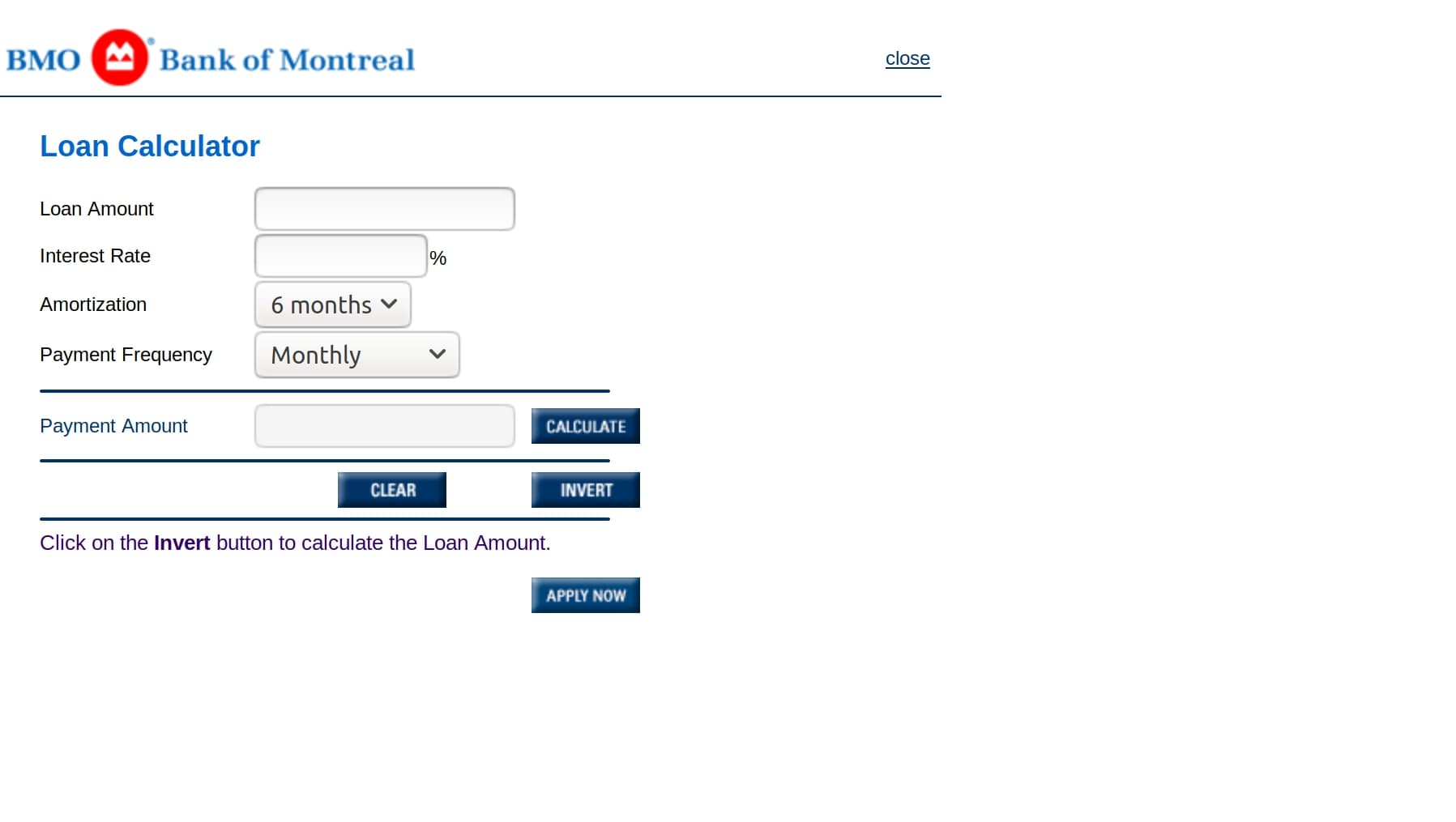} & \includegraphics[width=2.5cm,height=2.5cm,keepaspectratio] {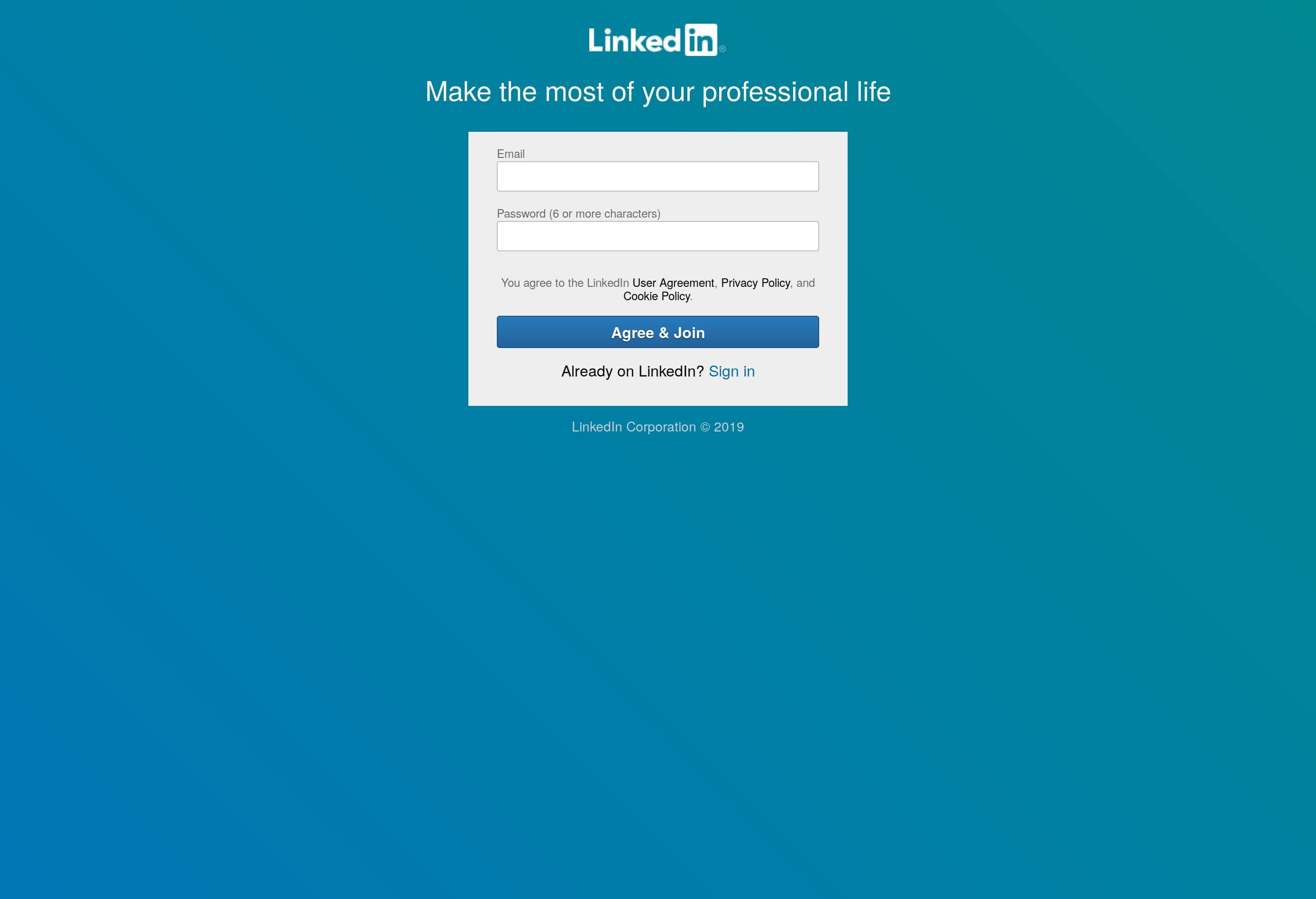} &
\includegraphics[width=2.5cm,height=2.5cm,keepaspectratio] {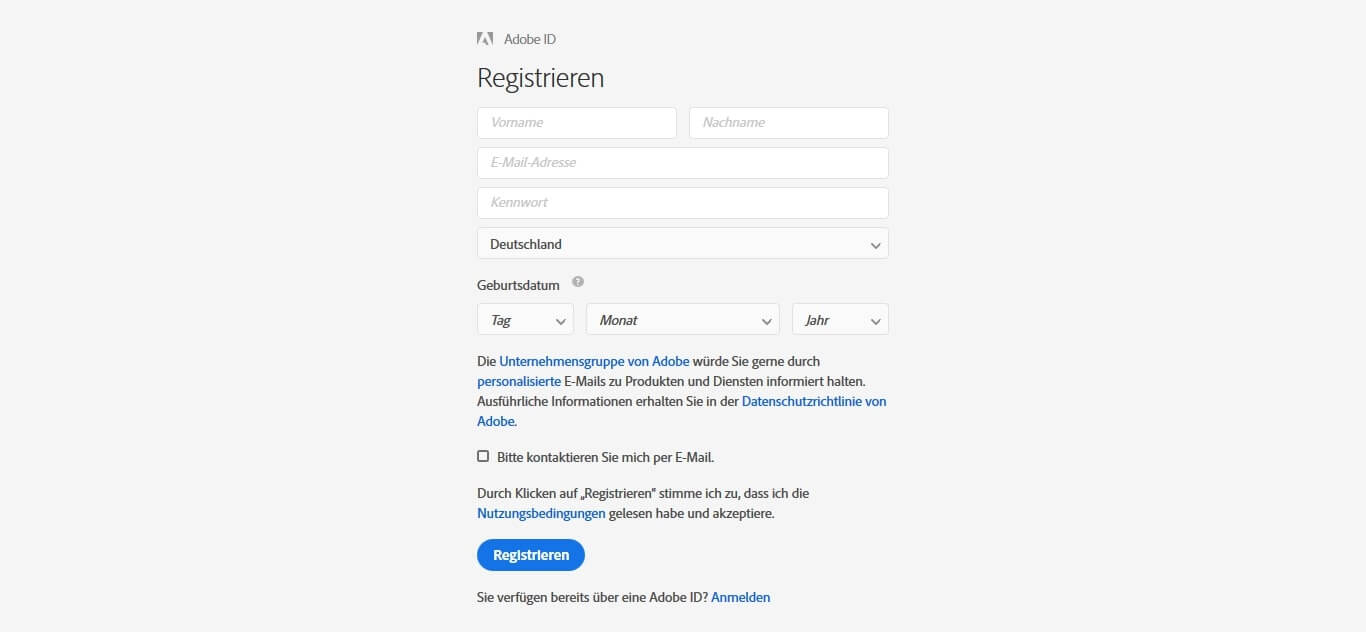} &
\includegraphics[width=2.5cm,height=2.5cm,keepaspectratio] {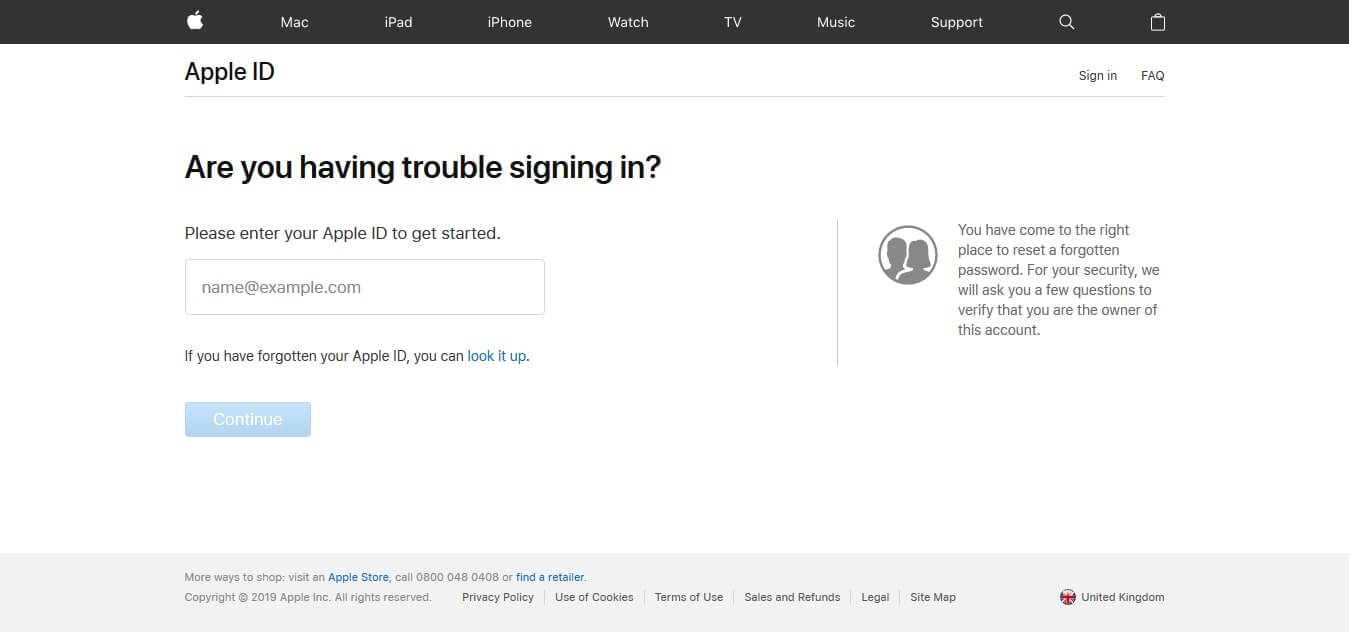} \\
\end{tabular}} 
\captionof{figure}{False positive examples of the top closest legitimate test pages to the training list.} \label{img:false_pos}
\end{table}

\subsection{Deployment Considerations} \label{deployment}
We here discuss practical considerations for the deployment of our system. First, regarding the required storage space and computation time, our system does not require storing all screenshots of the \new{trusted-list}, as it suffices to store the embedding vectors of screenshots (512-dimensional vectors). Also, the system is computationally feasible since the training \new{trusted-list} embeddings can be pre-computed, which at test time only leaves the relatively smaller computations of the query image embedding and the L2 distances. On a typical computer with 8 GByte RAM and Intel Core i7-8565U 1.80GHz processor, the average time for prediction was 1.1\rpm0.7 seconds which decreased to 0.46\rpm0.25 seconds on a NVIDIA Tesla K80 GPU.
If further speeding up is needed, the search for the closest point could be optimized. Besides, the decision could only be computed when the user attempts to submit information. We also show in our analysis of possible perturbations that the learned similarity is robust against partial removal of parts of the page, which suggests that a page could be detected even if it was partially loaded.
Other deployment issues are the browser window size variations at test time which could be solved by fixing the size of the captured screenshot. Another issue is the maintenance of the domain names of the \new{trusted-list} in case a website has changed its domain, which could be solved by rolling updates of the \new{trusted-list} without the need to retrain. Additionally, we observed that \method{} is robust against small changes or updates in the website logo's fonts or colors (e.g. see Yahoo examples with different versions that were still correctly detected in~\autoref{appendix_results}). Larger or more significant changes (that usually happen on long time intervals) might require retraining and updating. Moreover, the current system and dataset are focusing on Desktop browsers, however, the concept can be extended to other devices (e.g. smartphones) which may require re-training. Furthermore, our visual similarity model can either be used as a standalone phishing detection model or, \new{as the last defense mechanism for unseen pages} along with other (potentially faster) listing or heuristics approaches.
Regarding the \dataset{} dataset, we point out that the manual work in curating the dataset was mainly for constructing unbiased and non-duplicated test sets, however, it is less needed in collecting the training \new{trusted-list} of \new{trusted} websites. This enables the automatic update of the \new{trusted-list} to add new websites when needed. Nevertheless, detecting duplicity can be automated by finding the closest pages to the newly added one based on pixel-wise features (such as VGG features).

\section{Conclusion}
As visual similarity is a key factor in detecting zero-day phishing pages, in this work, we proposed a new framework for visual similarity phishing detection. We presented a new dataset (\dataset{}: 155 websites with 9363 screenshots) that covers the largest \new{trusted-list} so far and overcomes the observed previous limitation.

Unlike previous work, instead of only matching a phishing page to its legitimate counterpart, we generalize visual similarity to detect \new{unseen} pages targeting the \new{trusted} websites. To that end, we proposed \method{} that learns a visual profile of websites by learning a similarity metric between any two same-website pages despite having different contents. Based on our qualitative analysis of the successful cases of \method{}, our network identified easy phishing pages (highly similar to pages in training), and more importantly, phishing pages that were partially copied, obfuscated, or unseen. \method{} was found to be robust against the range of possible evasion attacks and perturbations that we studied, which makes our model less prone to the fierce arms race between attackers and defenders.

In conclusion, our work introduces important contributions to phishing detection research to learn a robust and proactive visual similarity metric that demonstrates a leap in performance over prior visual similarity approaches by an increase of 56 percent points in matching accuracy and 30 in the classification ROC area under the curve. 

\bibliographystyle{ACM-Reference-Format}
\bibliography{ccs-sample}

\appendix

\section{Extra Evaluation and Qualitative Results} \label{appendix_results}
\begin{figure}[!htbp]
    \centering
    \includegraphics [width =\linewidth, height=3.5cm,keepaspectratio]{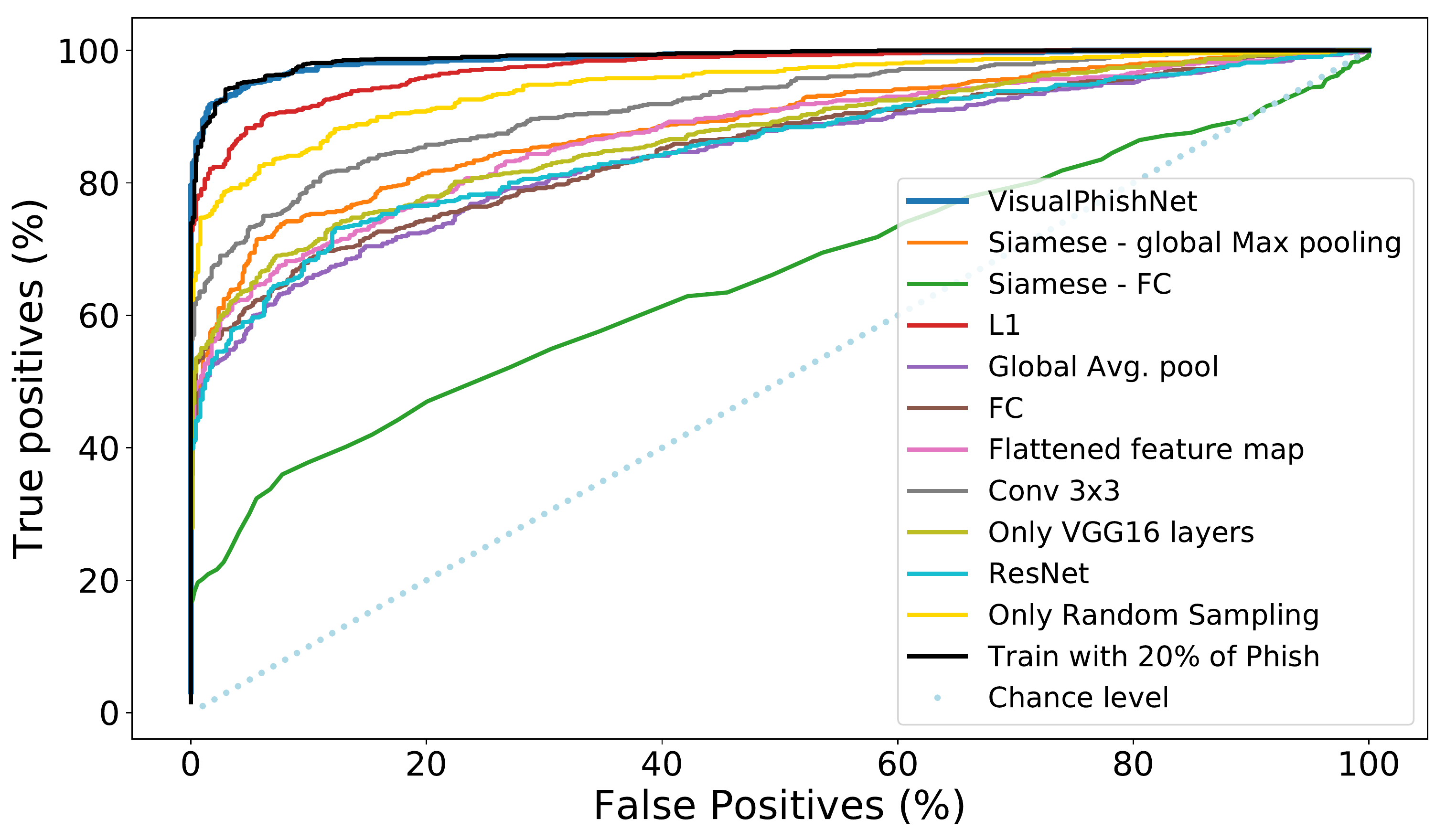}
    \caption{ROC curves for the ablation study in~\autoref{tab:summary}. The legend follows the same order of rows in~\autoref{tab:summary}. } \label{fig:ablation_roc}
\end{figure}

\begin{figure}[!htbp]
    \centering
    \includegraphics[width=\linewidth,height=4cm,keepaspectratio]{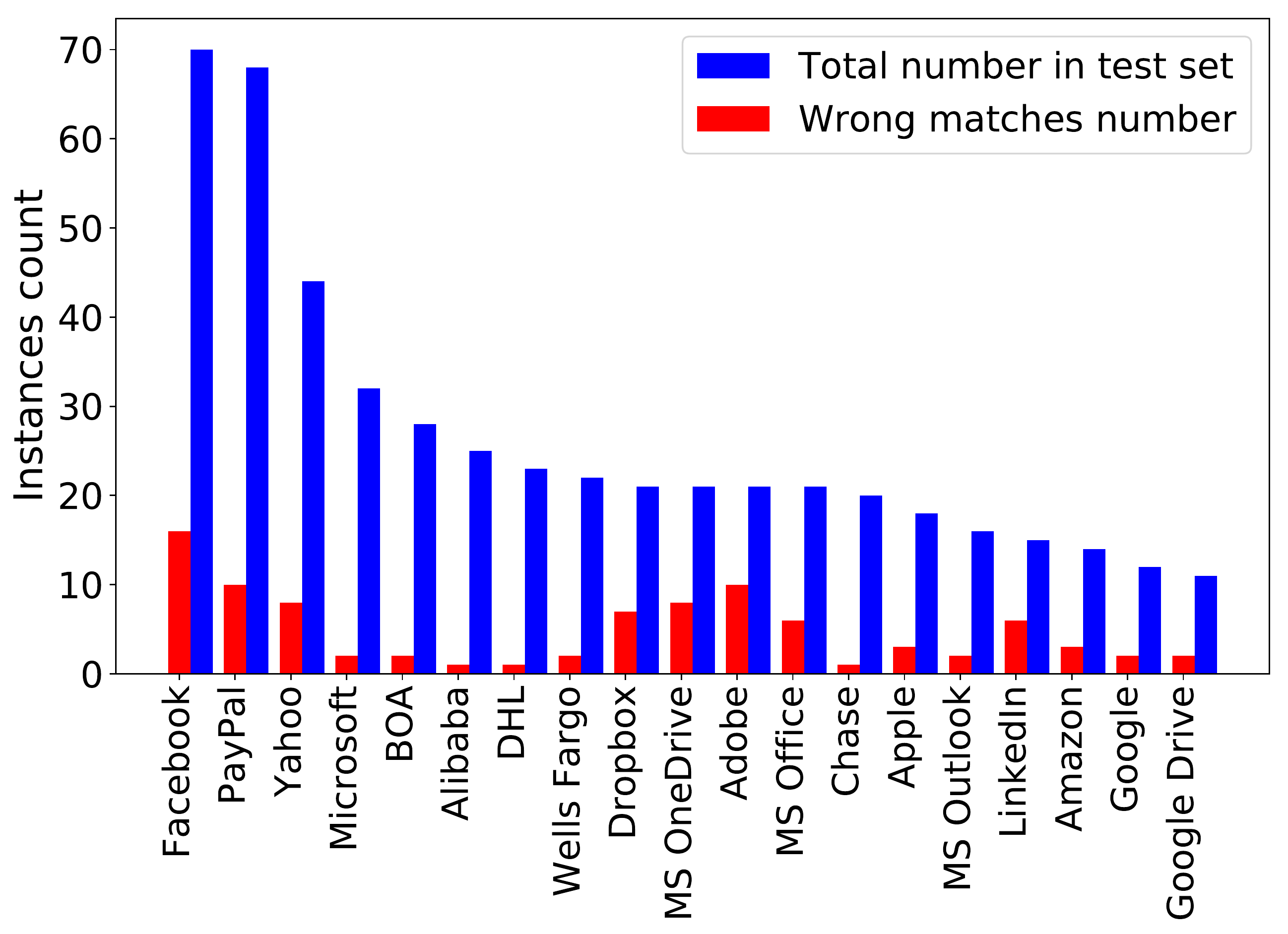} 
    \caption{Histogram of the wrong matches of phishing pages to their targeted website. The most frequent 19 websites are shown.}
    \label{fig:wrong_matches_hist}
\end{figure}

We here show supplementary results. In~\autoref{fig:ablation_roc}, we present the ROC curves of the binary classification task for each experiment in the ablation study (discussed in~\autoref{ablation}). In~\autoref{fig:wrong_matches_hist}, we show a histogram of the phishing pages false matches per website (\autoref{failures}). In Figure~\ref{tab:new_matches}, we show successful matching examples of the new crawled phishing pages (\autoref{new_data}). Moreover, in Figure~\ref{tab:sim_colors}, we show correct matches across different websites with similar colors that were still correctly distinguished from each other (\autoref{discuss_qual}). Finally, in Figure~\ref{tab:yahoo_diff}, we show successful examples where the website logo's colors and fonts were different than the \new{trusted-list} to test versions changes (\autoref{deployment}).

\begin{table}[!b]
\centering
\resizebox{\linewidth}{!}{%
\begin{tabular}{c|l|l|l}
\rot{New pages} &\includegraphics[width=2.5cm,height=2.5cm,keepaspectratio] {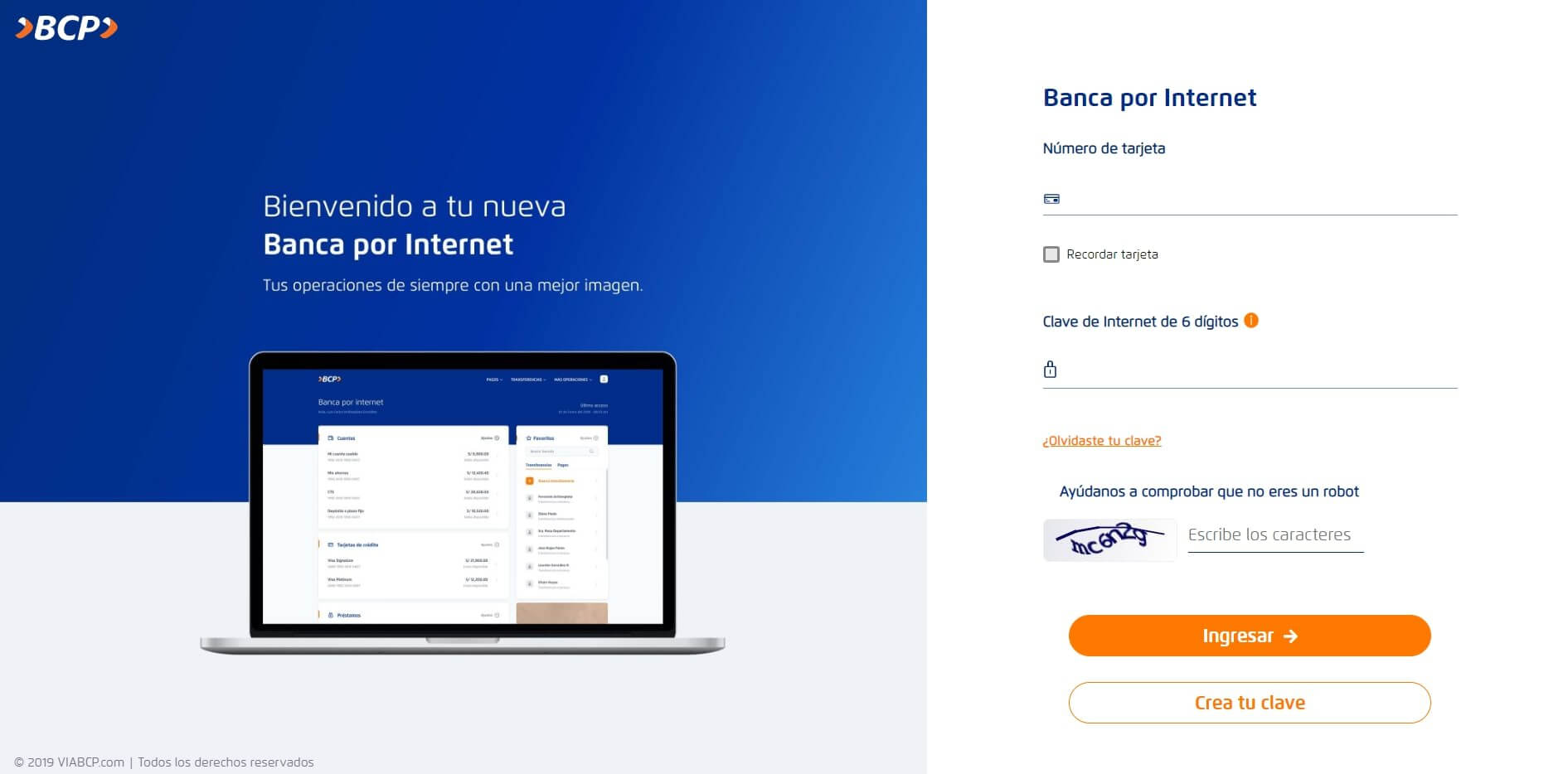} &
\includegraphics[width=2.5cm,height=2.5cm,keepaspectratio] {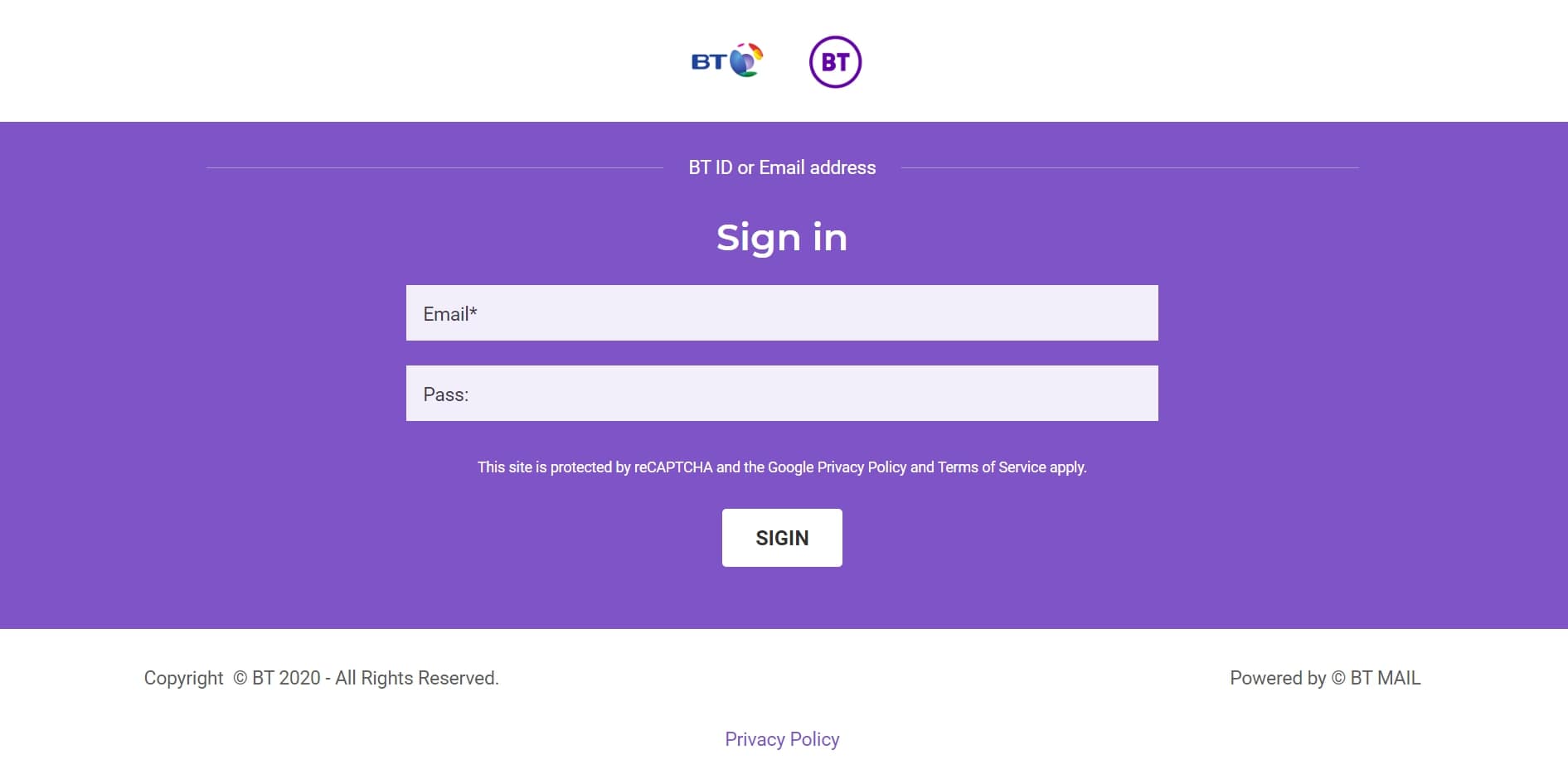} & \includegraphics[width=2.5cm,height=2.5cm,keepaspectratio] {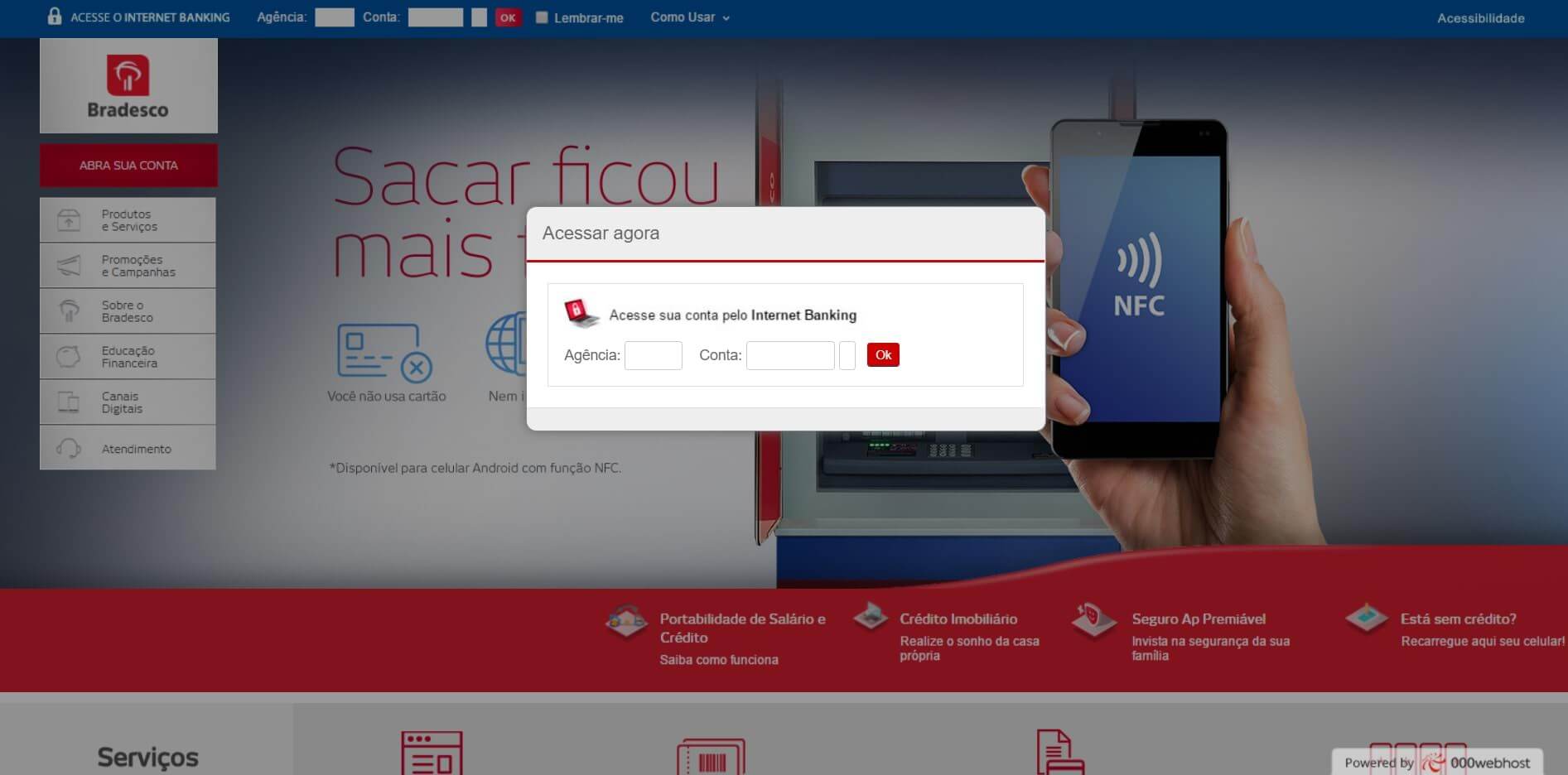}  \\ 
&&&\\
\rot{Closest match} &\includegraphics[width=2.5cm,height=2.5cm,keepaspectratio] {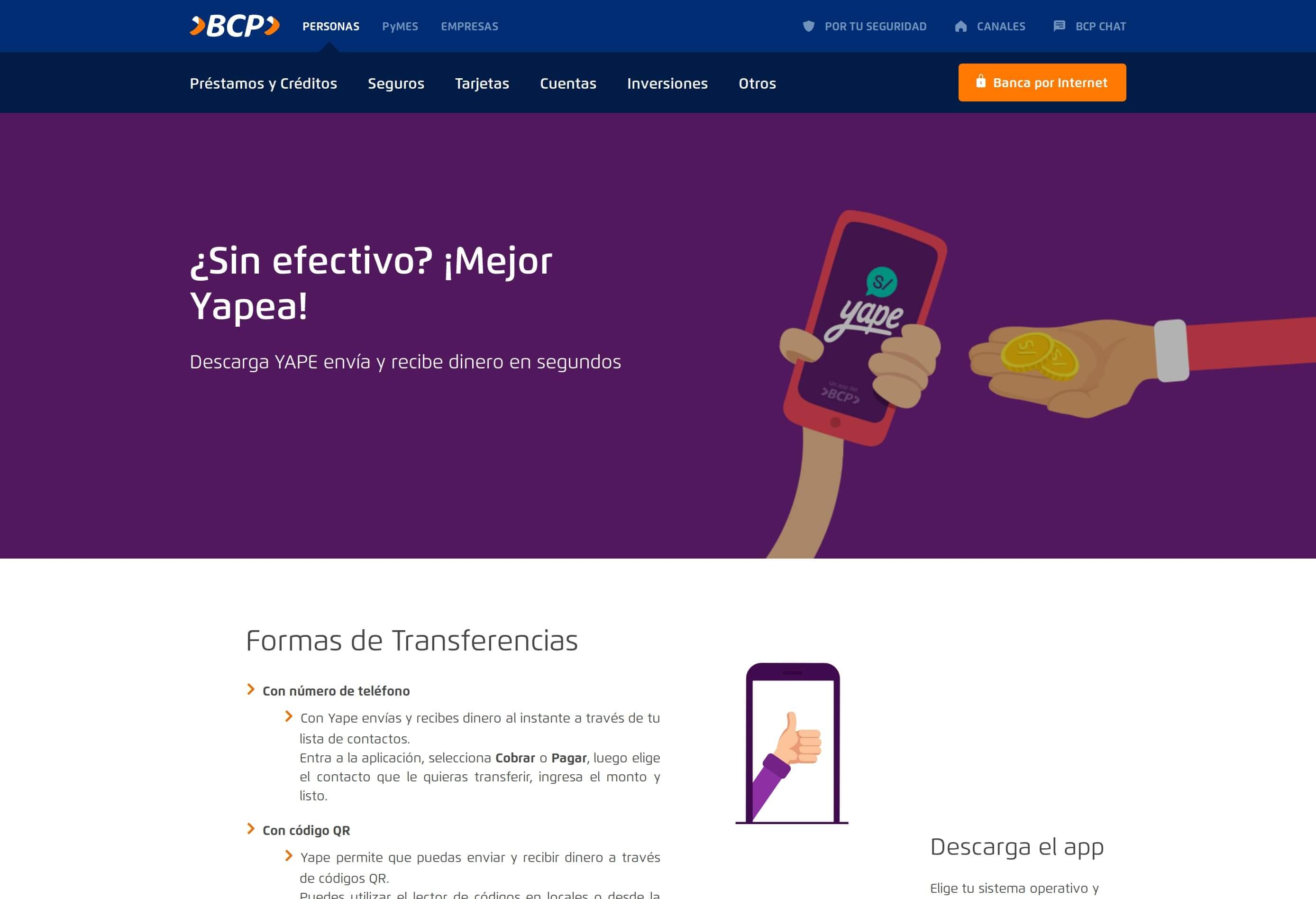} &
\includegraphics[width=2.5cm,height=2.5cm,keepaspectratio] {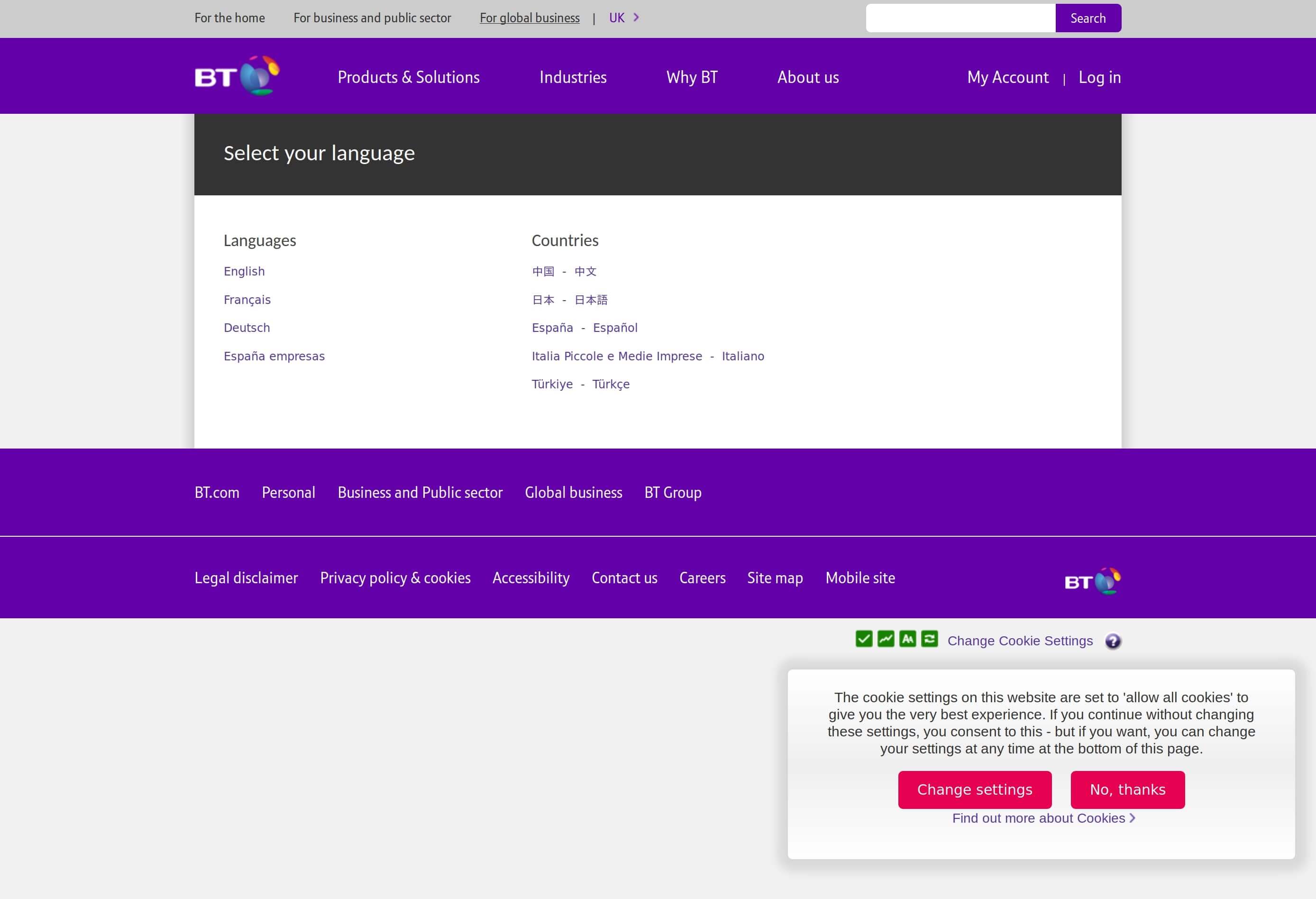} & \includegraphics[width=2.5cm,height=2.5cm,keepaspectratio] {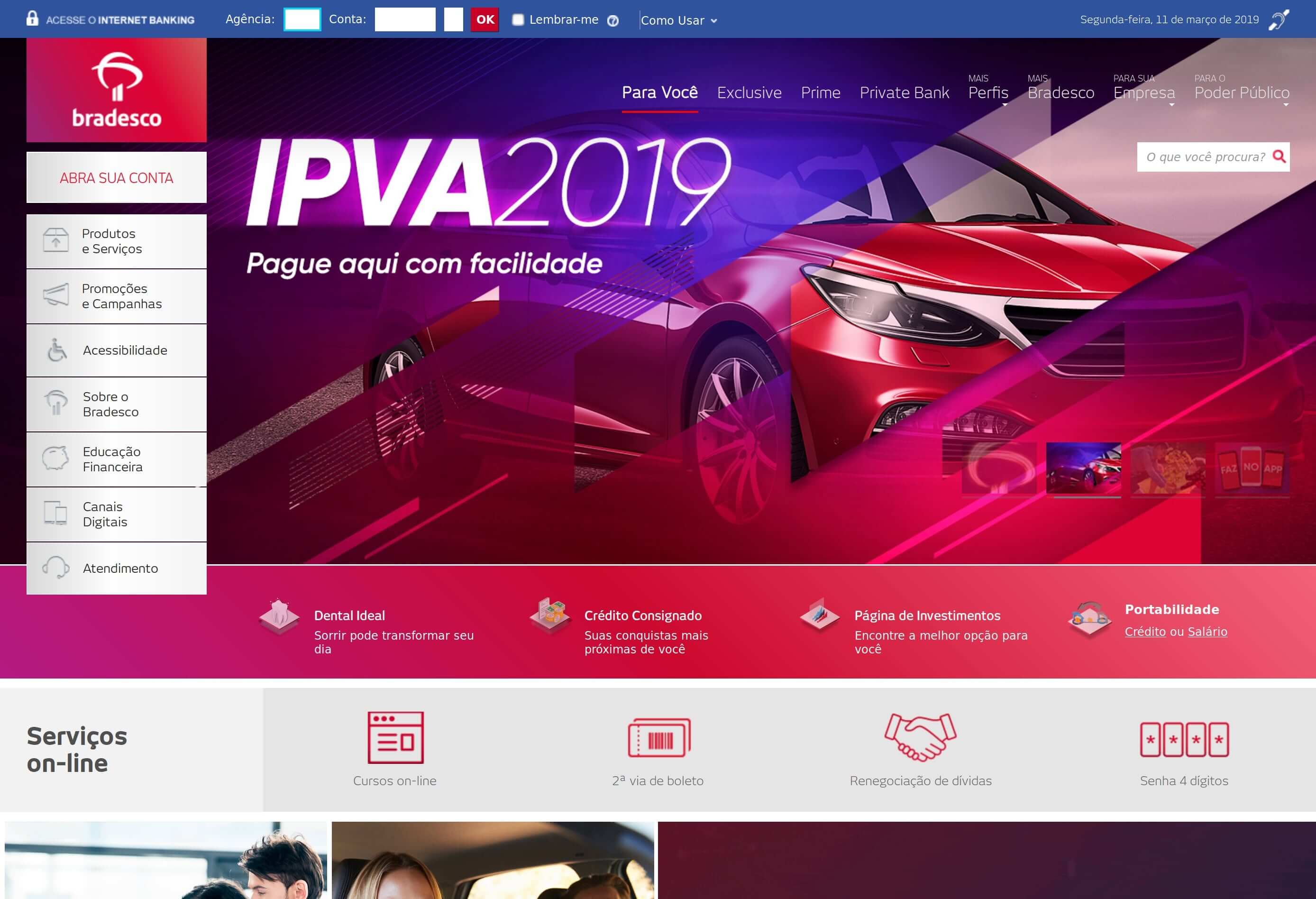}  \\
\end{tabular}} 
\captionof{figure}{Examples of the newly crawled phishing pages (row 1) that were correctly matched to the targeted website where the closest \new{trusted-list} screenshots are in row 2.} \label{tab:new_matches}
\end{table}

\begin{table}[!t]
\centering
\resizebox{\linewidth}{!}{%
\begin{tabular}{c|l l|l l}
\rot{Phishing test} &\includegraphics[width=2.5cm,height=2.5cm,keepaspectratio] {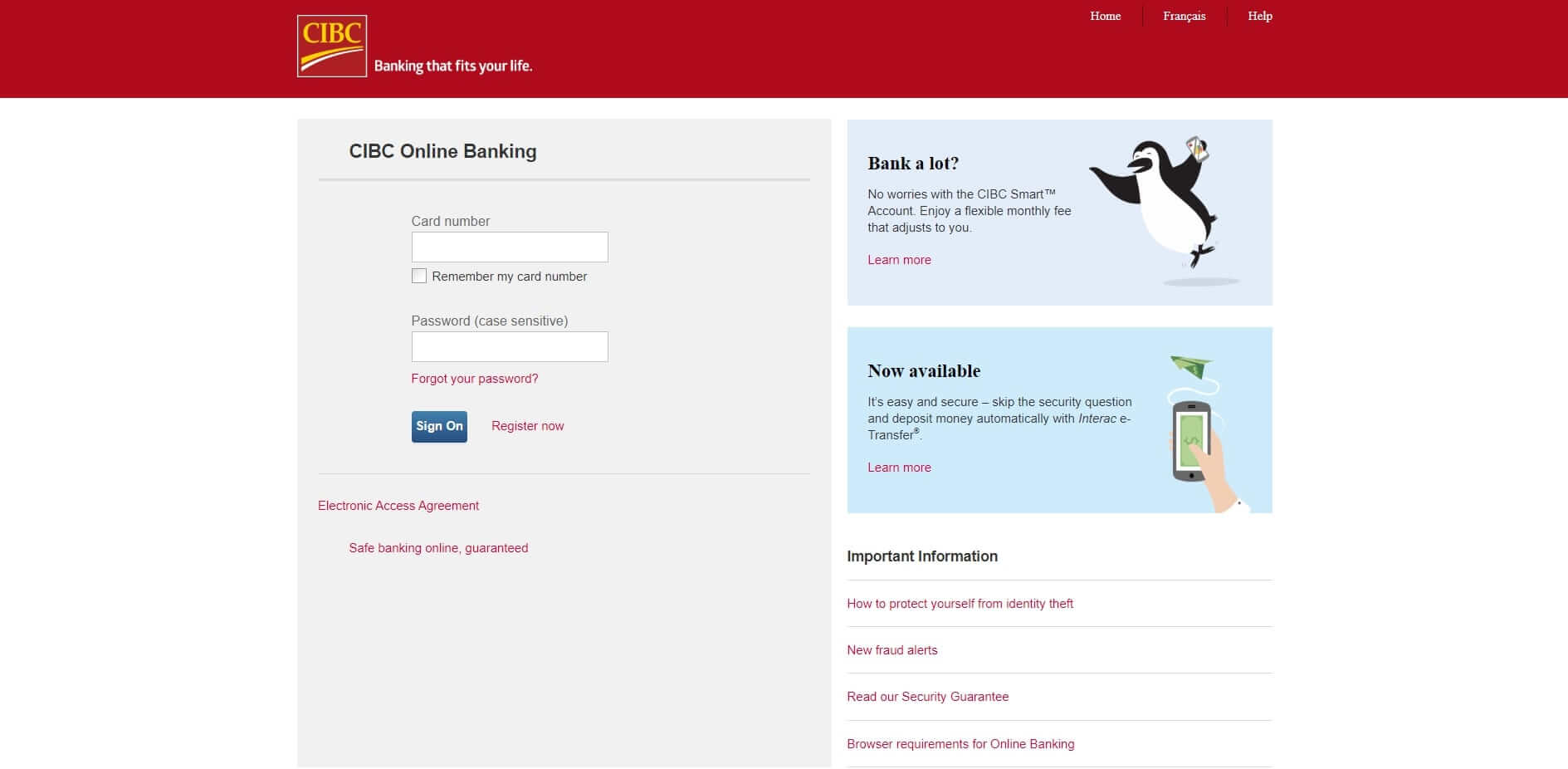} &
\includegraphics[width=2.5cm,height=2.5cm,keepaspectratio] {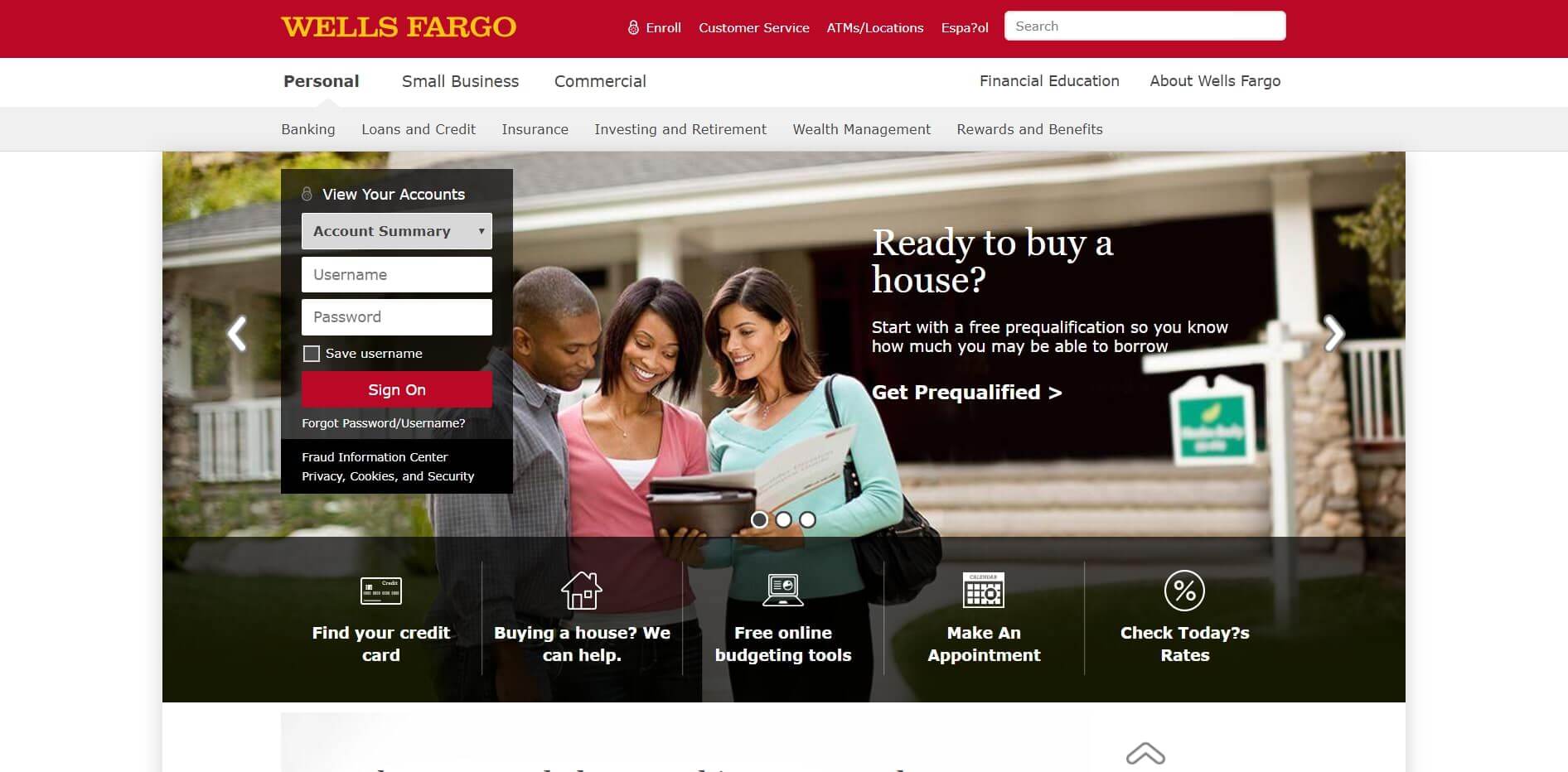} & \includegraphics[width=2.5cm,height=2.5cm,keepaspectratio] {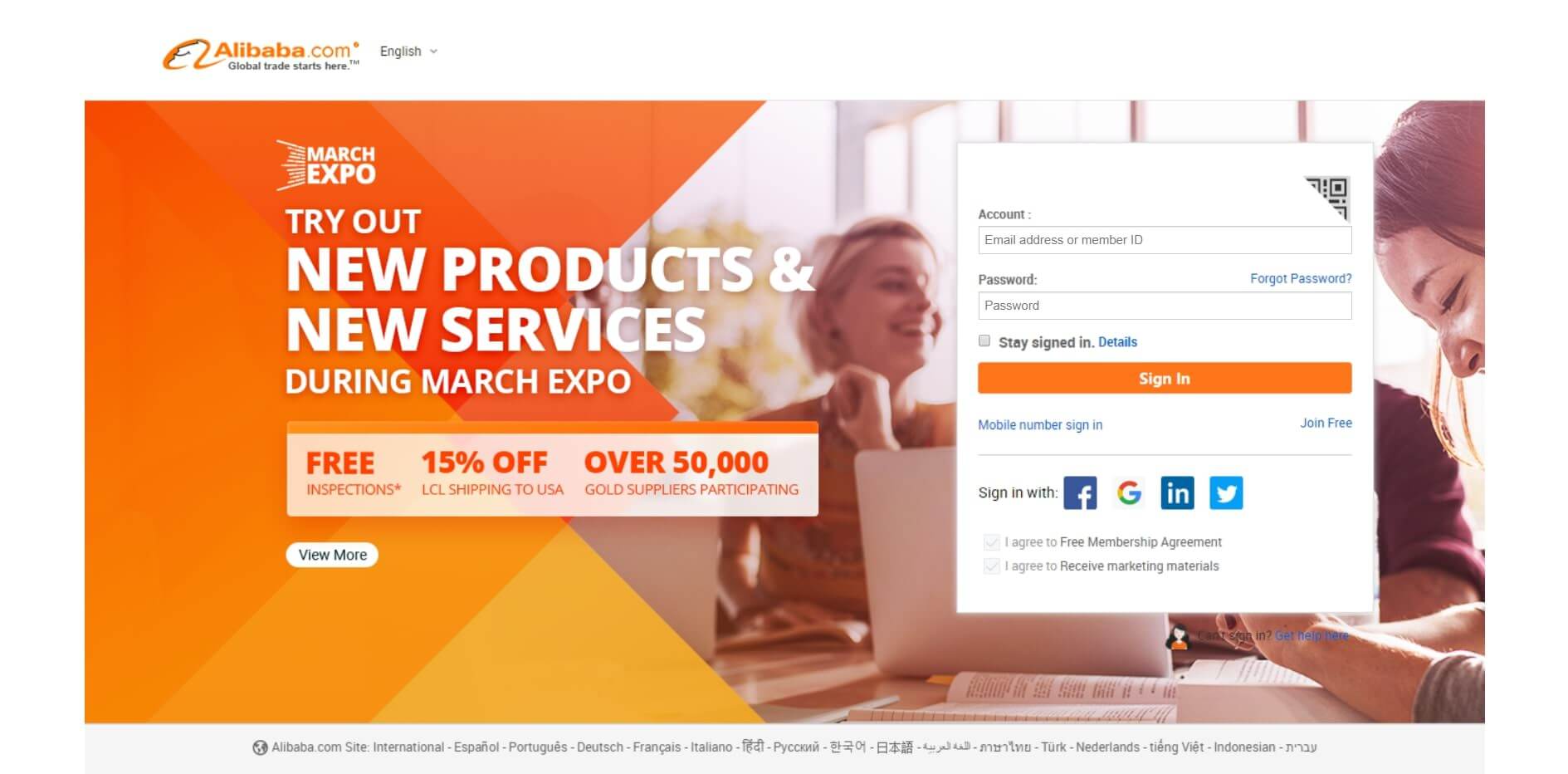} & \includegraphics[width=2.5cm,height=2.5cm,keepaspectratio] {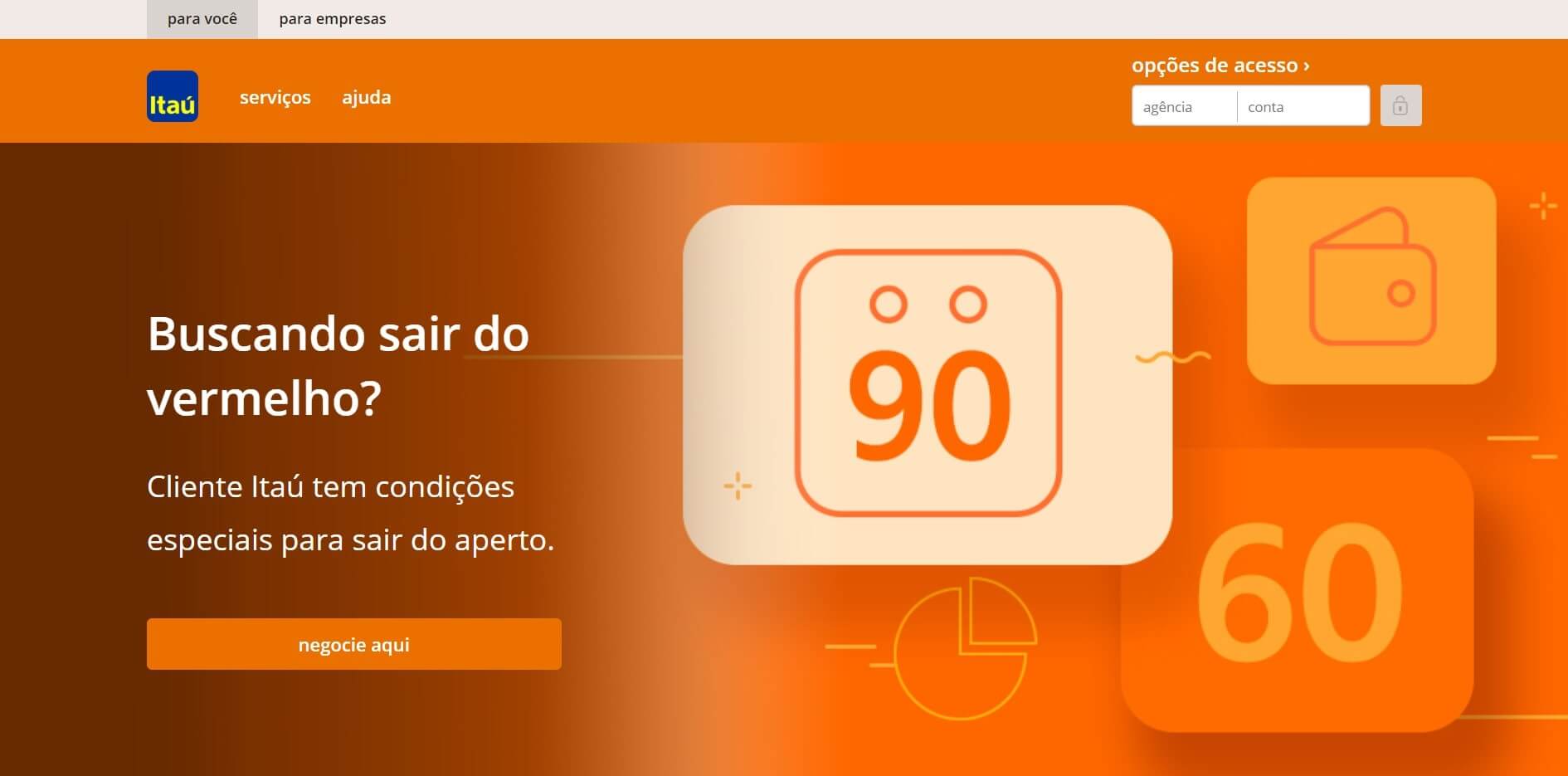}  \\ 
&&&&\\
\rot{Closest match} &\includegraphics[width=2.5cm,height=2.5cm,keepaspectratio] {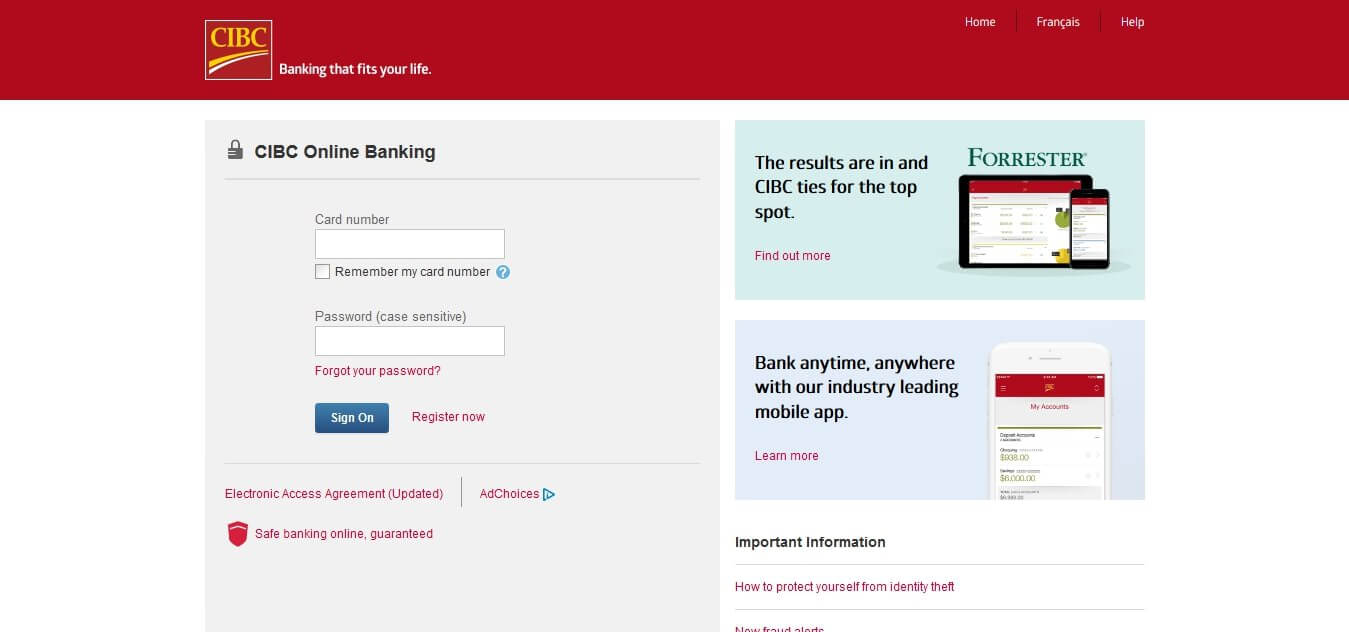} &
\includegraphics[width=2.5cm,height=2.5cm,keepaspectratio] {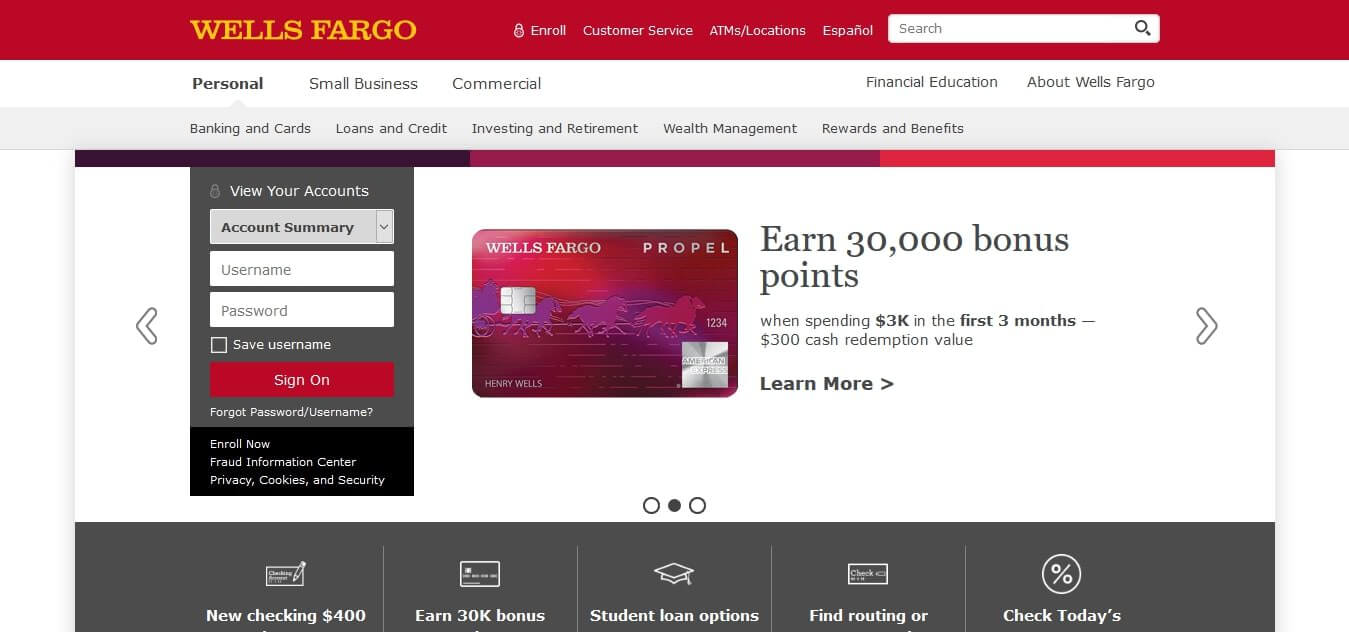} & \includegraphics[width=2.5cm,height=2.5cm,keepaspectratio] {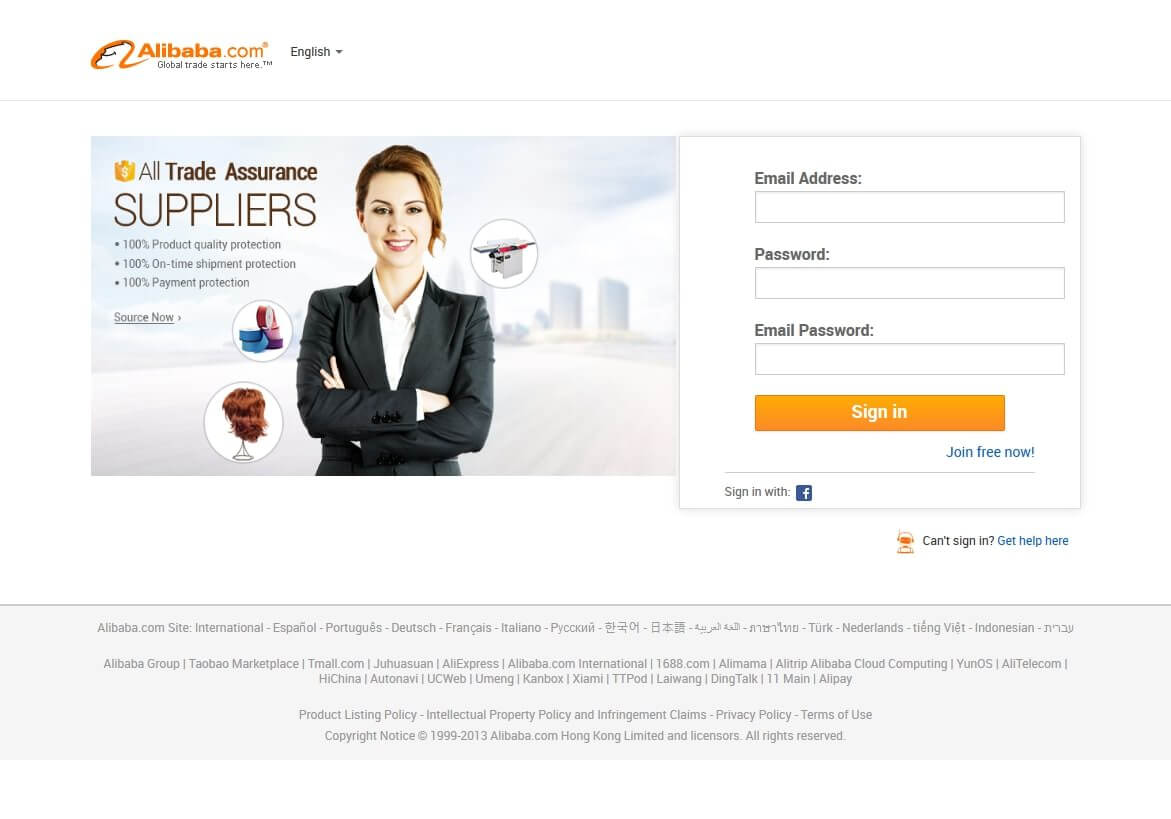} & \includegraphics[width=2.5cm,height=2.5cm,keepaspectratio] {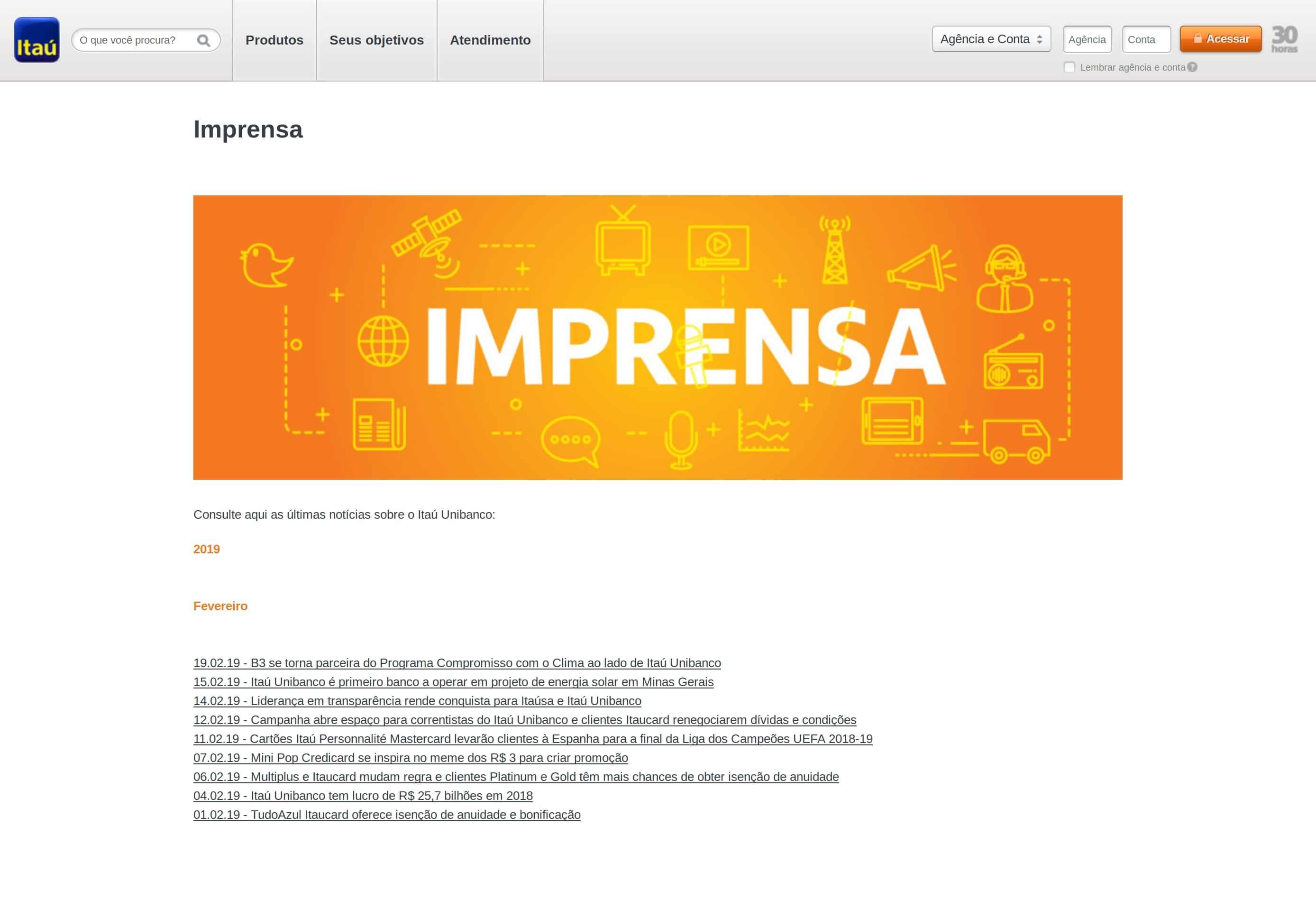}  \\
\end{tabular}} 
\captionof{figure}{Examples of websites with similar colors (Wells Fargo vs. CIBC, Alibaba vs. Banco Itau) that were correctly distinguished from each other.} \label{tab:sim_colors}
\end{table}

\begin{table}[!b]
\centering
\resizebox{\linewidth}{!}{%
\begin{tabular}{c|l|l|l}
\rot{Phishing test} &\includegraphics[width=2.5cm,height=2.5cm,keepaspectratio] {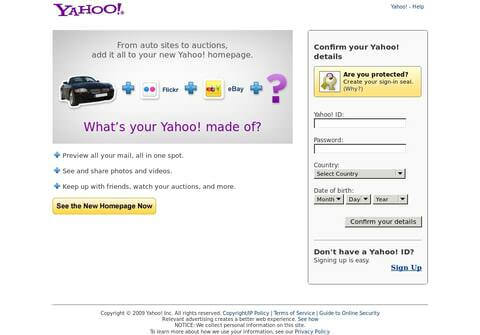} &
\includegraphics[width=2.5cm,height=2.5cm,keepaspectratio] {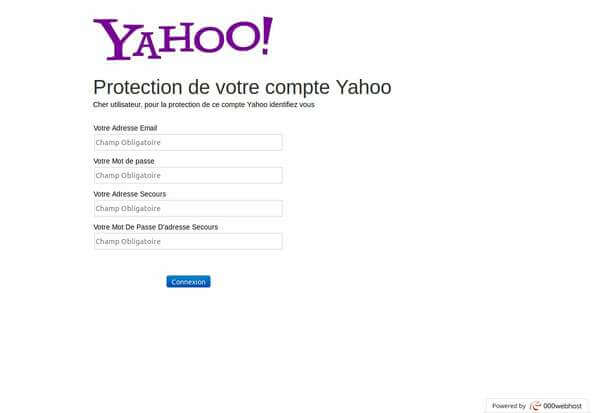} & \includegraphics[width=2.5cm,height=2.5cm,keepaspectratio] {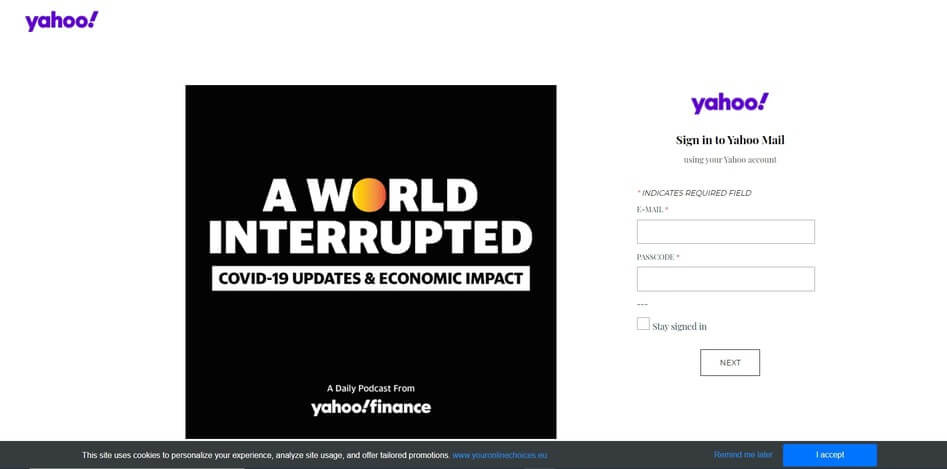}  \\ 
&&&\\
\rot{Closest match} &\includegraphics[width=2.5cm,height=2.5cm,keepaspectratio] {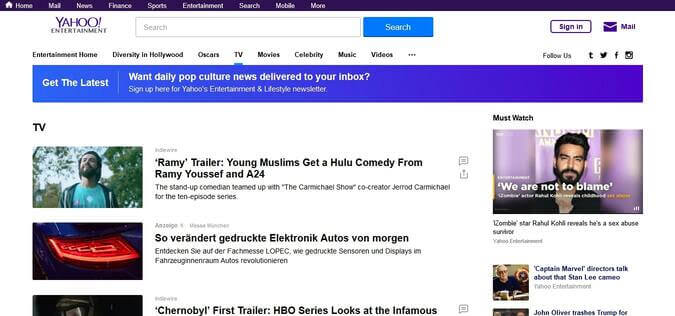} &
\includegraphics[width=2.5cm,height=2.5cm,keepaspectratio] {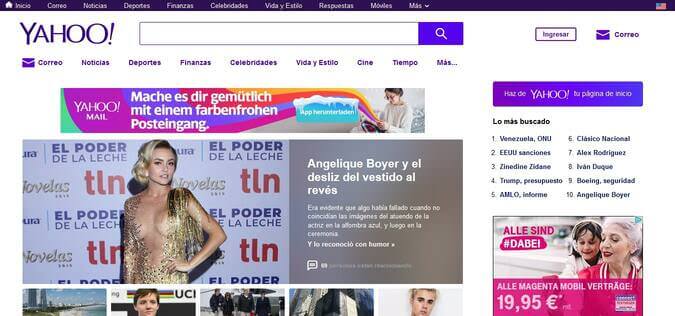} & \includegraphics[width=2.5cm,height=2.5cm,keepaspectratio] {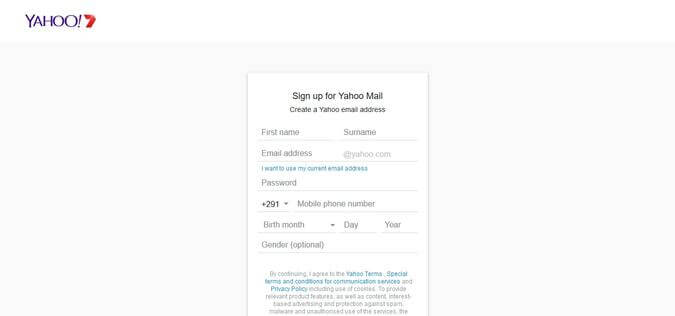}  \\
\end{tabular}} 
\captionof{figure}{Examples of successfully matched phishing pages where the website logo's fonts and colors are slightly different than the \new{trusted-list}. The first two examples contain an older version's logo, while the third example contains a newer version's logo (from the newly crawled data).} \label{tab:yahoo_diff}
\end{table}
\clearpage

\section{More Dataset Details and Examples} \label{appendix_dataset}
\begin{figure}[!htbp]
\begin{center}
\includegraphics [width=\linewidth,height=3.5cm,keepaspectratio]{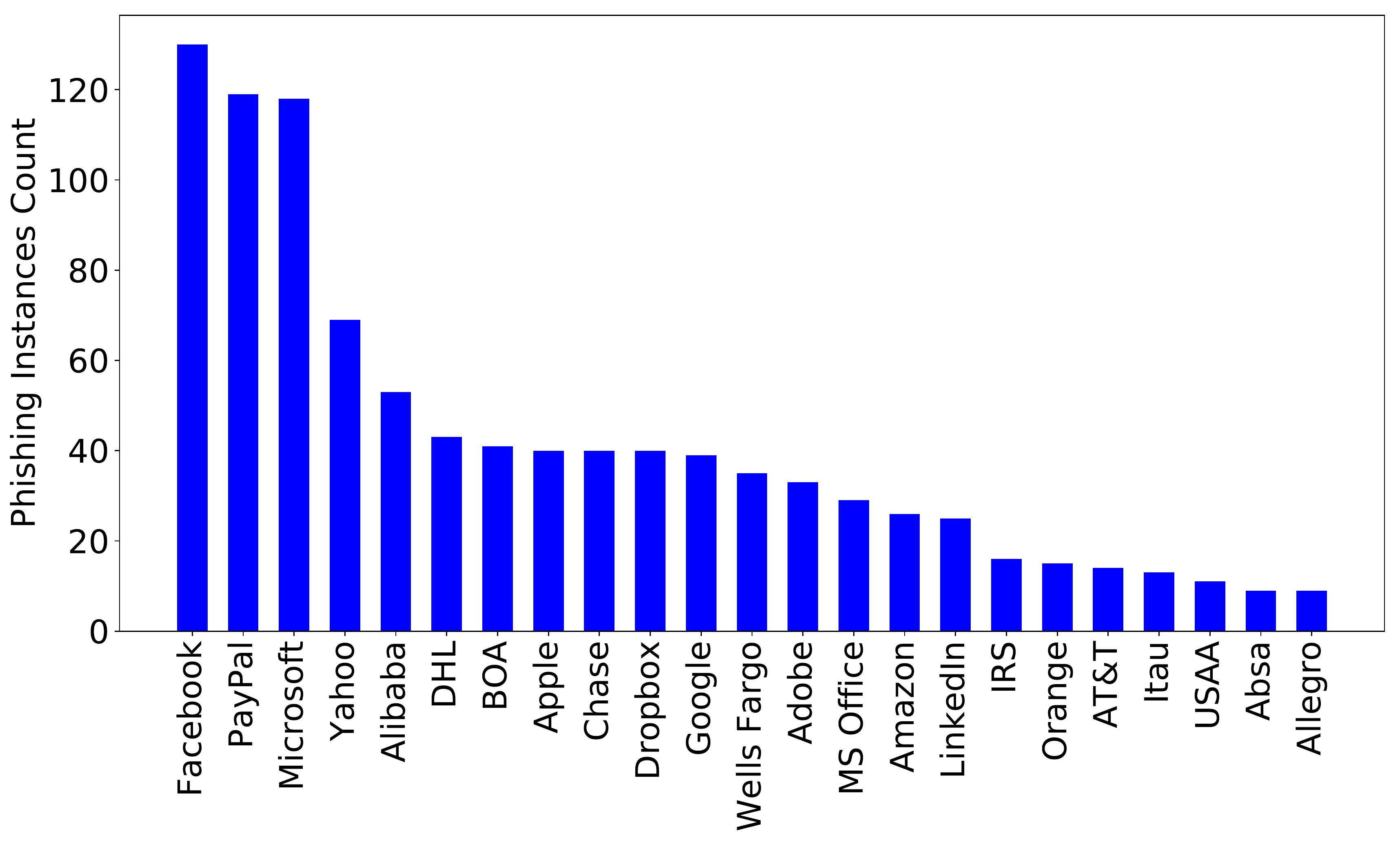}
\end{center}
   \caption{A histogram of the 23 most frequent websites that were found in the unique phishing set.}
\label{fig:phishing_hist}
\end{figure}

\begin{figure}[!htbp]
\begin{center}
\includegraphics [width=\linewidth,height=3.5cm,keepaspectratio]{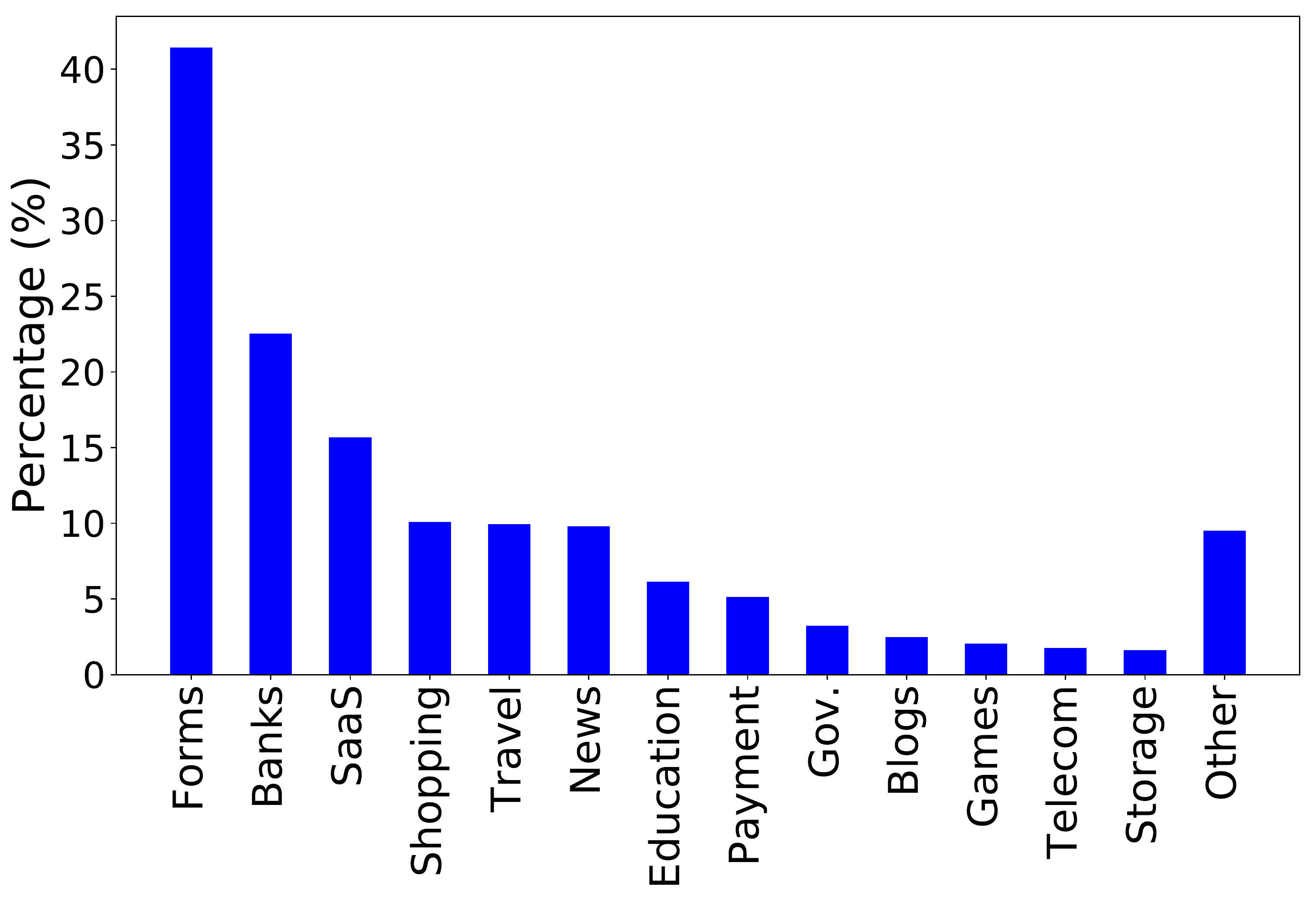}
\end{center}
   \caption{The categories in the legitimate test set.}
\label{fig:legit_test_hist}
\end{figure}

\begin{figure}[!htbp]
\centering
\begin{subfigure}{\columnwidth}
  \centering
  \includegraphics[width=0.65\textwidth]{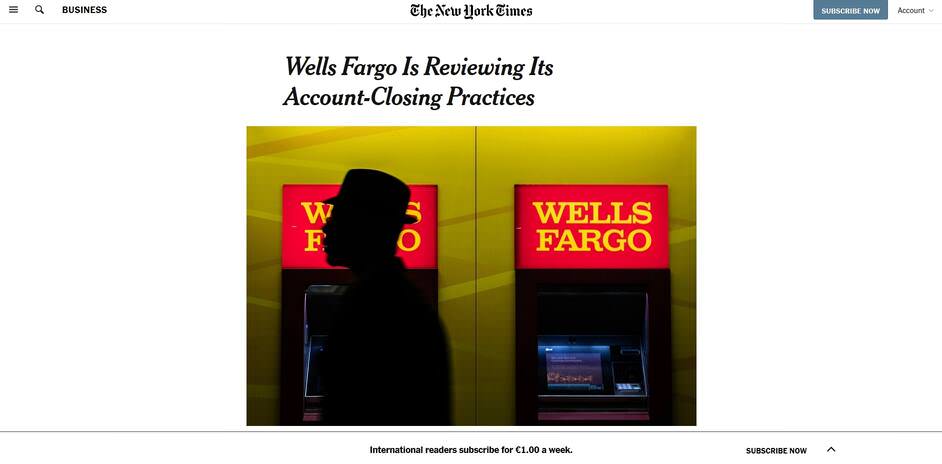}
  \caption{}
\end{subfigure}
\begin{subfigure}{\columnwidth}
  \centering
  \includegraphics[width=0.65\textwidth]{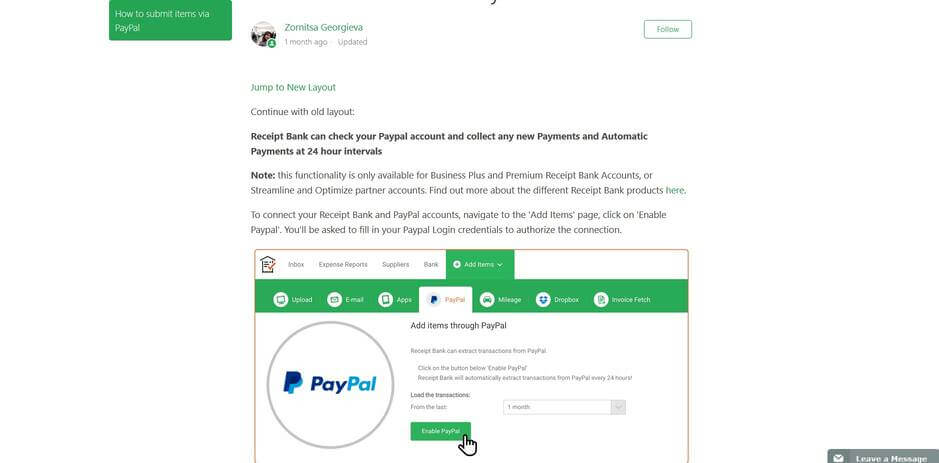}
  \caption{}
\end{subfigure}
\begin{subfigure}{\columnwidth}
  \centering
  \includegraphics[width=0.65\textwidth]{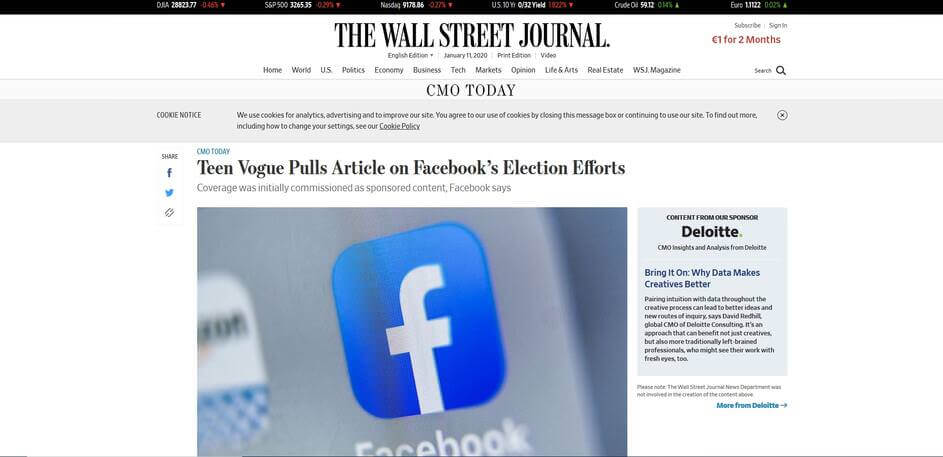}
  \caption{}
\end{subfigure}
\begin{subfigure}{\columnwidth}
  \centering
  \includegraphics[width=0.65\textwidth]{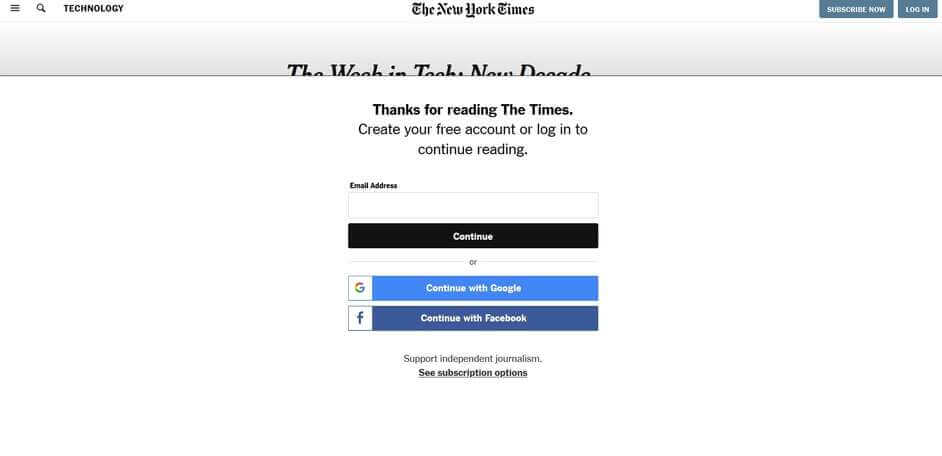}
  \caption{}
\end{subfigure}
\caption{Examples of the test set (consisting of 125 pages) used to evaluate the possible wrong matching to a trust-list's website whose logo appears in other benign pages (such as articles and login pages).}
\label{fig:logos}
\end{figure}

\begin{figure}[!t]
\centering
\begin{subfigure}{\columnwidth}
  \centering
  \includegraphics[width=0.5\textwidth]{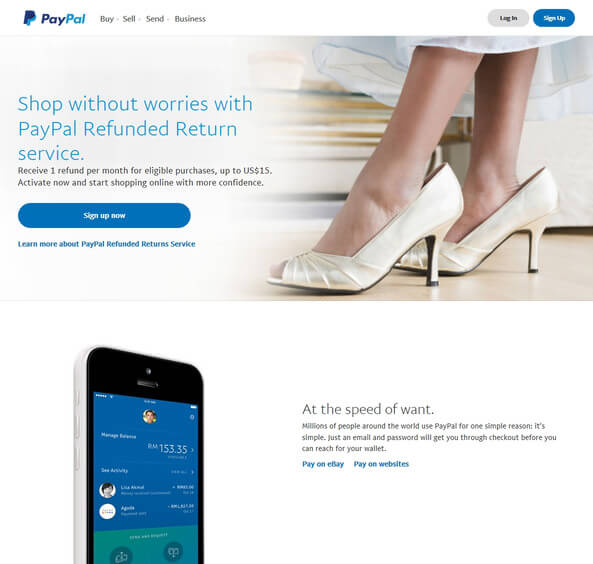}
  \caption{}
\end{subfigure}
\begin{subfigure}{\columnwidth}
  \centering
  \includegraphics[width=0.5\textwidth]{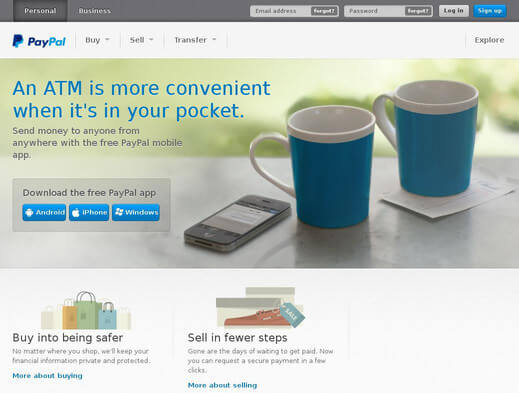}
  \caption{}
\end{subfigure}
\begin{subfigure}{\columnwidth}
  \centering
  \includegraphics[width=0.5\textwidth]{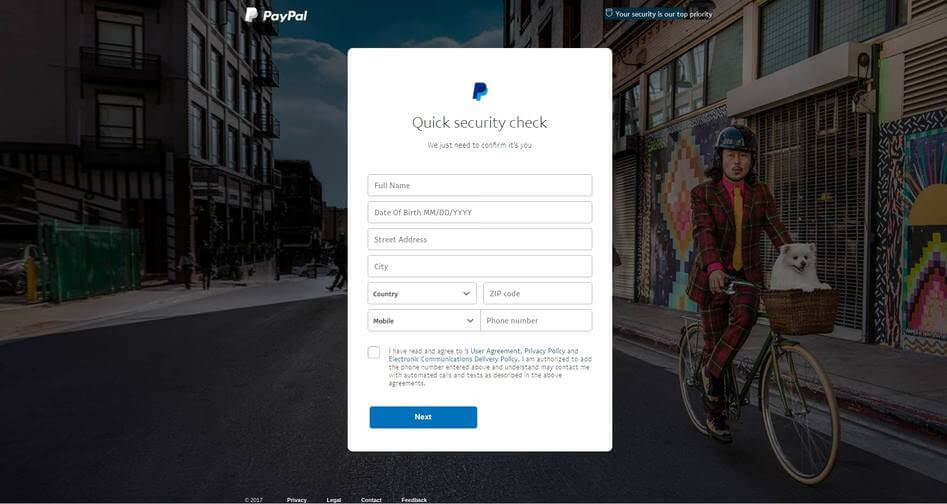}
  \caption{}
\end{subfigure}
\begin{subfigure}{\columnwidth}
  \centering
  \includegraphics[width=0.5\textwidth]{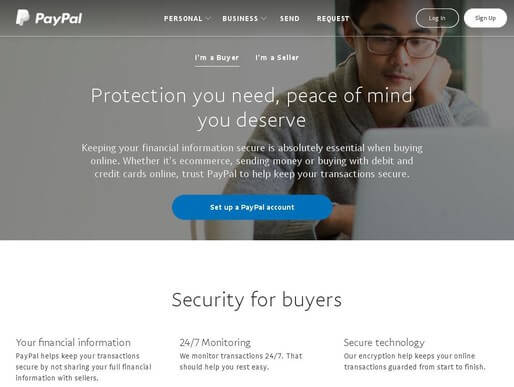}
  \caption{}
\end{subfigure}
\begin{subfigure}{\columnwidth}
  \centering
  \includegraphics[width=0.5\textwidth]{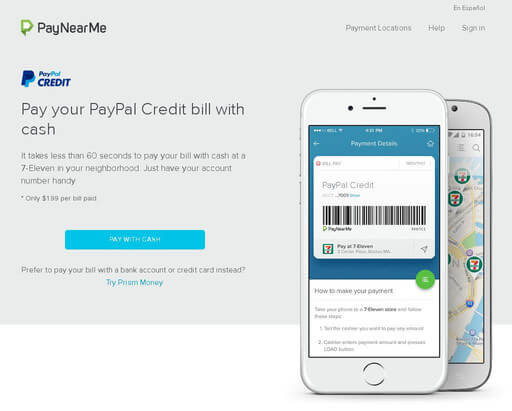}
  \caption{}
\end{subfigure}
\caption{Examples of the variations in the \dataset{} dataset of phishing examples targeting one website with no counterparts in the crawled legitimate examples (training list) of the same website.}
\label{fig:paypal_phish}
\end{figure}

\begin{figure}[!t]
\centering
\begin{subfigure}{\columnwidth}
  \centering
  \includegraphics[width=0.52\textwidth]{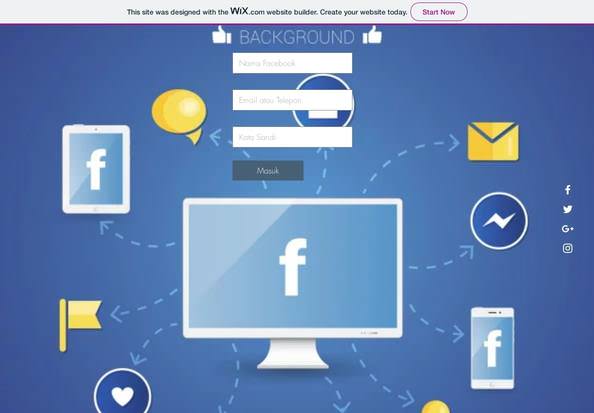}
  \caption{}
\end{subfigure}
\begin{subfigure}{\columnwidth}
  \centering
  \includegraphics[width=0.52\textwidth]{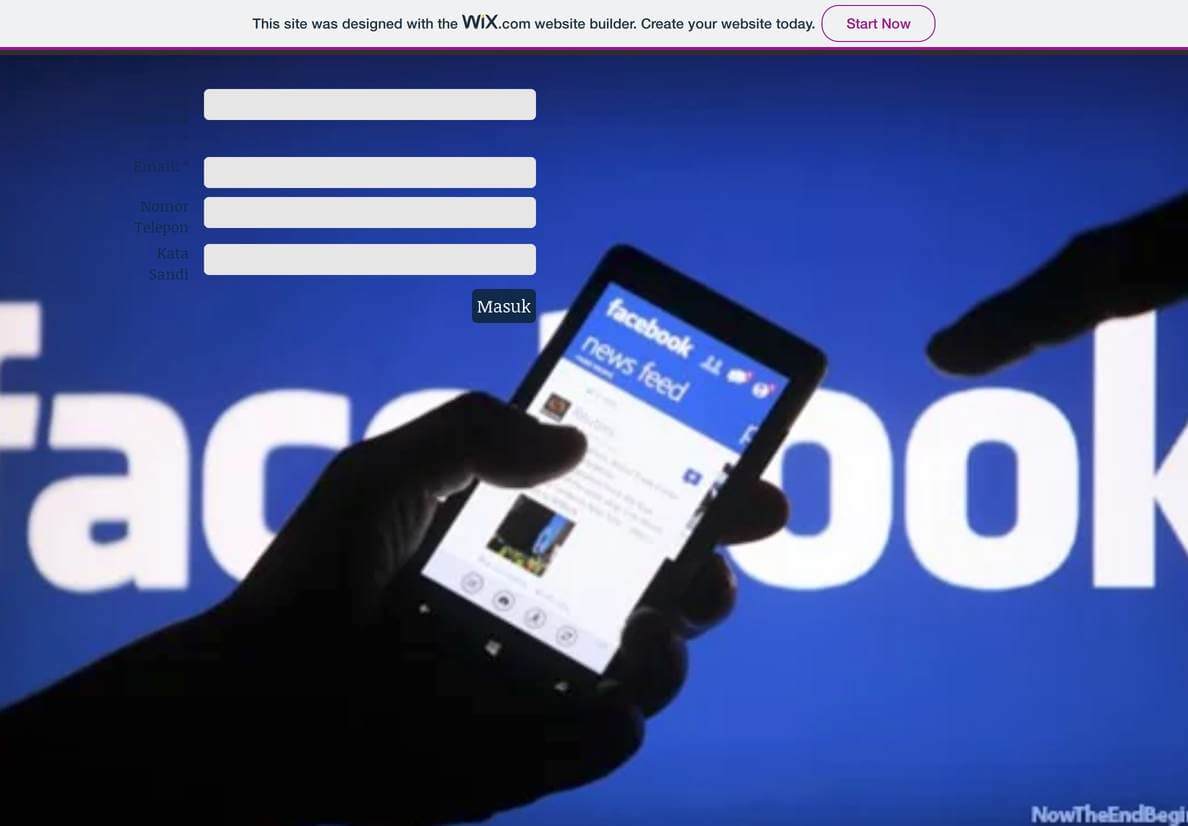}
  \caption{}
\end{subfigure}
\begin{subfigure}{\columnwidth}
  \centering
  \includegraphics[width=0.52\textwidth]{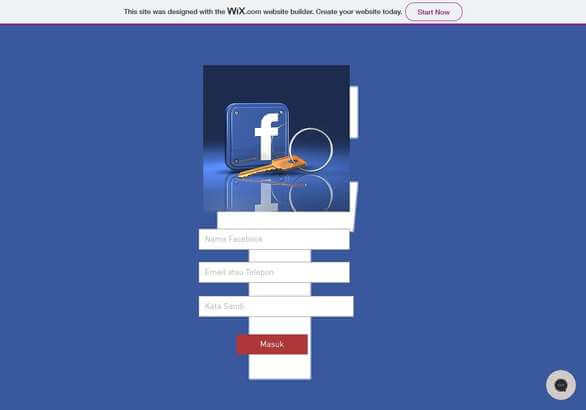}
  \caption{}
\end{subfigure}
\begin{subfigure}{\columnwidth}
  \centering
  \includegraphics[width=0.52\textwidth]{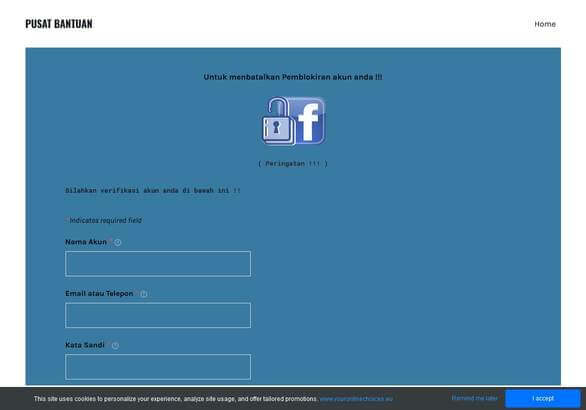}
  \caption{}
\end{subfigure}
\begin{subfigure}{\columnwidth}
  \centering
  \includegraphics[width=0.52\textwidth]{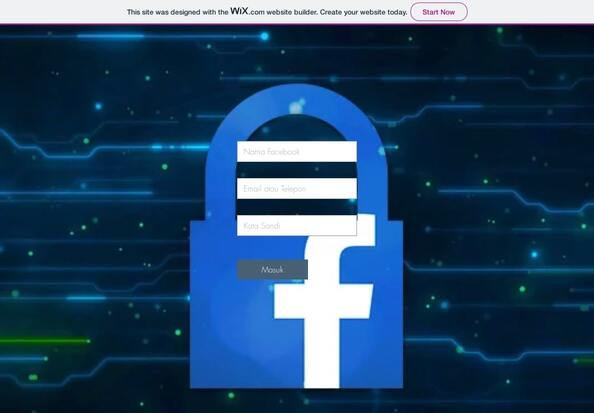}
  \caption{}
\end{subfigure}
\caption{Examples of phishing pages in the dataset that are not similar enough (either in colors or design) to the legitimate website which causes an increase in the mismatches when not partially train with a part of the phishing set.}
\label{fig:fb_phish}
\end{figure}

\begin{figure*}[!htb]
\centering
\begin{subfigure}{\columnwidth}
  \centering
  \includegraphics[width=0.5\textwidth, height=2.5cm, keepaspectratio]{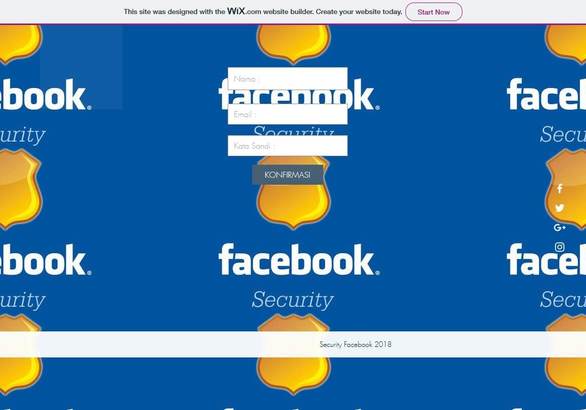}
  \caption{}
\end{subfigure}
\begin{subfigure}{\columnwidth}
  \centering
  \includegraphics[width=0.5\textwidth, height=2.5cm, keepaspectratio]{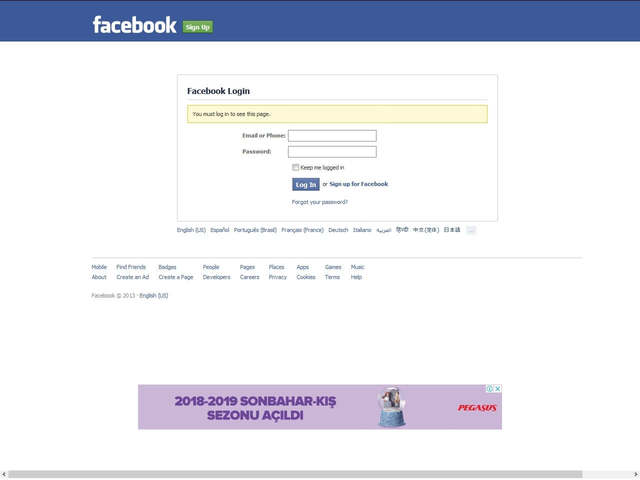}
  \caption{}
\end{subfigure}
\begin{subfigure}{\columnwidth}
  \centering
  \includegraphics[width=0.5\textwidth, height=2.5cm, keepaspectratio]{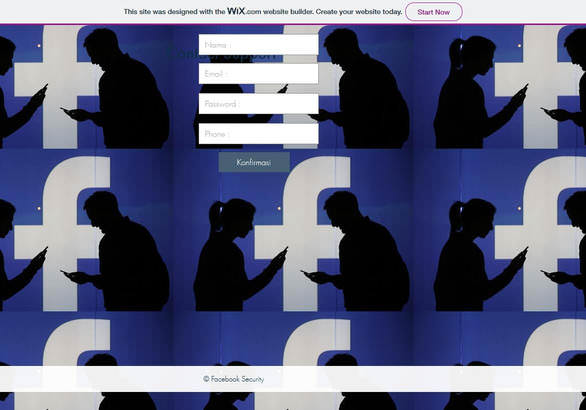}
  \caption{}
\end{subfigure}
\begin{subfigure}{\columnwidth}
  \centering
  \includegraphics[width=0.5\textwidth, height=2.5cm, keepaspectratio]{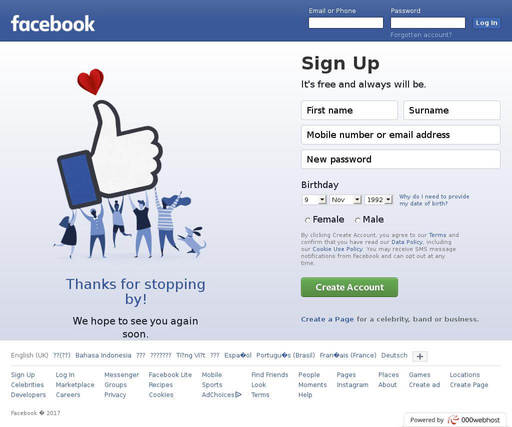}
  \caption{}
\end{subfigure}
\begin{subfigure}{\columnwidth}
  \centering
  \includegraphics[width=0.5\textwidth, height=2.5cm, keepaspectratio]{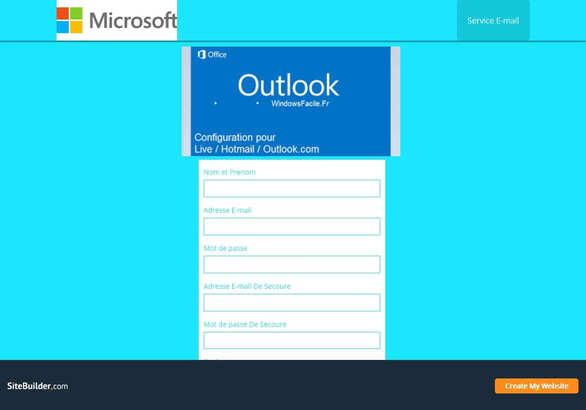}
  \caption{}
\end{subfigure}
\begin{subfigure}{\columnwidth}
  \centering
  \includegraphics[width=0.5\textwidth, height=2.5cm, keepaspectratio]{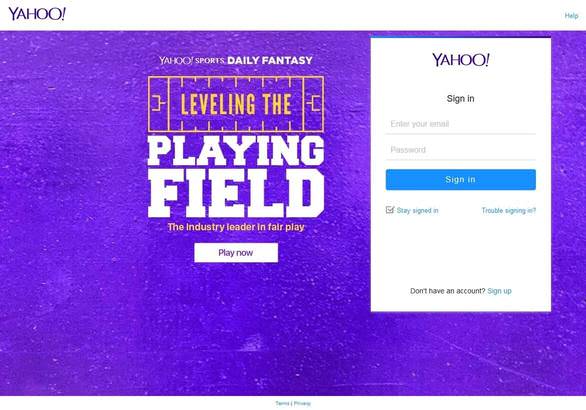}
  \caption{}
\end{subfigure}
\begin{subfigure}{\columnwidth}
  \centering
  \includegraphics[width=0.5\textwidth, height=2.5cm, keepaspectratio]{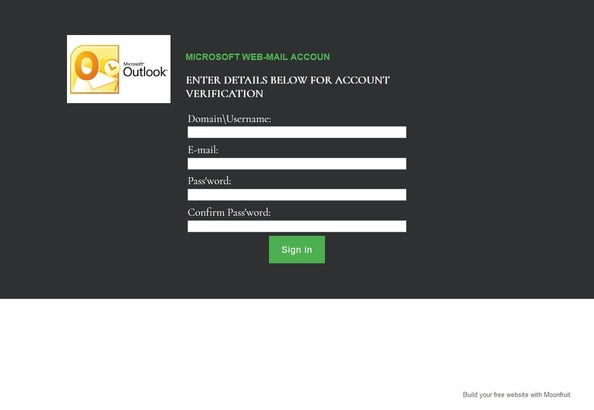}
  \caption{}
\end{subfigure}
\begin{subfigure}{\columnwidth}
  \centering
  \includegraphics[width=0.5\textwidth, height=2.5cm, keepaspectratio]{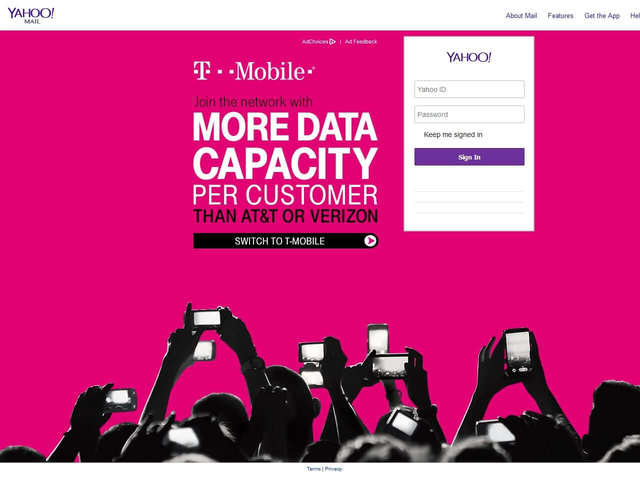}
  \caption{}
\end{subfigure}
\begin{subfigure}{\columnwidth}
  \centering
  \includegraphics[width=0.5\textwidth, height=2.5cm, keepaspectratio]{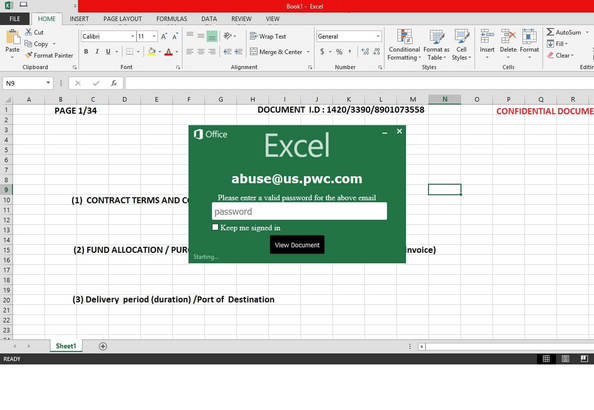}
  \caption{}
\end{subfigure}
\begin{subfigure}{\columnwidth}
  \centering
  \includegraphics[width=0.5\textwidth, height=2.5cm, keepaspectratio]{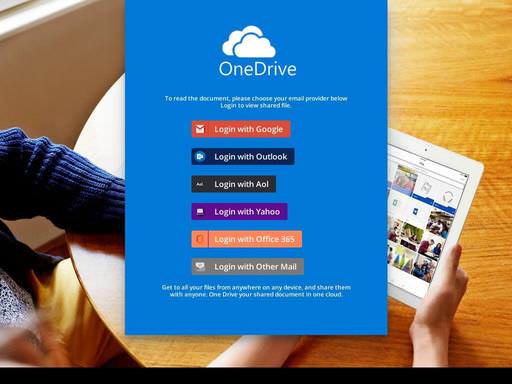}
  \caption{}
\end{subfigure}
\begin{subfigure}{\columnwidth}
  \centering
  \includegraphics[width=0.5\textwidth, height=2.5cm, keepaspectratio]{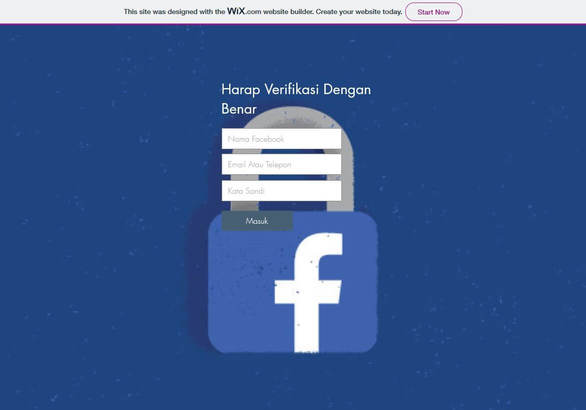}
  \caption{}
\end{subfigure}
\begin{subfigure}{\columnwidth}
  \centering
  \includegraphics[width=0.5\textwidth, height=2.5cm, keepaspectratio]{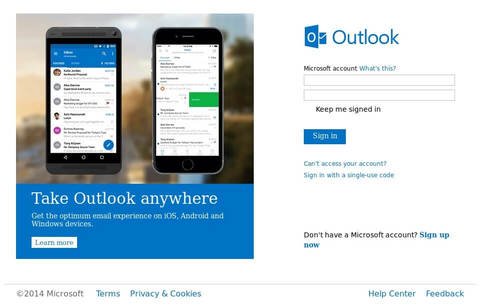}
  \caption{}
\end{subfigure}
\caption{Examples of the phishing pages used in the online study where participants were asked if they think the appearances of these pages are trustworthy. The first column screenshots are the dissimilar examples. Only 3.02\% of users (averaged on all screenshots) considered them trustworthy. The second column screenshots are the relatively more similar examples (with subtle differences) where 65.3\% of users considered them trustworthy.}
\label{fig:study2}
\end{figure*}

\begin{figure}[!htb]
\centering
\begin{subfigure}{\columnwidth}
  \centering
  \includegraphics[width=0.47\textwidth, height=1.8cm]{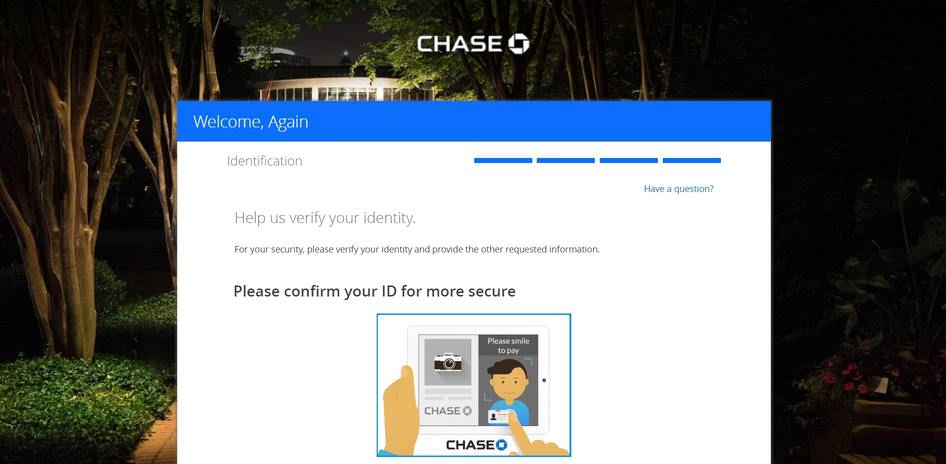}
  \caption{}
\end{subfigure}
\begin{subfigure}{\columnwidth}
  \centering
  \includegraphics[width=0.47\textwidth, height=1.8cm]{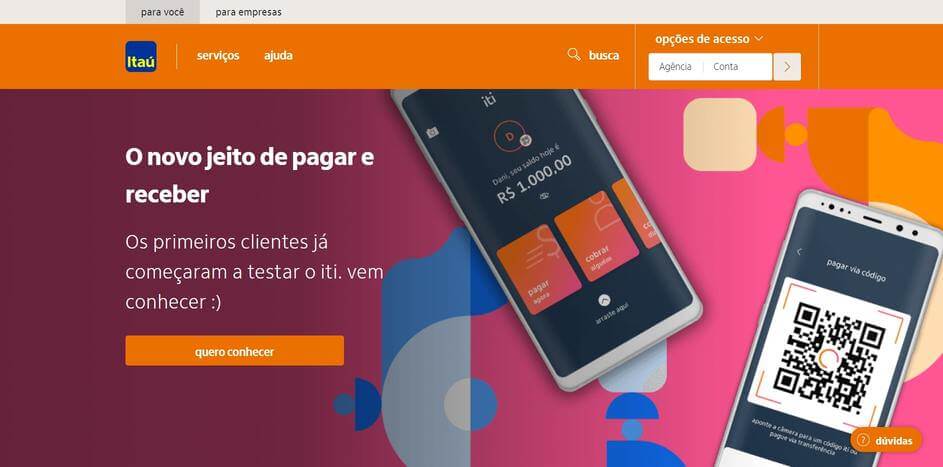}
  \caption{}
\end{subfigure}
\begin{subfigure}{\columnwidth}
  \centering
  \includegraphics[width=0.47\textwidth, height=1.8cm]{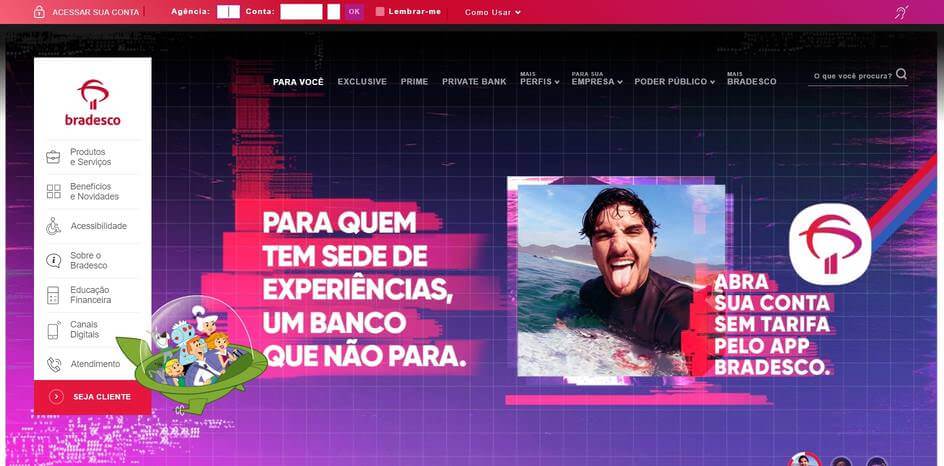}
  \caption{}
\end{subfigure}
\begin{subfigure}{\columnwidth}
  \centering
  \includegraphics[width=0.47\textwidth, height=1.8cm]{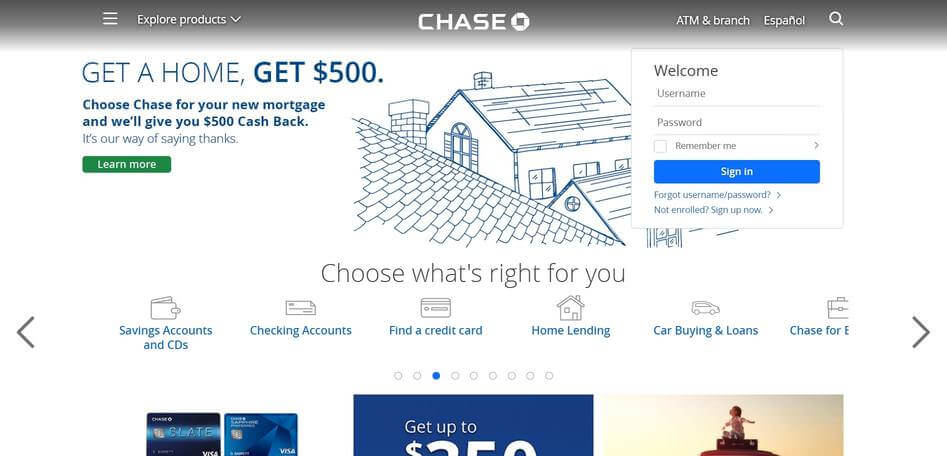}
  \caption{}
\end{subfigure}
\begin{subfigure}{\columnwidth}
  \centering
  \includegraphics[width=0.47\textwidth, height=1.8cm]{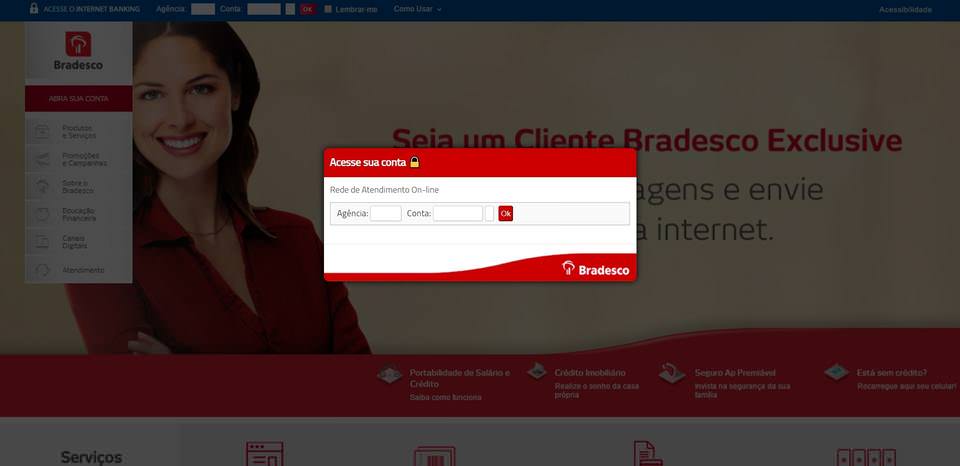}
  \caption{}
\end{subfigure}
\begin{subfigure}{\columnwidth}
  \centering
  \includegraphics[width=0.47\textwidth, height=1.8cm]{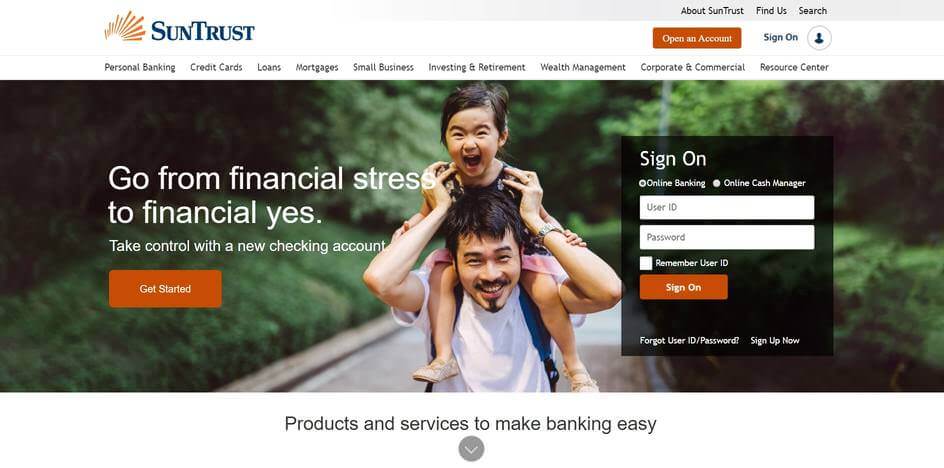}
  \caption{}
\end{subfigure}
\begin{subfigure}{\columnwidth}
  \centering
  \includegraphics[width=0.47\textwidth, height=1.8cm]{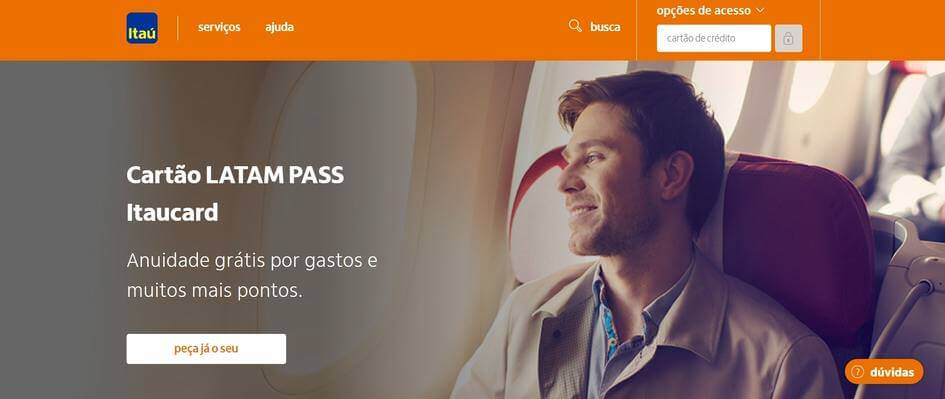}
  \caption{}
\end{subfigure}
\begin{subfigure}{\columnwidth}
  \centering
  \includegraphics[width=0.47\textwidth, height=1.8cm]{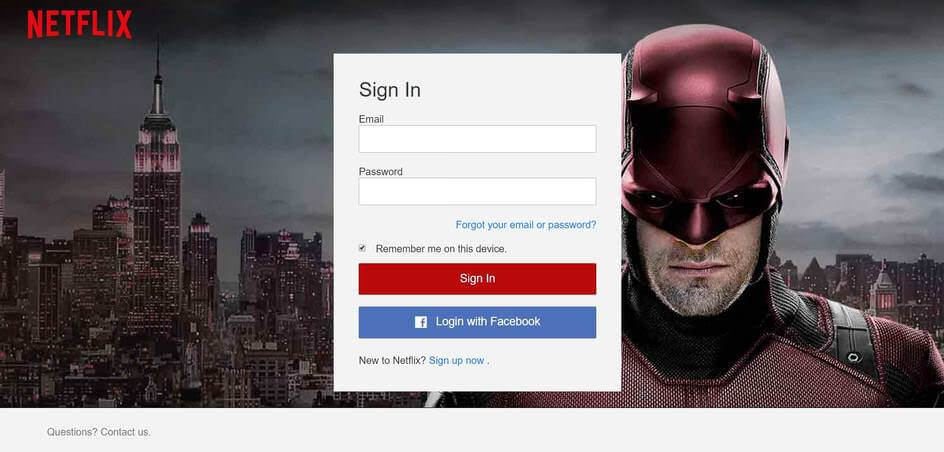}
  \caption{}
\end{subfigure}
\caption{Examples of the newly crawled phishing pages from PhishTank that are used to test zero-day pages. Additionally, all of these pages do not have counterparts in the targeted website's screenshots in the \new{trusted-list} and were not seen in training.}
\label{fig:new_phish}
\end{figure}

We show here more details about the \dataset{} dataset. In~\autoref{fig:phishing_hist}, we show a histogram of the most targeted websites by the crawled phishing pages (\autoref{dataset}).~\autoref{fig:legit_test_hist} shows the categories in the benign test set that we constructed to reduce bias by having similar categories to the \new{trusted} websites (\autoref{dataset}). Figure~\ref{tab:browsers_exampls} shows examples of the test set used to test browser differences (\autoref{robustness}).~\autoref{fig:logos} shows examples of the test set used to test false positives when \new{trusted} logos are found in the page (\autoref{discuss_qual}). Examples of the variations (e.g. designs, colors and layout) of the dataset's phishing pages targeting one website are demonstrated in~\autoref{fig:paypal_phish}. Examples of the poorly designed (i.e. dissimilar to their targets) phishing pages are in~\autoref{fig:fb_phish}. Also, ~\autoref{fig:study2} shows examples of the screenshots used in the online user study to evaluate dissimilar examples (discussed in~\autoref{failures}). Finally,~\autoref{fig:new_phish} shows examples of the newly crawled phishing pages to test the performance against zero-day pages (\autoref{new_data}).

\begin{table}[!htbp]
\centering
\resizebox{\linewidth}{!}{%
\begin{tabular}{c|l|l|l}
\rot{Browser 1} &\includegraphics[width=2.5cm,height=2.5cm,keepaspectratio] {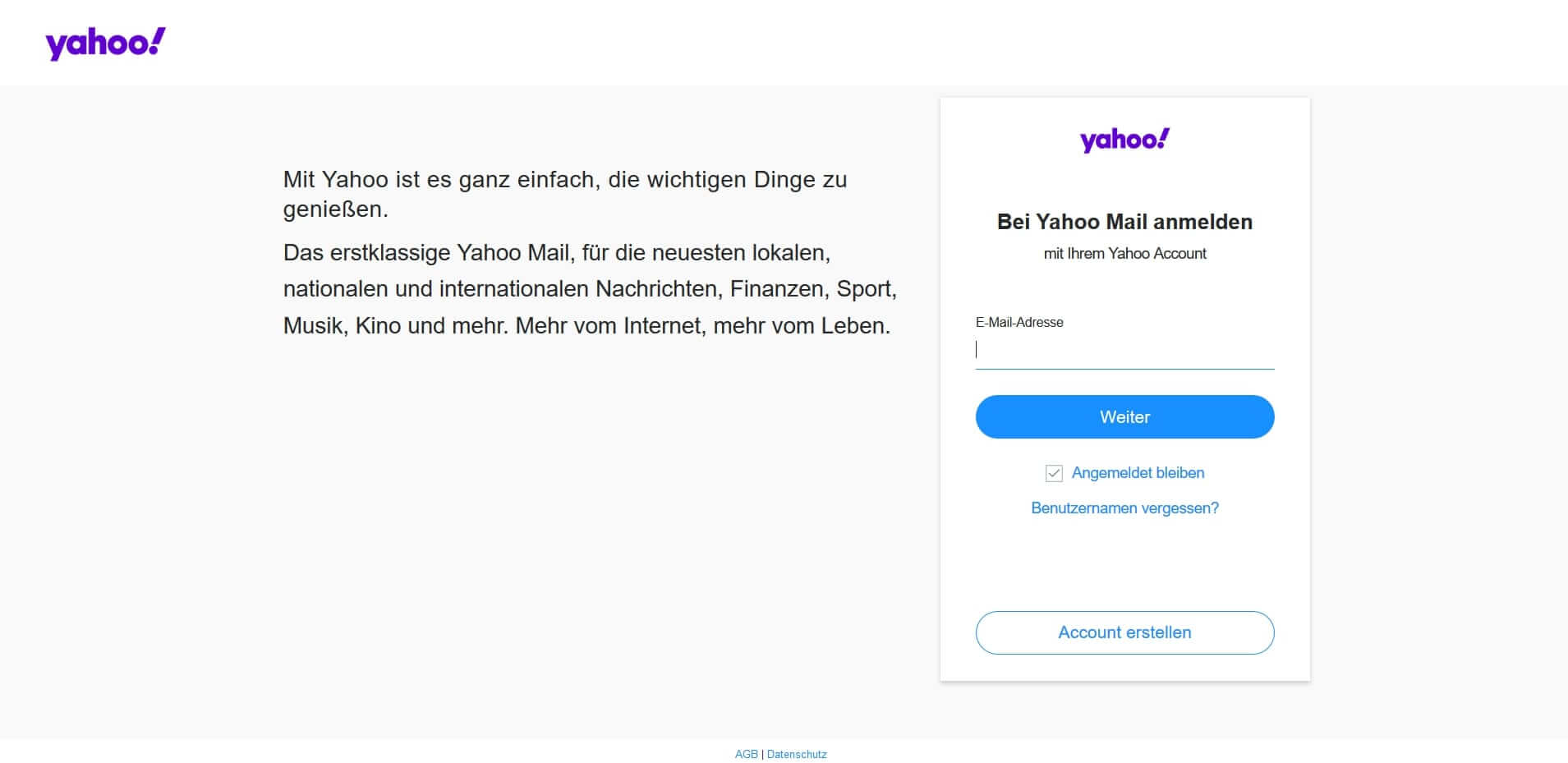} &
\includegraphics[width=2.5cm,height=2.5cm,keepaspectratio] {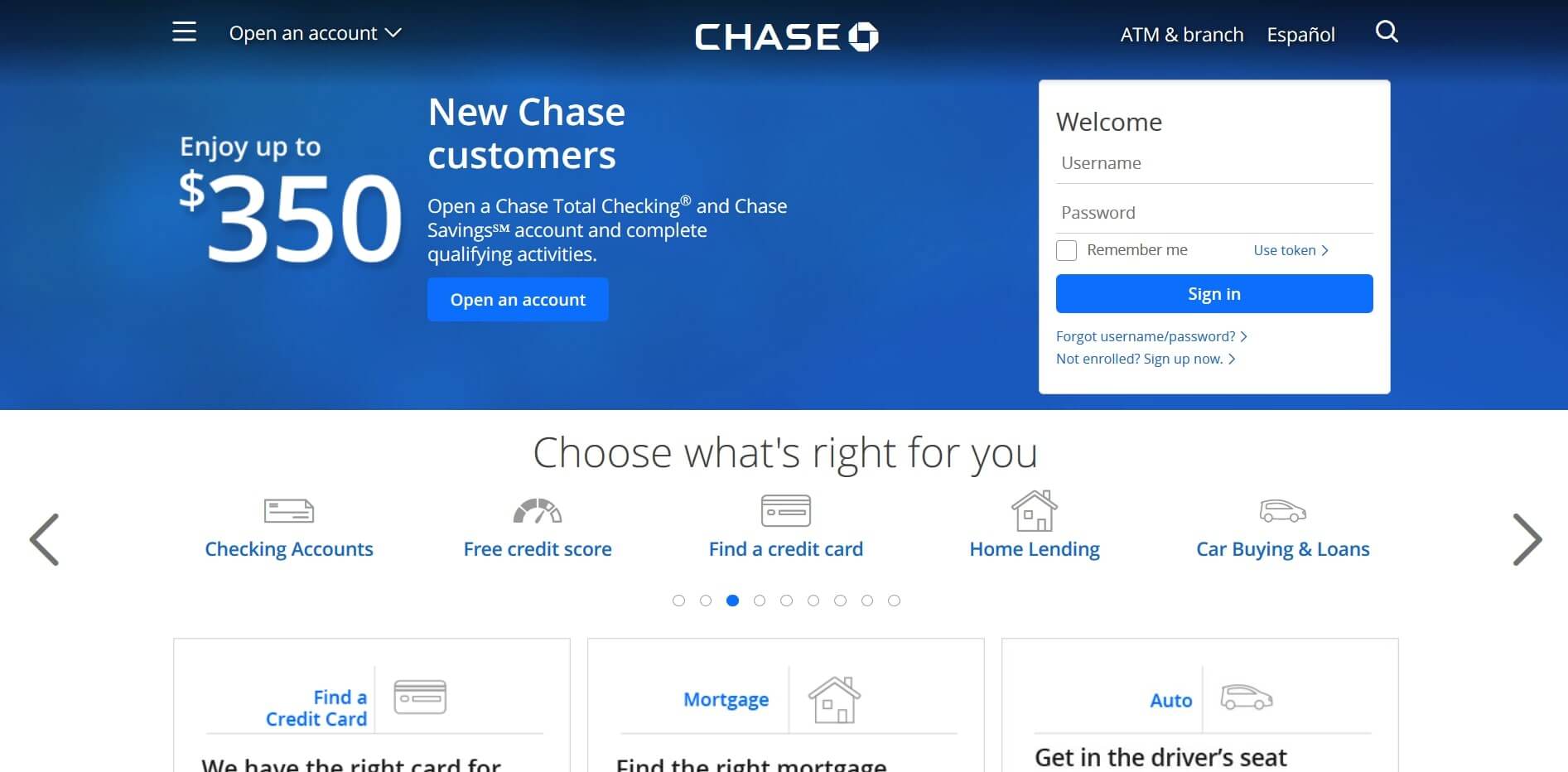} & \includegraphics[width=2.5cm,height=2.5cm,keepaspectratio] {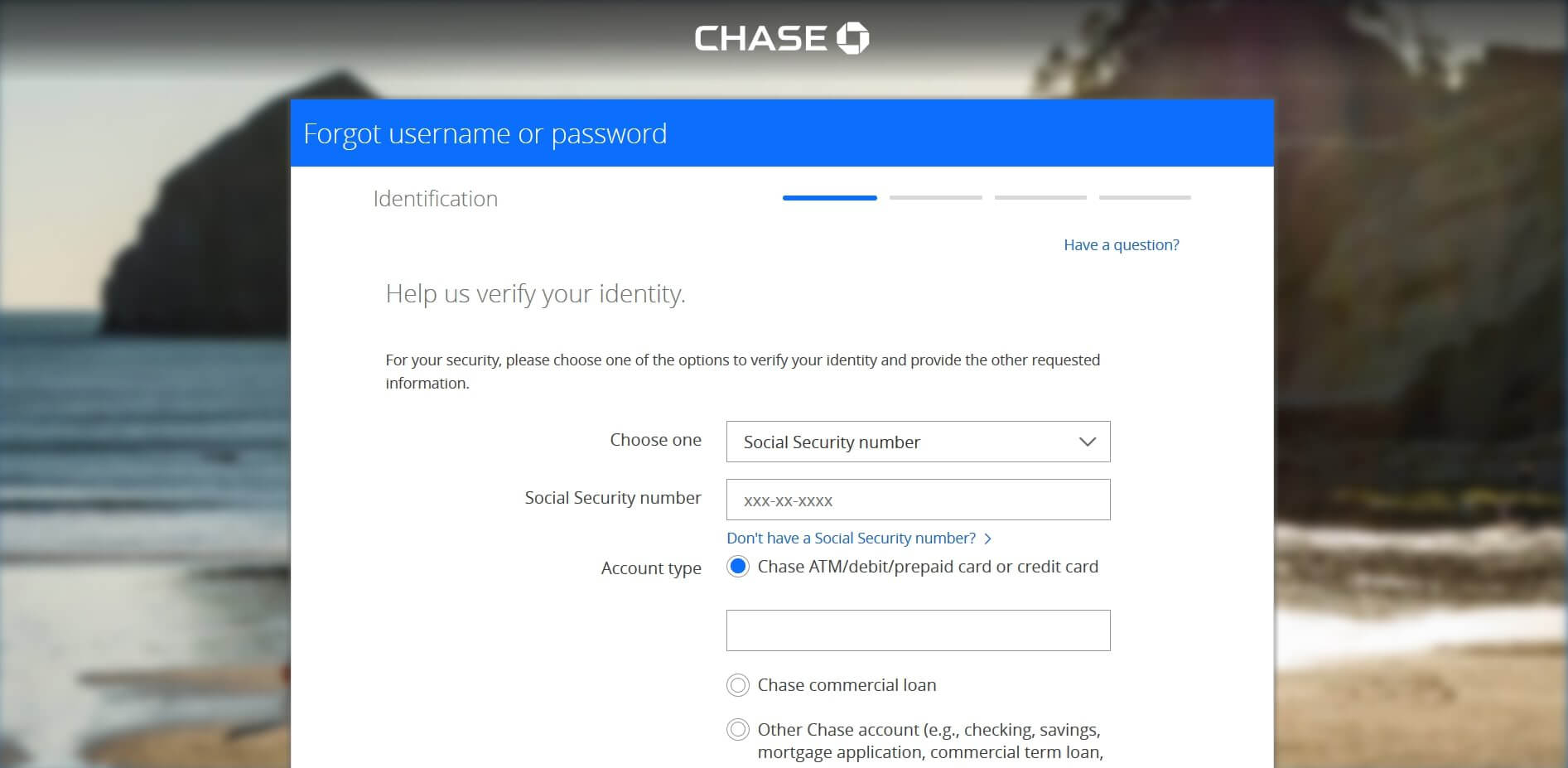}  \\ 
&&&\\
\rot{Browser 2} &\includegraphics[width=2.5cm,height=2.5cm,keepaspectratio] {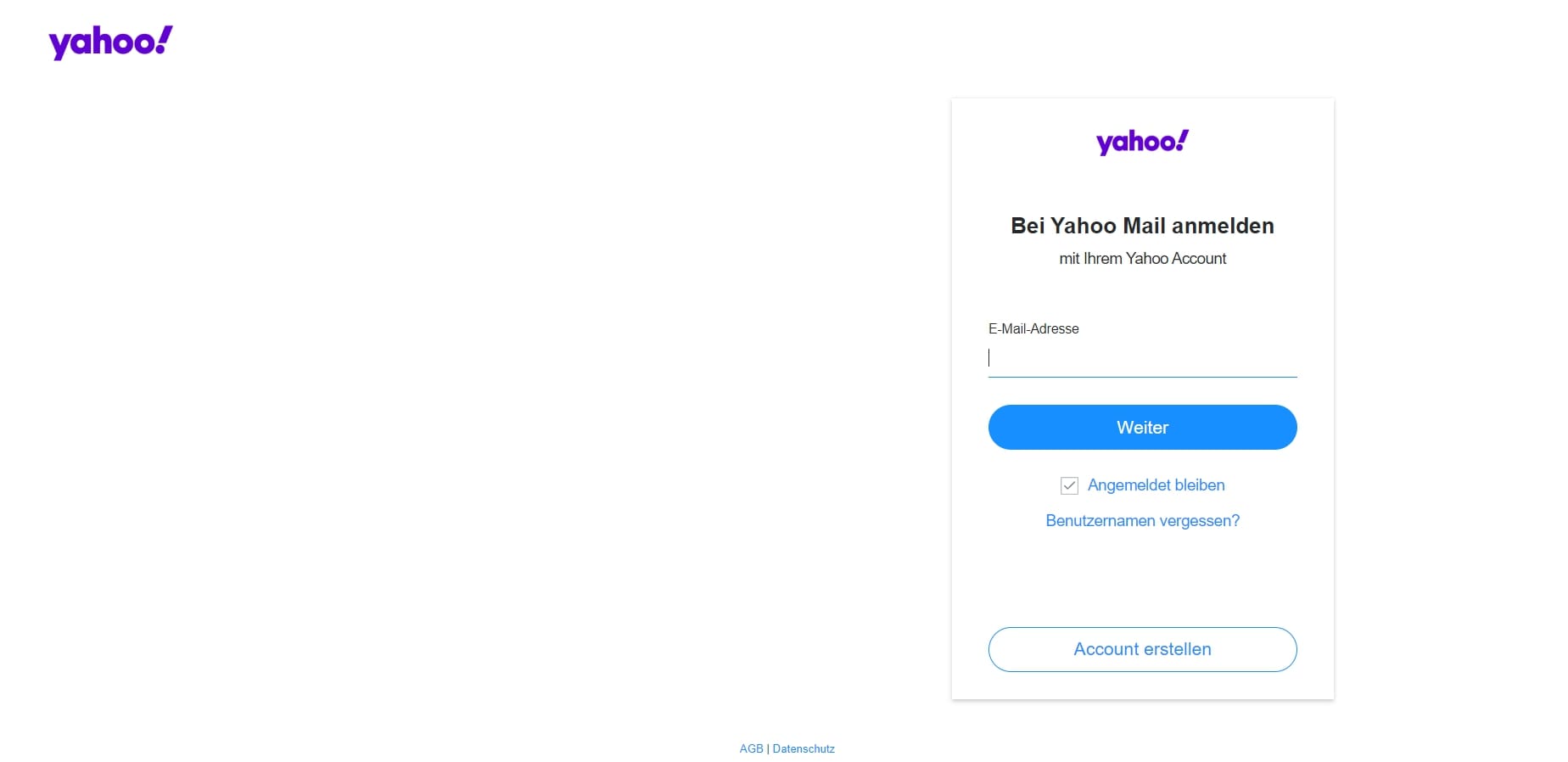} &
\includegraphics[width=2.5cm,height=2.5cm,keepaspectratio] {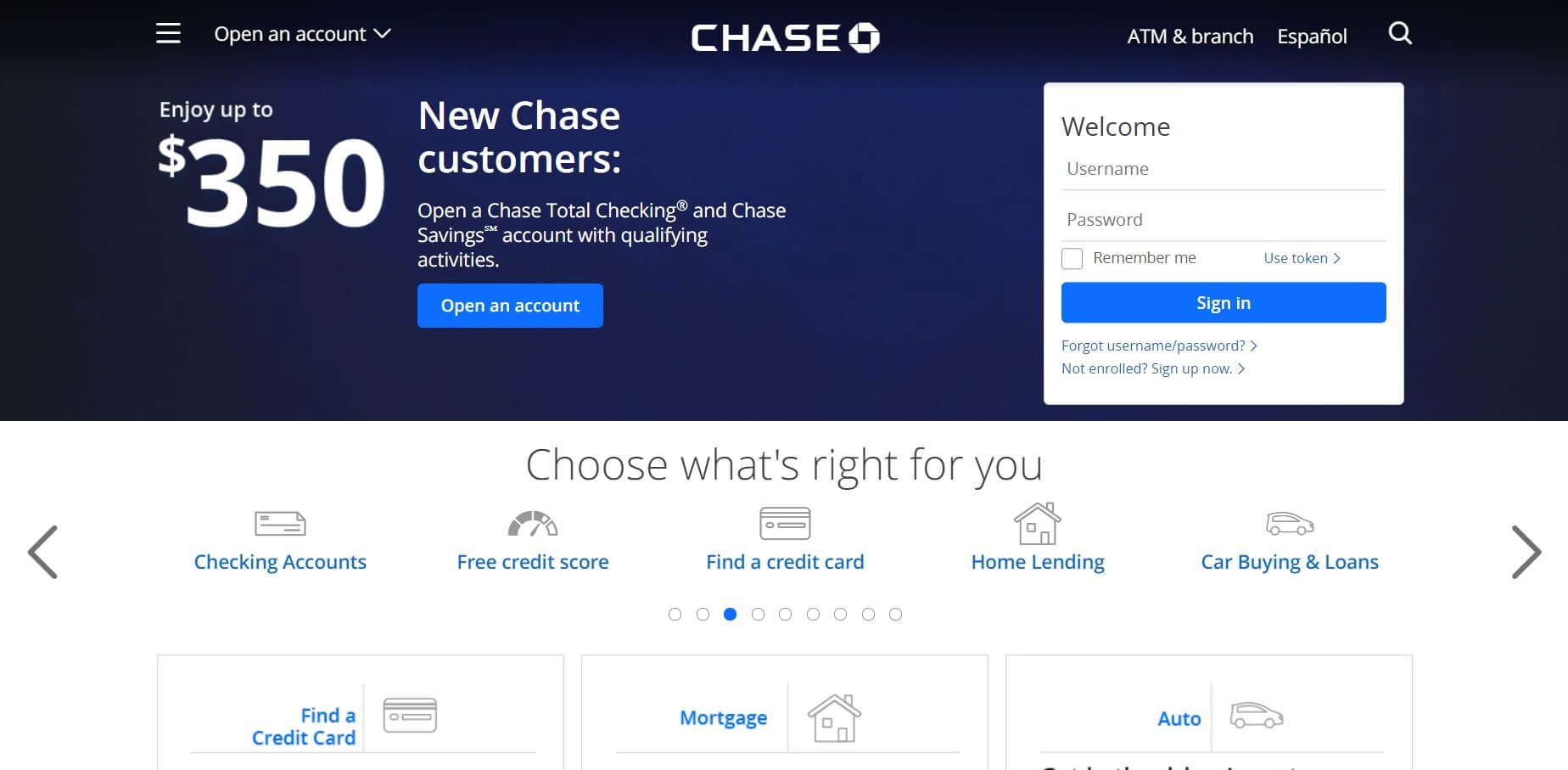} & \includegraphics[width=2.5cm,height=2.5cm,keepaspectratio] {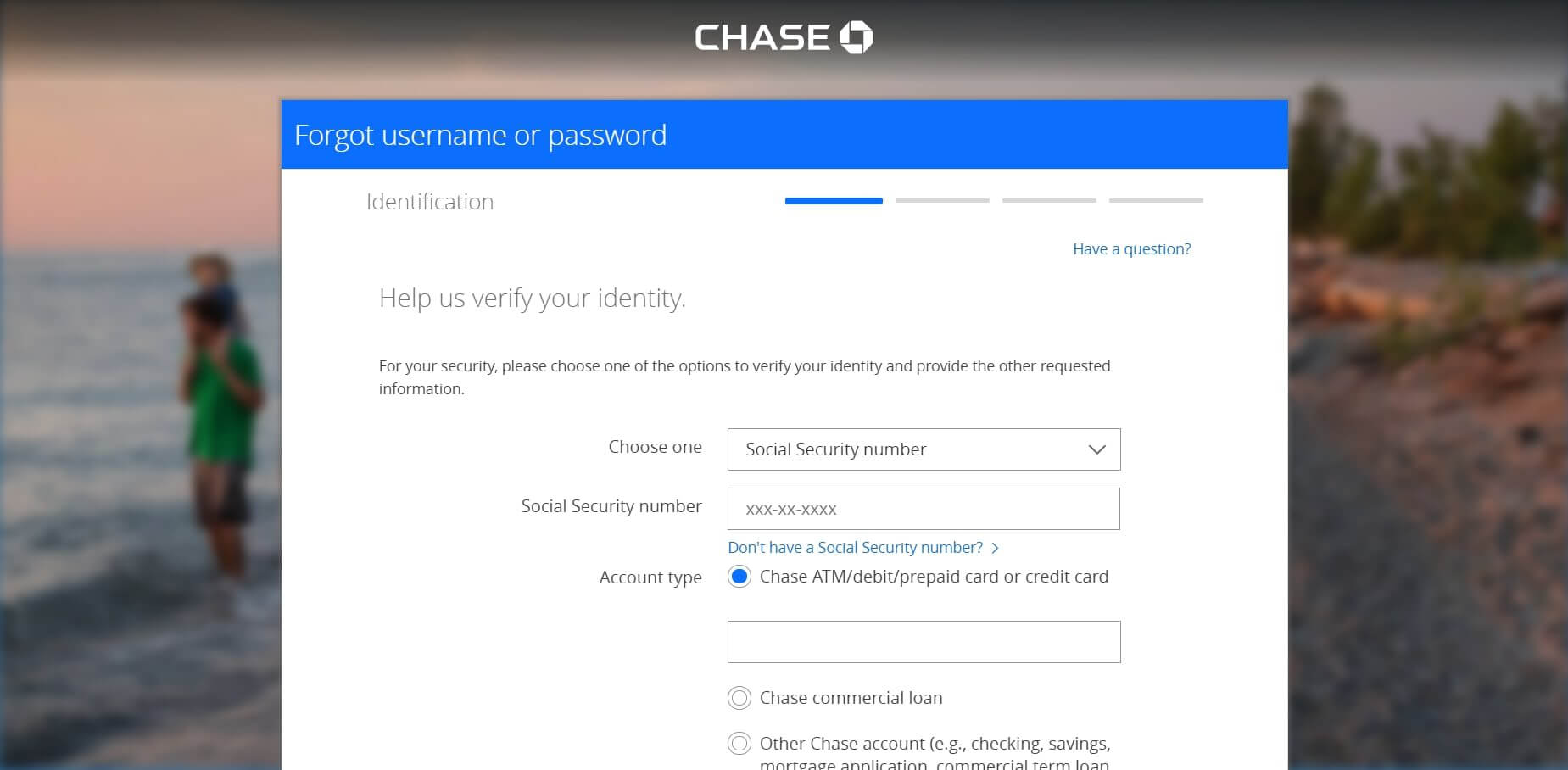}  \\
\end{tabular}} 
\captionof{figure}{Examples of the differences found between different browsers from the 50 pages used to evaluate the effect of browsers differences.} \label{tab:browsers_exampls}
\end{table}

\clearpage

\clearpage

\end{document}